\documentclass{aa}  
\usepackage{graphicx}
\usepackage{txfonts}
\usepackage[colorlinks=true, allcolors=blue, draft]{hyperref}
\usepackage[usenames, dvipsnames]{color}

\usepackage{amsmath}
\usepackage{amssymb}
\usepackage[colorinlistoftodos]{todonotes}
\usepackage{gensymb}

\usepackage{threeparttable}

\def\editorial{true}

\ifdefined\editorial
    \newcommand{\nb}[1]{{\color{red}TODO:#1}}
\else
    \newcommand{\nb}[1]{{}}
\fi

\newcommand{\Eq}[1]{Eq.~\ref{#1}}

\newcommand{\Fig}[1]{Fig.~\ref{#1}}

\newcommand{\Tab}[1]{Table~\ref{#1}}

\newcommand{\HI}{\ion{H}{I}}

\newcommand{\MgII}{\ion{Mg}{II}}

\newcommand{\OII}{\hbox{[{\rm O}{\sc \,ii}]}}
\newcommand{\Halpha}{\hbox{{\rm H}$\alpha$}}
\newcommand{\msun}{\hbox{M$_{\odot}$}}
\newcommand{\mpy}{\hbox{\msun~yr$^{-1}$}}
\newcommand{\Rhalf}{\hbox{$R_{\rm e}$}}

\def\galpak{\textsc{ GalPaK$^{\rm 3D}$}}
\def\galfit{\textsc{ GalFiT}}
\def\sersic{{S{\'e}rsic}}
\def\ngal{nine}

\begin{document}

   \title{ The MUSE Extremely Deep Field: Evidence for SFR-induced cores in dark-matter dominated galaxies at $z\simeq1$
   }

    \titlerunning{DM dominated SFGs}
    \authorrunning{Bouch\'e et al.}
 

   \author{Nicolas  F. Bouch\'e 
          \inst{1}\fnmsep\thanks{E-mail: nicolas.bouche@univ-lyon1.fr}
          \and
   	  Samuel Bera\inst{1}
          \and
          Davor Krajnovi\'c\inst{2}
          \and
          Eric Emsellem\inst{1,3}
              \and
          Wilfried Mercier\inst{4}
            \and
	  Joop Schaye\inst{5}
          \and
          Beno\^it Epinat\inst{6}
          \and
          Johan Richard\inst{1}
          \and
          Sebastiaan L. Zoutendijk\inst{5}
	  \and
	  Valentina Abril-Melgarejo\inst{6}
	  \and
           Jarle Brinchmann\inst{7,5}
          \and
         Roland Bacon\inst{1}
           \and
            Thierry Contini\inst{3}
          \and
         Leindert Boogaard\inst{8,5}
          \and          
          Lutz Wisotzki\inst{2}
          \and
          Michael Maseda\inst{6}
         \and
          Matthias Steinmetz\inst{2}
          }

      \institute{Univ Lyon, Univ Lyon1, Ens de Lyon, CNRS, Centre de Recherche Astrophysique de Lyon (CRAL) UMR5574, F-69230 Saint-Genis-Laval, France\\
              \email{nicolas.bouche@univ-lyon1.fr}
         \and
     Leibniz-Institut f\"ur Astrophysik Potsdam (AIP), An der Sternwarte 16, D-14482 Postdam, Germany
     \and 
     ESO, Karl-Schwarzschild-Strasse 2., D-85748 Garching b. M\"unchen, Germany
       \and
     Institut de Recherche en Astrophysique et Plan\'etologie (IRAP), Universit\'e de Toulouse, CNRS, UPS, F-31400 Toulouse, France
      \and
    Leiden Observatory, Leiden University, P.O. Box 9513, 2300 RA Leiden, The Netherlands
    \and
    Aix Marseille Univ, CNRS, CNES, LAM, Marseille, France
    \and
     Instituto de Astrofísica e Ci\^encias do Espa\c{c}o, Universidade do Porto, CAUP, Rua das Estrelas, 4150-762 Porto, Portugal 
     \and
     Max Planck Institute for Astronomy, K\"{o}nigstuhl 17, 69117 Heidelberg, Germany
            }

   \date{Received --; accepted --}

    \titlerunning{The dark-matter of SFGs}
     \authorrunning{Bouché et al.}

\abstract{Disc-halo decompositions  $z=1-2$ star-forming galaxies (SFGs) at $z>1$
  are often limited to massive galaxies ($M_\star>10^{10}$~\msun) and rely on either deep  integral field spectroscopy (IFS) data or  stacking analyses. 
}{We present a study of the dark-matter (DM) content of nine $z\approx1$ SFGs  selected among the brightest \OII\ emitters  in the deepest Multi-Unit Spectrograph Explorer (MUSE) field to date, namely the 140hr MUSE Extremely Deep Field. These SFGs
have low stellar masses, ranging from $10^{8.5}$ to $10^{10.5}$ M$_\odot$.}
{We analyzed the kinematics with a 3D modeling approach, which allowed us to measure individual rotation curves to $\approx3$ times the half-light radius $\Rhalf$.
We performed  disk-halo decompositions  on their \OII\ emission line with a 3D parametric model.
The disk-halo decomposition  includes a stellar, DM, gas, and occasionally a bulge component. 
The DM component  primarily uses the generalized $\alpha,\beta,\gamma$   profile or  a Navarro-Frenk-White (NFW)  profile.
}{The disk stellar masses $M_\star$ obtained from the \OII\ disk-halo decomposition    agree with the values inferred from the spectral energy distributions.
While the rotation curves show diverse shapes, ranging from rising to declining at large radii, the DM fractions within the half-light radius $f_{\rm DM}(<\Rhalf)$ are found to be 60\% to 95\%, extending to lower masses (densities) recent results who found low DM fractions in SFGs with $M_\star>10^{10}$~\msun. The DM halos show constant surface densities of $\sim100$~\msun~pc$^{-2}$.
For isolated galaxies, half of the sample shows a strong preference for cored over cuspy DM profiles.
The presence of  DM cores appears to be related to galaxies with  low stellar-to-halo mass ratio, $\log M_\star/M_{\rm vir}\approx-2.5$. In addition,
 the cuspiness of the DM profiles is found to be a strong function of the recent star-formation activity.
} {We  measured the properties of DM halos on scales from 1 to 15 kpc, put constraints on the $z>0$ $c_{\rm vir}-M_{\rm vir}$ scaling relation, and unveiled the cored nature of DM halos in some $z\simeq1$ SFGs. These results support feedback-induced core formation in the cold dark matter context.}
 

   \maketitle
%

\section{Introduction}
 
 The universe's matter content is dominated by elusive dark matter (DM), which has been one of the main topics in astronomical research. The idea of a dark or invisible mass was proposed numerous times based on the motions of stars in the Milky Way disk \citep{OortJ_32a}, the motion of galaxies in the Coma cluster \citep{ZwickyF_33a}, and by the lesser known argument made by \citet{PeeblesP_67a} using an upper limit on the mean mass density of galaxies from the average spectrum of galaxies (i.e., from the night-sky brightness). Nonetheless, the concept of DM became part of mainstream research only in the 1970s, based on the remarkable fact that the rotation curves (RC) of massive galaxies remain flat at large galactocentric distances \citep{RubinV_70a}. It was quickly realized that these flat rotation curves at large radii could not be explained by the Newtonian gravity of the visible matter alone, but instead implied the presence of an unobserved mass component attributed to a DM halo.
 
Today, the cold-DM (CDM) framework in which the large-scale structure originates from the growth of the initial density fluctuations \citep{PeeblesP_70a,PeeblesP_74a} is very successful in reproducing the large-scale structure \citep[e.g.,][]{SpringelV_06a}. However, understanding the nature and properties of DM on galactic scales remains one of the greatest challenges of modern physics and cosmology \citep[see ][for a review]{BullockJ_17a}.    

In this context, disentangling and understanding the relative distributions of baryons and dark matter in galaxies is still best achieved from a careful analysis of galaxies’ RCs on galactic scales. At redshift $z=0$, this type of analysis is   mature  with a wealth of studies  published in the past 20-30 years, using a variety of dynamics tracers such as \HI\ \citep[e.g.][]{deBlokW_97a,deBlokW_01a,VandenBoschF_00a},  \Halpha\ in the GHASP survey \citep{SpanoM_08a,KorsagaM_18a,KorsagaM_19b} or   a combination of \HI{} \& \Halpha\ as in the recent SPARC sample \citep{AllaertF_17a,KatzH_17a,LiLelli_20a} and the DiskMass survey \citep{BershadyM_10a,MartinssonT_13a}. These studies have shown that, in  low surface brightness (LSB) galaxies, the DM profiles  have a flat density inner ``core," contrary to the expectations from DM-only simulations that DM haloes ought to have a steep central density profiles or ``cusp" \citep[e.g.][NFW]{NavarroJ_97a}.
This cusp-core debate may be resolved  within CDM with feedback processes \citep[e.g.][]{NavarroJ_96b,PontzenA_12a,TeyssierR_13a,DiCintioA_14a,LazarA_20a,FreundlichJ_20a}  transforming cusps into cores~\footnote{Recently,    \citet{PinedaJ_17a} argued that NFW profiles can be mistaken as cores when the PSF/beam is not taken into account.}, a process that could be already present at $z=1$ \citep{TolletE_16a}.
DM-only simulations in the $\Lambda$CDM context have made clear predictions for the properties of DM halos, such as their concentration and  evolution  
 \citep[e.g.][]{BullockJ_01b,EkeV_01a,WechslerR_02a,DuffyM_08a,LudlowA_14a,DuttonA_14a,CorreaC_15c}, but the $c-M$ relation remains untested  beyond the local universe in SFGs \citep[e.g.][]{AllaertF_17a,KatzH_17a}. 
 
 At high redshifts,  where 21cm observations are not yet available, in order to measure the DM content of high-redshift galaxies, one must measure the kinematics in the outskirts of individual star-forming galaxies (SFGs) using nebular lines (e.g. \Halpha), at radii up to 10-15 kpc (2-3 times the half-light radius \Rhalf) where the signal-to-noise ratio (S/N) per spaxel drops approximately exponentially and quickly falls below unity. 
Disk-halo decompositions have proven to be possible at $z\simeq2$ in the pioneering work of \citet{GenzelR_17a} using very deep ($>30$ hr) near-IR integral field spectroscopy (IFS) on a small sample of six massive star-forming galaxies (SFGs).  Exploring lower mass SFGs, this exercise requires a stacking approach \citep[as in][]{LangP_17a,TileyA_19b} or deep IFS observations  \citep[as in][]{GenzelR_20a}. These studies of massive SFGs with $M_\star>10^{11}\msun$ showed that RCs are declining at large radii, indicative of a low DM fraction within \Rhalf; see also \citet{WuytsS_16a,UblerH_17a,AbrilV_21a} for dynamical estimates of DM fractions.
 
 Recently, 3D algorithms such as \galpak\ \citep{BoucheN_15a} or \textsc{$^{\rm 3D}$Barolo} \citep{DiTeodoroE_15a} have pushed the limits of what can be achieved at high-redshifts.
 For instance, one can study the kinematics of low mass SFGs, down to $10^8$ \msun\  \citep[as in][]{BoucheN_21a} in the regime of low S/Ns
 or study the kinematics of SFGs at large galactic radii $\sim3\times \Rhalf$ as in \citet{SharmaG_21a}, when combined with stacking techniques. Most relevant for this paper,
disk-halo decompositions of distant galaxies have been performed with \textsc{$^{\rm 3D}$Barolo} at $z\simeq4$ on bright submm [CII] ALMA sources \citep{RizzoF_20a,NeelemanM_20a,FraternaliF_21a}.
In addition, when used in combination with stacking or lensing,   3D algorithms are powerful tools to extract resolved kinematics at very high-redshifts as in \citet{RizzoF_21a}.

 This paper aims to show that a disk-halo decomposition can be achieved for {\it individual} low-mass SFGs at intermediate redshifts ($0.6<z<1.1$) 
 using the \galpak\ algorithm combined with the deepest (140hr) Multi-Unit Spectroscopic Explorer \citep[MUSE][]{BaconR_10a} data obtained on the {\it Hubble} Ultra Deep Field  (HUDF) and presented in \citet{BaconR_21a}.
 We show that rotation curves can be constrained up to 3 \Rhalf\, thanks to the 3D modeling approach on these deep IFU data. 
 This paper is organized as follows.
 In section~\ref{section:sample}, we present the sample used here.
In section~\ref{section:methodology}, we present our methodology.
In section~\ref{section:results}, we present our results. 
 Finally, we present our conclusions in section~\ref{section:conclusions}.
 
  Throughout this paper, we use a `Planck 2015'  cosmology \citep{Planck2015} with $\Omega_{\rm M}=0.307$, $\Lambda=0.693$, $H_0=67.7$ km/s/Mpc, yielding 8.23 physical kpc/arcsec at $z=1$, and $\Delta_{\rm vir}=157.2$. We also consistently use `log' for the base-10 logarithm. Error bars are 95\%\ confidence intervals (2$\sigma$), unless noted otherwise.

\section{Sample}
\label{section:sample}

In this paper, we selected \ngal\ \OII\ emitters from the recent MUSE eXtremely Deep Field (MXDF) region of the HUDF. The MXDF  
consists of a single MUSE field observed  within the MUSE observations of the HUDF \citep{BaconR_17a} taken in 2018-2019 (PI. R. Bacon; 1101.A-0127) with the dedicated VLT GALACSI/Ground-Layer Adaptive Optics (AO) facility for a total of 140 hours of integration. 
The MXDF field  is thus located within the 9sq. arcmin mosaic observations (at 10hr depth) and overlaps with the deep 30~hr  `UDF-10' region, as described in \citet{BaconR_21a} (their Fig.1).
The MXDF was observed with a series of 25 min exposures, each rotated by a few degrees  yielding a  final field of view that is approximately circular with radius  41\arcsec, where the deepest 140hr are contained within the central  31\arcsec\ \citep[see][for details]{BaconR_21a}. Thanks to the AO, the resulting point-spread function (PSF) full-width-at-half-max (FWHM) ranges from $\approx 0.6$\arcsec\ at 5000\AA\ to 0.4\arcsec at 9000\AA.

The sample of  \OII\ emitters was selected from the mosaic catalogue \citep{InamiH_17a} where we chose galaxies with the highest S/N per spaxel in \OII\  that were not face-on  \citep[using the \OII\ inclinations estimated in][]{BoucheN_21a}. From the catalog, we found 9 galaxies matching these criteria,  listed in Table~\ref{tab:galaxies}, with some 
reaching S/N$_{\rm pix}\sim100$ in the central spaxel.
All galaxies but one are contained within the deepest 140hr MXDF circle of 31\arcsec. One galaxy, ID3, has only 24hr of integration in the MXDF dataset, but is fortuitously located in a deep stripe of the UDF10 region \citep{BaconR_17a} leading to a total of 42hr integration. 

These galaxies have redshifts ranging from 0.6 to 1.1, and have stellar masses   from $ M_\star=10^{8.9}~\msun$ to $M_\star=10^{10.3}~\msun$ with SFRs from 1 to 5 \mpy. The stellar masses and SFR were determined from spectral energy distribution (SED) fits with the \textsc{Magphys} \citep{daCunha_15a} software
on the HST photometry \citep[as in][]{MasedaM_17a,BaconR_17a,BaconR_21a} using a \citet{ChabrierG_03a} initial mass function (IMF). 
 Uncertainties on these quantities are obtained from the marginalized posterior probability distributions given by Magphys under the assumption of smooth star formation histories with additional random bursts  \citep[][and references therein]{daCunha_08a}.
The main properties of these galaxies are listed in Table~\ref{tab:galaxies}.

\begin{figure}
\centering
\includegraphics[width=0.49\textwidth]{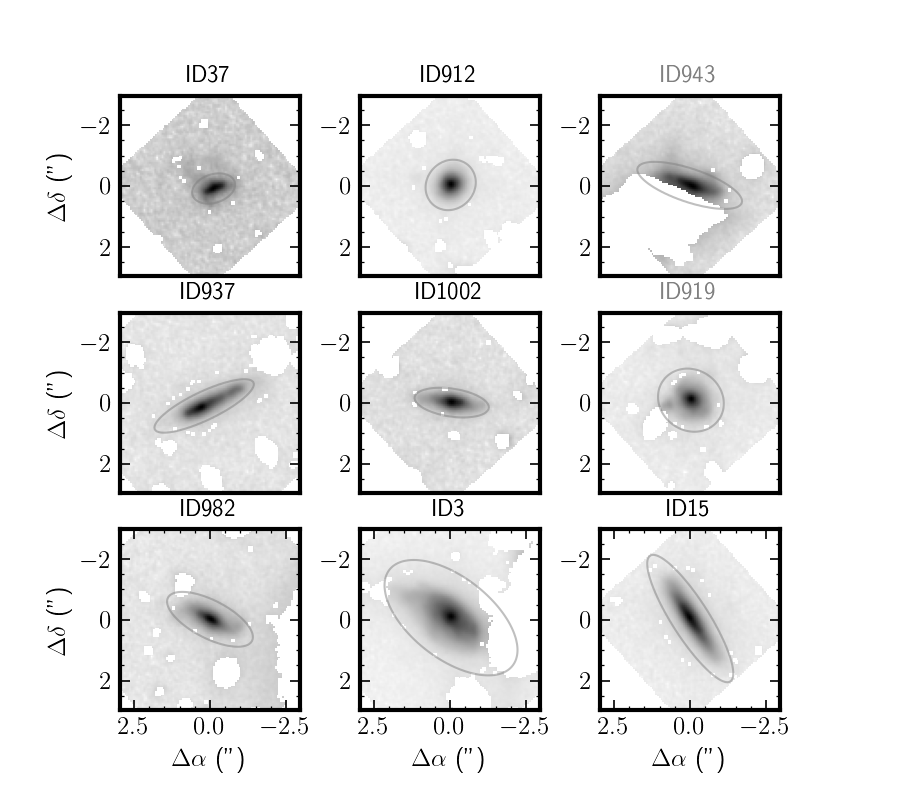}
\caption{{\it HST}/F160W  postage stamps of \ngal\ \OII\ emitters from the MXDF used in this study  (ordered by increasing $M_\star$ from top left to bottom right).
Background and foreground objects have been masked.
The ellipse shows a constant isophote
 on the models.
}
\label{fig:f160w}
\end{figure}

In \Fig{fig:f160w}, we show the {\it HST}/F160W images of the \ngal\ SFGs where the background and foreground objects have been masked.
This figure shows that not all galaxies are regular and axisymmetric. In particular,  ID943 has a large companion 1" away (masked) and ID919 has a small satellite at the same redshift, as seen in \Fig{fig:examples}, and both of these galaxies show signs of tidal tails. ID3 has also a companion $2\farcs2$ away (masked).

Using  \citet{SersicJ_63a} fits  with the \galfit\ tool \citep{PengC_02a} on the HST/F160W WFC3 images, we find that these galaxies have surface brightness profiles consistent with an exponential. More specifically, we modeled the flux distribution using a single Sérsic profile with its total magnitude, effective radius, Sérsic index $n$, Position Angle (PA) and axis ratio ($b/a$) as free parameters, in combination with a sky component to take into account the sky background in the HST images. We used the F160W WFC3 images since these probe older stellar populations which better trace the underlying mass distribution and also because they have the best spatial resolution available. In order to improve the fits, we additionally masked the nearby objects appearing in the HST  segmentation maps.

In order to get a measure of the galaxies bulge to total ratio (B/T), we remodeled them performing a multicomponent decomposition. This time, we used a combination of an exponential disk (with fixed $n=1$) with a de Vaucouleurs bulge (with fixed $n=4$, PA and $b/a$) on the same masked HST images.

\begin{table*}
\centering
\caption{Sample of star-forming galaxies selected in the HUDF observed in the MXDF. Columns are
(1) Galaxy MUSE ID from \citet{InamiH_17a}; (2) redshift; (3) Exposure time (hr); (4) Maximum S/N in the brightest \OII\ spaxel; (5) Stellar mass $M_\star$ from SED fitting with \textsc{Magphys}; (6) SFR from SED fitting; (7) F775W magnitude; 
 (8) Sersic $n_\star$ from \citep{vanderWelA_14a} and from our \galfit\ fits both on HST/F160W WFC3;
 (9) Inclination $i_{\star}$ from HST/F160W WFC3;
 (10) B/T ratio at 2 \Rhalf\  from a two component \galfit\ fit to HST/F160W;
 (11) \Rhalf\ in kpc measured from HST/F160W from a single component \galfit\ fit to the HST/F160W data;
(12) ID in the \citet{RafelskiM_15a} catalog. The quoted erros are $1\sigma$ (68\%).  
}
\label{tab:galaxies} 
\begin{threeparttable}
\begin{tabular}{rrrrrrrrrrrrr}
ID & $z$ &  $t_{\rm exp}$  & S/N$_{\rm max}$ & $\log M_{\star}$ & SFR & m$_{\rm F775W}$ & $n_\star$ & $i_{\star}$ & B/T & \Rhalf$_\star$\ & RAFID\\
& & [hr]  & & [\msun] & [\mpy] & & & [deg.] & & [kpc] & \\
(1) & (2) & (3) & (4) & (5) & (6) & (7) & (8) & (9) & (10) & (11)& (12) \\
\hline
3 & 0.62 &  42\tnote{a} & 15 & 10.08$^{+0.02}_{-0.06}$ & 2.62$^{+0.80}_{-1.2}$ & 21.68 & 0.87/0.81$\pm0.06$ & 64$\pm5$ &  0.04 & 5.61$\pm0.04$ &24353 \\
15 & 0.67 & 136 & 9 & 10.23$^{+0.02}_{-0.10}$ 	      &   0.99$^{+2.53}_{-0.29}$    & 23.30 & 0.86/0.74$\pm0.06$ & 86$\pm5$  & $<0.01$ & 5.75$\pm0.03$ & 10345 \\
37 & 0.98 & 136 & 61 & 8.87$^{+0.06}_{-0.17}$ 		& 0.80$^{+1.22}_{-0.18}$ & 24.65 & 1.15/0.97$\pm0.08$& 56$\pm5$ & $<0.01$ &  3.54$\pm0.06$ & 9791 \\
912 & 0.62 & 136 & 124 & 9.19$^{+0.19}_{-0.06}$ 	& 1.29$^{+1.60}_{-0.30}$ & 22.99 & 0.97/0.93$\pm0.06$ & 27$\pm5$ & $<0.01$ &  1.87$\pm0.01$ & 5082\\
919 & 1.10 & 136 & 98 & 9.83$^{+0.02}_{-0.06}$ 	& 4.28$^{+0.52}_{-0.55}$ & 23.23 & 1.54/1.20$\pm0.06$ & 28$\pm5$ & $0.21$ & 3.28$\pm0.03$ &  23037\\
937 & 0.73 & 44 & 18 & 9.35$^{+0.08}_{-0.06}$ 		& 1.50$^{+0.34}_{-0.36}$ & 23.49 & 1.13/0.96$\pm0.06$ & 82$\pm5$ & $0.16$ & 5.39$\pm0.06$ & 7734\\
943 & 0.66 & 136 & 32 & 9.28$^{+0.06}_{-0.06}$ 	& 1.00$^{+0.57}_{-0.40}$ & 23.63 & 0.86/0.59$\pm0.06$ &  77$\pm5$ & $<0.01$ & 5.09$\pm0.07$ & 22950\\
982 & 1.10 & 92 & 22 & 9.75$^{+0.09}_{-0.07}$ 		& 3.25$^{+0.84}_{-0.97}$ & 23.95 & 1.47/1.02$\pm0.06$ &70$\pm5$ & $0.43$ & 4.79$\pm0.05$ & 22509 \\
1002 & 0.99 & 47 & 35 & 9.43$^{+0.02}_{-0.075}$	& 1.18$^{+0.02}_{-0.26}$ & 24.15 & 1.29/1.05$\pm0.06$ & 73$\pm5$ & $0.21$ & 4.15$\pm0.03$ & 25458\\
\end{tabular}
\begin{tablenotes}
\item[a] This galaxy is located on the edge of the MXDF deep footprint, and we use the deeper data in the UDF10 pointing \citep{BaconR_17a}.
\end{tablenotes}
\end{threeparttable}
\end{table*}

\begin{table*}[h]
\centering
\caption{Kinematics results   from our \galpak\ fits with our disk-halo decomposition.
(1) Galaxy ID; (2)  \sersic\ index from MUSE data (\OII); (3) Inclination from \OII; (4) Half-light radius from \OII; 
 (5) Model used for the disk-halo decomposition; 
 (6) Velocity dispersion $\sigma_0$ (see text);
 (7) Virial velocity $V_{\rm vir}$ for the DM halo component;
  (8) Halo concentration parameter $c_{\rm vir}$;
(9) Stellar mass $M_\star$ from \galpak;
(10) Halo mass $M_{\rm vir}$;
(11) logarithm of the evidence $\cal Z$.
The quoted errors are $2\sigma$ (95\%).
\label{tab:results} 
}
\begin{threeparttable}
\begin{tabular}{rrrrrrrrrrrr}
ID &    $n_{\rm O2}$ & $i_{\rm O2}$ & $\Rhalf_{\rm O2}$ & model &  $\sigma_0$ & $V_{\rm vir}$  & $c_{\rm vir,-2}$ & $\log M_{\star}/\msun$ & $\log M_{\rm vir}/\msun$ &   $\ln \cal Z$ \\
& & [deg.] &   [kpc] &  & [km/s] &  [km/s] & &  &  \\
(1) & (2) & (3) & (4) & (5) & (6) & (7) & (8) & (9) & (10) & (11) \\
\hline
3 &     0.75$\pm0.05$ & 65$\pm1$ & 6.0$\pm0.1$ &  DC14.MGE & 27$\pm3$      & 122$\pm6$ 		& 12$\pm2$  & 10.02$\pm0.12$ & 11.71$\pm 0.06$  &17317\tnote{d} & \\
15 &  0.6$\pm0.1$  &  63$\pm1$ &   6.4$\pm0.2$ &  DC14.MGE  & 35$\pm2$ 	& 115$\pm7$ 			&  15$\pm2$     & 10.19$\pm0.15$  & 11.61$\pm 0.09$ &  8019 & \\
37 &    0.9$\pm0.1$ & 55$\pm1$ &  3.7$\pm0.1$ &   DC14.MGE & 25$\pm3$   &  143$\pm20$			 & 6.6$\pm1$      & 8.89$\pm 0.15$ & 11.80$\pm 0.19 $ & 9514 &   \\
912\tnote{b} &    0.5$\pm0.1$ & 35 & 2.4$\pm0.1$ & DC14.MGE & 27$\pm3$ 	& 144$\pm16$ 			& 17$\pm2$   & 9.34$\pm 0.11$ & 11.92$\pm 0.14$  &  8829 &\\
919\tnote{b} &  0.6$\pm0.1$ &  35 & 3.6$\pm0.1$  & DC14.Freeman &  39$\pm$2 & 142$\pm$5				 & 16$\pm1$ & 9.01$\pm0.14$ & 11.76$\pm0.05$ &  27552\tnote{d} &  \\
937 &  0.5$\pm0.1$ & 84$\pm1$ & 7.0$\pm0.1$  & NFW.MGE  & 18$\pm3$ 	          & 96$\pm4$			  &  9.0$\pm0.8$  & 9.29$\pm0.11$\tnote{c}  & 11.62$\pm0.08$ &  8632 & \\
943 &  1.0$\pm0.1$  & 72$\pm1$ & 4.8$\pm0.1$ & DC14.MGE & 48$\pm2$        & 89$\pm3$ 			 & 14.5$\pm3.7$       & 9.97$\pm0.13$  & 11.27$\pm0.08$  & 15374\tnote{d} & \\
982 &   1.2$\pm0.1$ & 63$\pm1$ & 5.4$\pm0.3$ &   DC14.MGE   & 35$\pm3$      &  203$\pm22$ 			& 7.0$\pm0.4$   & 9.54$\pm0.14$  &  12.22$\pm0.09$ &  6736 & \\
1002 &  0.95$\pm0.5$  & 70$\pm1$ &  3.9$\pm0.1$ & DC14.Freeman & 36$\pm2$      &   122$\pm15$    	&   7.1$\pm1.3$ &  9.66$\pm0.13$ & 11.59$\pm0.16$  & 8151 &    \\
\end{tabular}
\begin{tablenotes}
\item[b] The inclination for this galaxy ($i_{\OII}$ was $\sim45$\degree) is restricted to $i_\star<35$\degree.
\item[c] The disk mass was restricted to the SED mass $\log M_\star/{\rm M_{\odot}}\pm0.2$dex.
\item[d] Large residuals are  associated with galaxies with companions.
\end{tablenotes}
\end{threeparttable}
\end{table*}

\section{Methodology}
\label{section:methodology}


In order to measure the DM content of high-redshift galaxies, one must measure individual RCs in the outskirts of individual SFGs, at radii up to 10-15 kpc (2-3 Re) where the S/N per spaxel falls below unity. This is possible thanks to the combination of the deep MUSE data and  3D analysis tools such as \galpak\ \citep{BoucheN_15a}.
In \S~\ref{section:kinematics}, we describe the 3D algorithm and our parameterization designed to analyze the shape of the RCs in order to characterize the outer slope of RCs.
In \S~\ref{section:disk:halo}, we describe our methodology for performing a full disk-halo decomposition directly to the 3D MUSE data-cubes.

\subsection{Simultaneous measurements of the morphology and kinematics from 3D modeling}
\label{section:kinematics}

The \galpak\ algorithm \citep{BoucheN_15a}  compares 3D parametric models directly to the 3D data, taking into account the instrumental resolution and PSF\footnote{See \url{http://galpak3d.univ-lyon1.fr}.}.  Briefly, \galpak\  
performs a parametric fit of the 3D emission line data, simultaneously fitting the morphology and kinematics using a 3D ($x,y,\lambda$) disk model, which specifies
the  morphology and kinematic parametric profiles.
\galpak\ convolves the 3D model with the Point Spread Function and Line Spread Function, which implies that all the fitted parameters are ``intrinsic'' (i.e., corrected for beam smearing and instrumental effects).

For the morphology, the model assumes a \citet{SersicJ_63a} surface brightness profile $\Sigma(r)$, with \sersic\ index $n$.
The disk model is inclined to any given inclination $i$ and orientation or positional angle (P.A).
The thickness profile is taken to be Gaussian whose scale height $h_z$ is  0.15$\times\Rhalf$. 
For \OII\ emitters, as in this analysis, we add a global \OII\  doublet ratio $r_{\rm O2}$.

For the kinematics, the 3D model uses a parametric form for the rotation curve $v(r)$ and the dispersion $\sigma(r)$ profile as discussed in \citet{BoucheN_21a}. 
In order to assess the shape of the RCs,  we can use several  RC models which allow for a rising or declining RC, such as in  \citet{RixH_97a}, \citet{CourteauS_97a} or the 
universal RC  (URC)  of \citet[][PSS96]{PersicM_96a}. After experimentation, the latter is often our preferred choice because it has fewer parameter degeneracies. 
The URC of PSS96 has three parameters, the core radius $r_t$, the velocity $V_{\rm 2}$ (at $R_{\rm opt}\simeq2\Rhalf$) and the outer slope $\beta$ of $v(r)$. 

Finally, as described in \citet{BoucheN_15a,BoucheN_21a}, the   velocity dispersion profile  $\sigma_{\rm t}(r)$ consists of the combination of a thick disk $\sigma_{\rm thick}$, defined from the identity $\sigma_{\rm thick}(r)/v(r) =h_z/r$ \citep{GenzelR_06a,CresciG_09a} where $h_z$ is the disk thickness (taken to be $0.15\times\Rhalf$) and a dispersion floor, $\sigma_{0}$,   added in quadrature    \citep[similar to $\sigma_0$ in][]{GenzelR_06a,GenzelR_08a,ForsterSchreiberN_06a,ForsterSchreiberN_18a,CresciG_09a,WisnioskiE_15a,UblerH_19a}.

Altogether, this 3D model (hereafter `URC' model) has 13 parameters: $x_{\rm c}$, $y_{\rm c}$, $z_{\rm c}$,  $f_{\rm O2}$, $\Rhalf_{|\rm O2}$, $n_{\rm O2}$, $i_{\rm O2}$, P.A.$_{|\rm O2}$,  $r_t$, $V_{\rm 2}$ , $\beta$, $\sigma_{0}$ and the \OII\ doublet ratio $r_{\rm O2}$. We use flat priors on these parameters and fit them simultaneously with a   Bayesian Monte-Carlo algorithm.
 \galpak\ can use a variety of  Monte-Carlo algorithms and here we use the python version of \textsc{MultiNest} \citep{multinest} from \citet{BuchnerJ_14a} because it  is found to be very robust and insensitive to initial parameters. Moreover, it also provides the model evidence $\cal Z$. 
 
 There are several advantages to note here. The 3D algorithm \galpak\ allows to fit the kinematics and morphological parameters {\it simultaneously} and thus no prior information is required on the inclination~\footnote{The traditional  $i-V_{\rm max}$ degeneracy is broken using the morphological information, specifically the axis $b/a$ ratio.}.
 The agreement between HST-based  and MUSE-based inclinations is typically better than 7$^{\circ}$(rms) for galaxies with $25<i<80$ as demonstrated in  \citet{ContiniT_16a} and with mock data-cubes \citep{BoucheN_21a} derived from the Illustris ``NewGeneration 50 Mpc'' (TNG50) simulations \citep{NelsonD_19a,PillepichA_19a}. Nonetheless, we checked that the inclinations $i$ and \sersic\ $n$ parameters obtained from the \OII\ MUSE data  are consistent with those obtained from the {\it HST}/F160W images (see Tables~\ref{tab:galaxies}-\ref{tab:results}). We find good agreement  except for ID912 and ID919, which have the most face-on inclination with $i_{\star}<30^{\circ}$. For these two galaxies, we restrict the \galpak\ fits to $i_{\rm O2}<35^{\circ}$.  

\begin{figure*}
\centering
\includegraphics[width=0.9\textwidth]{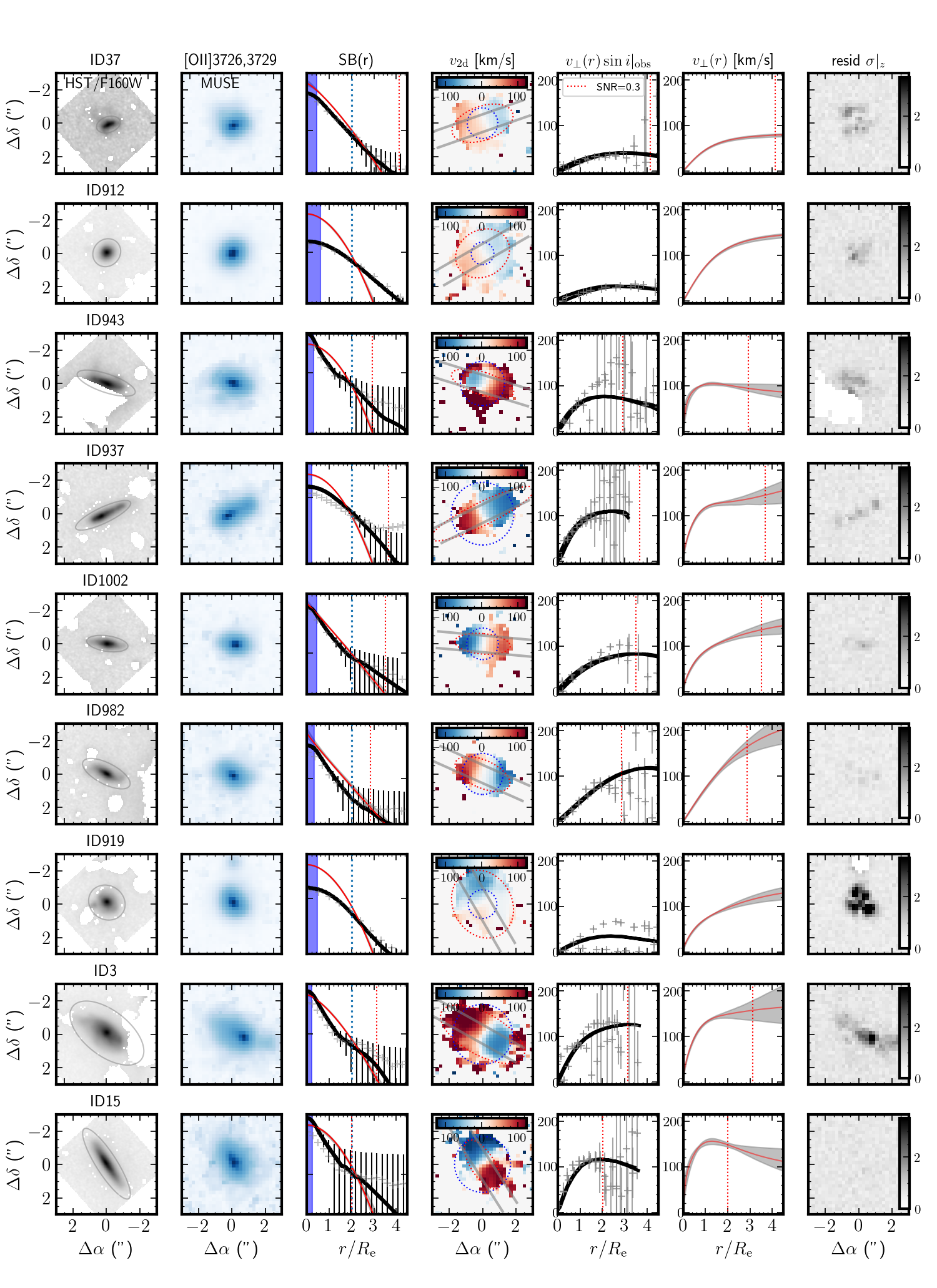}
\caption{For each galaxy, we show the stellar continuum from {\it HST}/F160W, the \OII\ flux map from MUSE, the \OII\ surface brightness profile (SB$(r)$), 
the observed projected velocity field ($v_{\rm 2d}$),
the observed 1d velocity profile $v_{\perp}\sin i$,
the  intrinsic  (i.e., deprojected, corrected for beam smearing) modeled rotation curve ($v_{\perp}$) using the URC model  of PSS96 
(see \S~\ref{section:kinematics}),
 and the residuals map obtained from the residual cube (see text).
The red solid lines show the intrinsic SB($r$) and $v_{\perp}$ model.
The solid black lines show the convolved SB profile and modeled 1d velocity profile.
The gray symbols represent the data extracted from the flux and velocity maps.
The blue vertical dotted lines represent $2\Rhalf$, while the red dotted lines  show the radius at which the S/N per spaxel reaches 0.3. 
\label{fig:examples}
}
\end{figure*} 

 \subsection{Disk-halo decompositions}
 \label{section:disk:halo}
 
In most general terms, a disk-halo decomposition of the rotation curve $v(r)$ is made of the combination of a dark-matter $v_{\rm dm}$, a stellar disk $v_{\star}$, a molecular $v_{\rm g, H_2}$  and an atomic a gas $v_{\rm g,\HI}$  component:
 \begin{equation}
 v^2_{\rm c}(r)=v_{\rm dm}^2(r)+v_{\star}^2(r)+v_{\rm g,H_2}^2+v_{\rm g,\HI}^2(r)\label{eq:diskhalo}.
 \end{equation}
 The mass profile for the molecular gas $\rm H_2$ component is negligible at low redshifts (due to the low gas fraction) \citep[e.g.][]{FrankB_16a}. At high-redshifts, the molecular gas follows the SFR profile (or \OII) \citep{LeroyA_08a,NelsonE_16a,WilmanD_20a}, and approximately the stellar component, and thus can become significant. However, because of the similar mass profile in molecular gas and stars, the two are inherently degenerate without direct CO measurements, currently inaccessible at our mass range. Depending on the molecular gas fractions, this component could be significant, but given that the molecular gas fractions in our redshift range $0.6-1.1$ are typically 30-50\%\ \citep[e.g.][]{FreundlichJ_19a}, this amounts to a systematic uncertainty of 0.1-0.15~dex on the mass of disk component.
 
 The neutral gas profile $v_{\rm g,\HI}$, however, is important given that (i), locally, it extends much further than the stellar component \citep[e.g.][]{MartinssonT_13a,WangJ_16a,WangJ_20a} and can extend up to 40 kpc using stacking techniques \citep{Ianja_18a}, and (ii) at $z\approx1$ several absorption line surveys show extended cool structures that extend up to 80 kpc \citep[e.g.][]{BoucheN_13a,BoucheN_16a,HoS_17a,HoS_19a,ZablJ_19a} traced by quasar \MgII\ absorption lines. 
 Because of the roughly constant surface density of \HI\ gas, this neutral gas contribution to $v(r)$ is important at large distances \citep[e.g.][]{AllaertF_17a}.
 
 Within the context of this paper, we have implemented a 3D disk-halo decomposition in \galpak, where the rotation curve is made of the combination of a dark-matter $v_{\rm dm}$, a disk $v_{\star}$,  a neutral gas $v_{\rm g}$  component (hereafter $v_{\rm g}\equiv v_{\rm g,\HI}$):
 \begin{equation}
 v^2_{\rm c}(r)=v_{\rm dm}^2(r)+v_{\star}^2(r)+v_{\rm g}^2(r)\label{eq:Vc}
 \end{equation}
 and in some cases   an additional central or `bulge' component $v_{\rm bg}$. For instance, ID982  has a B/T greater than 0.2 from the HST/F160W photometry (see Table~\ref{tab:galaxies}).
  For these, we add a bulge to the flux profile and a \citet{HernquistL_90a} component $v_{\rm bg}(r)$ to Eq.~\ref{eq:Vc} whose parameters are the \sersic\ index for the bulge $n_{\rm b}$ (taken between $n_{\rm b}=2$ and $n_{\rm b}=4$), the bulge kinematic mass ($v_{\rm max,b}$), the bulge radius $r_{\rm b}$ and the bulge-to-total (BT) ratio.

 The disk component $v_{\star}$ can be  modeled as a   \citet{FreemanK_70a}  disk suitable for exponential mass profiles
 and most of our galaxies have stellar \sersic\ indices $n_\star$ close to $n_\star\simeq1$ (see Table~\ref{tab:galaxies}).
  For a mass profile of any \sersic\ $n$, the rotation curve $v_{\star}(r)$ can be derived analytically \citep[e.g.][]{LimaNetoG_99a} or approximated  using the Multi-Gaussian Expansion (MGE) approach of the  spatial distribution \citep{EmsellemE_94a}, assuming axisymmetry, with a sufficiently high number of gaussian components to ensure a given accuracy (e.g. $< 1$\%) within $0.1<r/\Rhalf<20$. Here, we use the MGE approach where the shape of $v_\star(r)$ is determined by the \sersic\ $n_{\rm O2}$ index from the \OII\ SB profile, and the normalization of $v_\star$ is given by the disk mass $M_{\star}$, the sole free parameter.  Naturally,  this assumes that \OII\ traces mass, which  might not be appropriate.
Hence, when preferred by the data (i.e., with a better evidence), we relax this constraint and use a \citet{FreemanK_70a} disk ($n\equiv 1$) together with $n_{\rm O2}$ different than unity for the disk $v_{\star}$ component. For two galaxies, we selected this option (`DC14.Freeman`  in Table~\ref{tab:results}).

 The DM component $v_{\rm dm}(r)$ can be modeled as a generalized $\alpha-\beta-\gamma$  double power-law model \citep[][hereafter DC14]{JaffeW_83a, HernquistL_90a, ZhaoH_96a,DiCintioA_14a}, hereafter the Hernquist-Zhao profile:
  \begin{equation}
 \rho(r;\rho_s,r_s,\alpha,\beta,\gamma)=\frac{\rho_s}{{\left(\frac{r}{r_s}\right)^\gamma\left(1+\left(\frac{r}{r_s}\right)^\alpha\right)^{(\beta-\gamma)/\alpha}}} \label{eq:rho}
 \end{equation}
 where $r_s$ is the scale radius, $\rho_s$   the scale density,
 and $\alpha,\beta,\gamma$ are the shape parameters, with $\beta$ corresponding to the outer slope, $\gamma$ the inner slope and $\alpha$ the transition sharpness.
 
 The density  $\rho_s$ is set by the halo virial velocity $V_{\rm vir}$ (or halo mass $M_{\rm vir}$) and   following   DC14 $r_s$ can be recasted as
  \begin{equation}
 r_{-2}\equiv \left(\frac{2-\gamma}{\beta-2}\right)^{1/\alpha} r_s \label{eq:r-2}
 \end{equation}
 where  $r_{-2}$ the radius at which the logarithmic DM slope is $-2$~\footnote{For a NFW profile $r_{-2}$ is equal to $r_s$, and $c_{\rm vir}=R_{\rm vir}/r_s$.}.
 The concentration $c_{\rm vir}$ is defined as $c_{\rm vir,-2}\equiv R_{\rm vir}/r_{-2}$, where $R_{\rm vir}$ is the halo virial radius \cite[using the virial overdensity definition of][]{BryanG_98a}.

The shape parameters $\alpha,\beta,\gamma$ in Eq.~\ref{eq:rho} are a direct function of  the disk-to-halo mass ratio $\log X\equiv \log (M_{\star}/M_{\rm vir})$, 
in simulations with supernova feedback \citep[e.g. DC14,][]{TolletE_16a,LazarA_20a}. This parameter $X$ then uniquely determines the shape of the DM halo profile and its associated $v_{\rm dm}(r)$.
Hence, this DM profile $v_{\rm dm}(r)$ has three free parameters, namely $\log X, V_{\rm vir}$ and $ c_{-2}$.
Since we used the $\alpha(X),\beta(X),\gamma(X)$ parametrisation with $\log X$ from \citet{DiCintioA_14a} (their Eq.3), we refer to this model as `DC14', and  refer the reader to DC14, \citet{AllaertF_17a}, and \citet{KatzH_17a} for the details.

We also use a NFW  DM profile, which is a special case of Hernquist-Zhao profiles with $\alpha,\beta,\gamma=(1,3,1)$~\footnote{It is important to note that a pseudo-isothermal profile has $\alpha,\beta,\gamma=(2,2,0)$,
 the modified NFW  \citep[used in][]{SonnenfeldA_15a,WassermanA_18a,GenzelR_20a} has $\alpha,\beta,\gamma=(1,3,\gamma$) and the \citet{DekelA_17a} profile has $\alpha,\beta,\gamma=(0.5,3.5,a)$ \citep[see][]{FreundlichJ_20b}. Other DM profiles include the \citet{BurkertA_95a}, the \citet{EinastoJ_65a} profiles and the core-NFW profile of \citet{ReadJ_16a}. }.  In Appendix~C, we  relax the DC14 assumption and explore Hernquist-Zhao DM profiles with unconstrained $\alpha,\beta,\gamma$ parameters (\Fig{fig:DC14:Zhao}), and refer this model as the `Zhao' models. The shape of these DM profiles are not linked to $\log (M_{\star}/M_{\rm vir})$ as in DC14, and thus require a prior input for $M_\star$.

The gas component $v_{\rm g}$ is made of a velocity profile $v_{\rm g}(r)\propto \sqrt{\Sigma_{\rm g}\,r}$, appropriate for a gas distribution with a constant surface density   $\Sigma_{\rm g}$. This constant $\Sigma_g$ is appropriate for \HI\ gas profiles in the local universe \citep[e.g.][]{MartinssonT_13a,WangJ_16a,WangJ_20a}.
 Empirically, it has been shown that $v_{\rm g}(r)$ can be well approximated with $\propto\sqrt{r}$ at $z=0$ \citep[e.g.][]{AllaertF_17a}.
Here, $\Sigma_{\rm g}$ is an additional parameter which can be marginalized over.   $\Sigma_{\rm HI}$ is typically  $\sim5$~\msun~pc$^{-2}$ \citep{MartinssonT_13a} which is a consistent with the well-known size-mass $D_{\HI}-M_{\HI}$ $z=0$ relation \citep{BroeilsA_97a,WangJ_16a,MartinssonT_16a,LelliF_16a}. We allowed $\Sigma_{\rm HI}$ to range over $0-12.5$~\msun~pc$^{-2}$. The maximum gas surface density at $\sim10$~\msun~pc$^{-2}$ can be thought as of a consequence of molecular gas formation \citep[e.g.][]{SchayeJ_01b}.

Finally, we include the correction for pressure support (often called asymmetric drift correction)  namely $v_{\rm AD}^2$
following  \citet{WeijmansAM_08a}, \citet{BurkertA_10a}, and many others (see Appendix~\ref{appendix:asym}), such that the circular velocity in \Eq{eq:Vc} is $v^2_{\rm c}(r) = v_{\perp}^2(r)+v^2_{\rm AD}(r)$ where $v_{\perp}$ is the observed rotation velocity. Since the ISM pressure $P$ is approximately linearly dependent on the gas surface density $\Sigma_{\rm g}$ \citep[e.g.][]{BlitzL_06a,LeroyA_08a}, and can be described with $P\propto\Sigma_{\rm g}^{0.92}$ \citep[e.g][]{DalcantonJ_10a}, one has
\begin{eqnarray}
v^2_{\rm c}(r) &=& v_{\perp}^2(r)+0.92\sigma_0^2\left(\frac{r}{r_d}\right)
\end{eqnarray}
where $r_d$ is the disk scale length, (see \Eq{eq:AD:Dalcanton} in Appendix~\ref{appendix:asym}).

To summarize, the disk-halo  3D-model has 14 parameters: $x_c$, $y_c$, $z_c$, $f_{\rm O2}$, $\Rhalf$, $i$, $n_{\rm O2}$, P.A., $V_{\rm vir}$, $c_{-2}$, $\log X$, $\sigma_0$, $\Sigma_{\rm g}$ and the $\OII$ doublet ratio $r_{\rm O2}$. For the halo component, we can use a \citet{DiCintioA_14a} (`DC14') or a NFW model, for which, we use directly  $\log M_\star$ instead of $\log X$ as a parameter. For the `DC14' halo model, we restrict $\log X$ to [-3.0, -1.2] to ensure a solution in the upper branch~\footnote{The lower branch is  appropriate for dwarfs.} of the core-cusp vs. $\log X$ parameter space \citep[see Fig. 1 of][]{DiCintioA_14a}.
In the cases with a bulge component, there are 4 additional parameters: $r_{\rm b}$, $n_{\rm b}$, $v_{\rm max,b}$ and $B/T$.

\subsection{Parameter optimization and model selection}

Having constructed disk-halo models in 3D($x,y,\lambda$) within \galpak, we optimize the 14 parameters simultaneously with \galpak\ using the  python \textsc{pyMultiNest} package \citep{BuchnerJ_14a} against the MUSE data where the stellar continuum was removed taking into account the PSF and LSF. 
As in \S~\ref{section:kinematics}, we use flat priors on each parameter. 
We also do not use the stellar mass from SED  as input/prior  on  the disk mass (via $\log X$) because  the traditional disk-halo degeneracy is broken from the shape ($\alpha$-$\beta$-$\gamma$) of the DM halo profile \Eq{eq:rho} which  depends on the disk-to-halo mass ratio $X=M_\star/M_{\rm vir}$ as discussed in the previous section. 
We do not use priors on the inclination~\footnote{Except for the  two low-inclination galaxies (ID912, 919).} because
the $i-V$ degeneracy is broken from the simultaneous fit of the kinematics with the morphology, as discussed in \S~\ref{section:kinematics}.

Regarding model selection between DC14 or NFW DM profiles, we  choose the preferred model by comparing the evidence
$\ln \cal Z$ or marginal probability $\ln P(y|M_1)$ (namely the integral of the posterior over the parameters space)
\citep{JeffreysH_61a,KassR_95a,RobertsC_09a,JenkinsC_18a} for the DC14 DM model ($M_1$) against the NFW model ($M_2$) and using the Bayes factor defined as the ratio of the marginal probabilities $B_{12}\equiv P(y|M_1)/P(y|M_2)$.  Throughout this paper, following \citet{KassR_95a} \citep[see also][]{GelmanA_14a},
 we rescale the evidence by -2 such that it is on the same scale as the usual 
information criterion (Deviance, Bayesian Information Criterion, etc.). 
With this factor in mind, as discussed in \citet{JeffreysH_61a} and \citet{KassR_95a}, 
positive (strong) evidence against the null hypothesis (that the two models are equivalent) occurs when
the Bayes factor is $>3$ ($>20$), respectively. This corresponds to a logarithmic difference $\Delta \ln \cal Z$ of 2 and 6, respectively. 
Thus, we use a minimum $\Delta \ln \cal Z$ of 6 as our threshold to discriminate between models.
Table~\ref{table:evidence} shows   the logarithmic difference of the Bayes factors, $\Delta\ln\cal Z$,  for the NFW DM models with respect to the fiducial DC14 models.

\subsection{Stellar rotation from HST photometry}
\label{section:mge}

{In order to independently estimate the contribution of the stellar component to the RC, we parameterized the light distribution of HST/F160W images with the MGE method \citep{MonnetG_92a,EmsellemE_94b}\footnote{An implementation of the method \citep{CappellariM_02a} is available at \url{https://www-astro.physics.ox.ac.uk/ mxc/software/}}. For each galaxy we made an MGE model by considering the PSF of the HST/F160W filter, removing the sky level and masking any  companion galaxies or stars. Each MGE model consists of a set of concentric 2D Gaussians defined by the peak intensity, the dispersion and the axial ratio or flattening. The Gaussian with the lowest flattening is critical as it sets the lower limit to the inclination at which the object can be projected \citep{MonnetG_92a}. Therefore, following the prescription from \citet{ScottN_13a}, we also optimize the allowed range of axial ratios of all MGE models until the fits become unacceptable. { In practice, convergence is achieved when the mean absolute deviation of the model 
for a given axial ratio pair increases by less than 10 per cent over the previous step.} Finally, we convert  the Gaussian peak counts to surface brightness using the WFC3 zeropoints from the headers, and then to surface density (in L$_\odot$ pc$^{-2}$) adopting 4.60 for the absolute magnitude for the Sun in the F160W \citep{WillmerC_18a}.

We follow the projection formulas in \citet{MonnetG_92a}  and the steps outlined in \citet{EmsellemE_94a,EmsellemE_94b}
  to determine the gravitational potential for our MGE models \cite[see also Appendix A of][]{CappellariM_02b}. 
The critical parameters here are the distance, inclination, and the mass-to-light ratio of the galaxy. The distances are simply calculated from the  redshifts and our assumed Planck 2015 cosmology. 

As we assume that the stellar component is distributed in a disk, we use the axial ratio of galaxies measured from the HST/F160W images to derive the inclinations of galaxies. An alternative approach would be to use the inclinations returned from the \galpak models, which lead to almost identical results.

We estimate the mass-to-light ratios of galaxies combining the stellar masses obtained from photometric SED fits (see \S~\ref{section:sample}) and the total light obtained from the MGE models. Finally we use the module {\tt mge\_vcirc} from the JAM code \citep{CappellariM_08a} to calculate the circular velocity in the equatorial plane of each galaxy.

\begin{figure}
\centering
\includegraphics[width=0.45\textwidth]{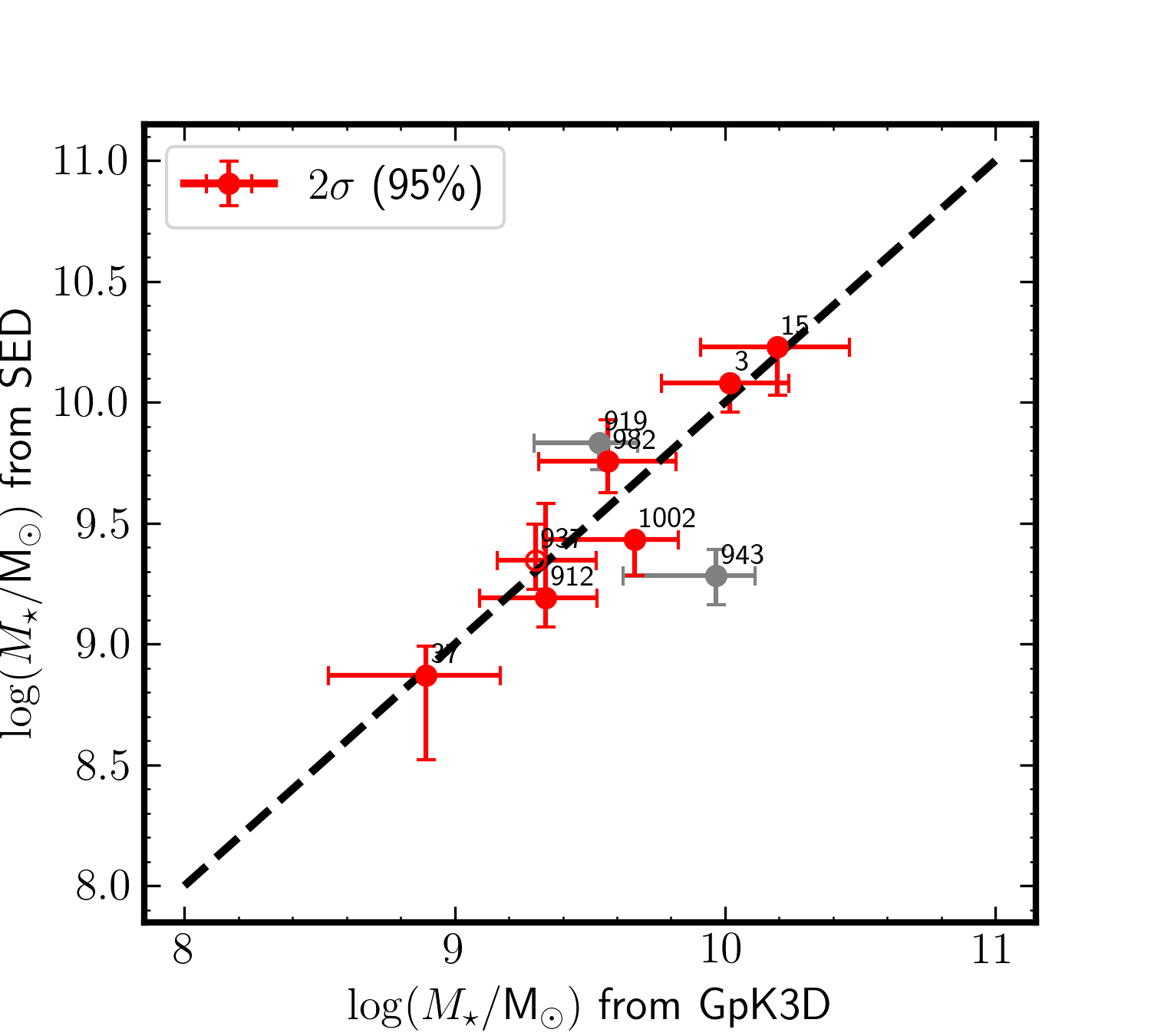}
\caption{Galaxies stellar masses. Comparison between the stellar mass $M_{\star}$ obtained from \galpak\ disk-halo fits and the SED-based $M_{\star}$ derived from the {\it HST} photometry.  The error bars represent the 95\% confidence intervals.
The  $M_{\star}$ obtained with \galpak\ (one of the 14 free parameters in \S~\ref{section:disk:halo}) and from {\it HST} photometry are completely independent, except for ID912 (open circle). The dashed line shows the 1:1 line and this figure shows the two are in excellent agreement, except for ID919 and ID943.
}
\label{fig:mdisk}
\end{figure}

\begin{figure*}
\centering
\includegraphics[width=0.9\textwidth]{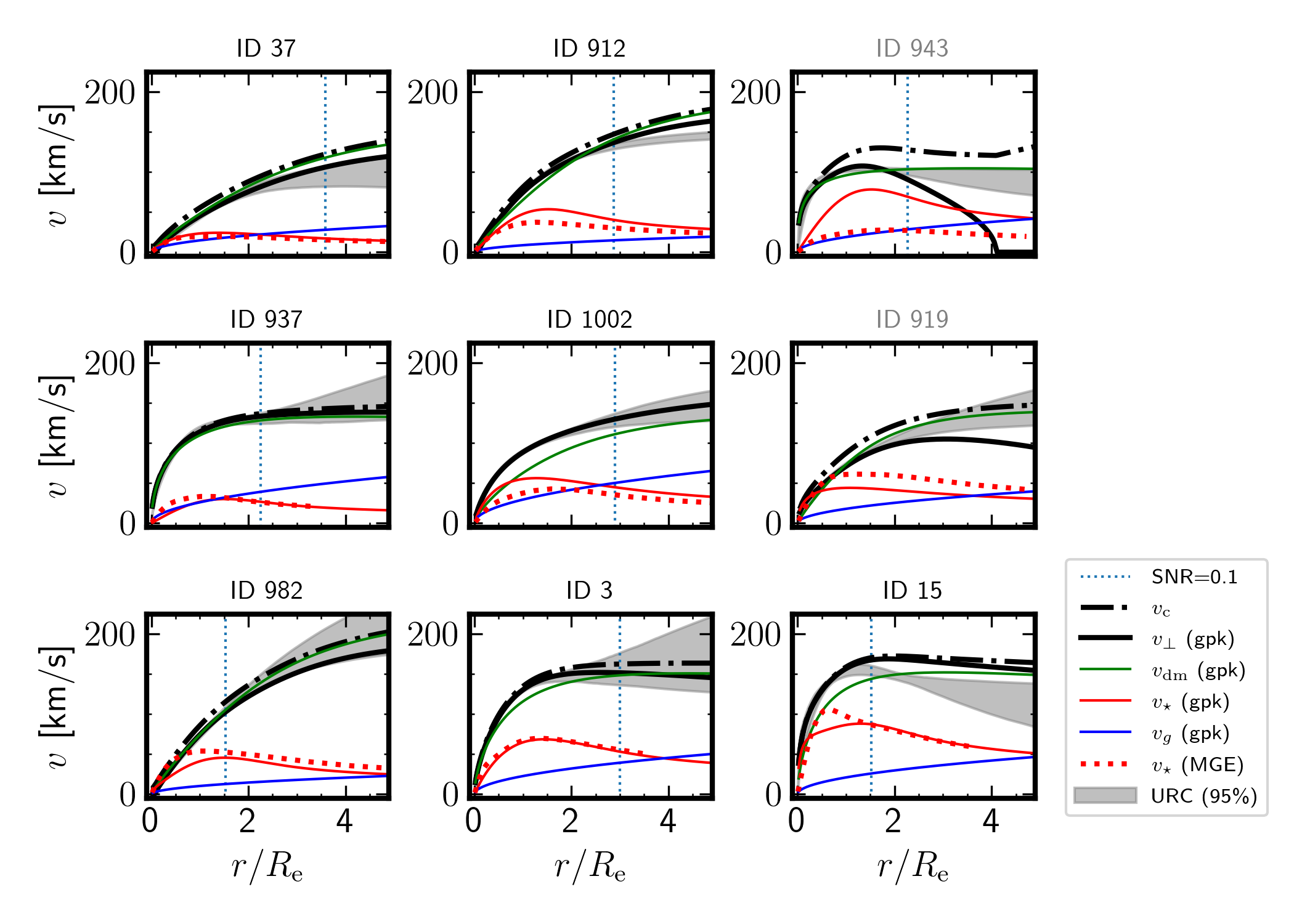}
\caption{Disk-halo decompositions for the \ngal\ galaxies in our sample (ordered by increasing $M_\star$).
The solid black line represents the total rotation velocity $v_\perp(r)$. All velocities are `intrinsic', that is corrected for inclination and instrumental effects.
The dot-dashed line represents the circular velocity $v_{\rm c}(r)$, that is $v_\perp(r)$ corrected for asymmetric drift.
The gray band represents the intrinsic universal rotation curve (URC) using the parameterization of PSS96 as in \Fig{fig:examples}.
The solid red (blue) line represents the stellar (gas)  component $v_\star(r)$ obtained from \galpak\ modeling of the MUSE \OII\ data.
The dotted red line represents the stellar component obtained using a MGE decomposition of the {\it HST}/F160W stellar continuum images.
The green line represents the DM component.
The vertical dotted lines are as in \Fig{fig:examples}.
\label{fig:diskhalo}
}
\end{figure*}

\section{Results}
\label{section:results}

\subsection{The diversity of rotation curve shapes}

In Figure \ref{fig:examples}, we show the morpho-kinematics of the galaxies used in this study.
The first column shows the  stellar continuum from {\it HST}/F160W.
The second column shows  the \OII\ flux map obtained from the \textsc{CAMEL}\footnote{Available at \url{https://gitlab.lam.fr/bepinat/CAMEL}} algorithm \citep{EpinatB_12a}.
The third column shows the \OII\ surface brightness profile as a function of radius $r$, in units of $\Rhalf$. 
The fourth column shows the observed 2D velocity field $v_{\rm 2d}$ obtained from  \textsc{CAMEL}.
The fifth column shows the intrinsic rotation velocity $v_{\perp}(r)$ corrected for inclination and instrumental effects (beam smearing, see \S~\ref{section:methodology}), using the parametric model of PSS96 (see \S~\ref{section:kinematics}). The vertical dotted lines represent  the radius at which the S/N per spaxel reaches 0.3, and indicates the limits of our data.
The last column shows the residual map, obtained from computing the standard deviation in the residual cube along the wavelength direction.

This figure shows that $z=1$ RCs have diverse shapes \citep[as in][]{TileyA_19b,GenzelR_20a} with mostly increasing  but some presenting declining RCs at large radii as in \citet{GenzelR_17a}. The diversity, albeit for a smaller sample, is similar to the diversity observed at $z=0$ \citep[e.g.][]{PersicM_96a,CatinellaB_06a,MartinssonT_13b,KatzH_17a}.

\subsection{The disk-halo decomposition}

We now turn to our disk-halo decomposition using the method described in \S~\ref{section:disk:halo}.
For each SFG, we ran several combinations of disk-halo models, such as different halo components (DC14/NFW), different disk components (Freeman/MGE), with or without a bulge, with various  asymmetric drift corrections and chose the model that best fit the data for each   galaxy according to the model evidence. 
 We find that  the DC14 halo model is generally preferred over a NFW profile and the resulting model parameters are listed in Table~\ref{tab:results}.
The evidence for the DC14 models is discussed further in \S~\ref{section:cores}.

 Before showing the disk-halo decompositions, we compare the disk stellar mass $M_\star$  ($M_\star$ being one of the 14 free parameters) obtained from the 3D fits  with the SED-derived $M_\star$.
 This comparison is performed in \Fig{fig:mdisk} where  the total $M_\star$ (disk$+$bulge from our fits) is plotted along the $x$-axis. 
This figure shows that there is relatively good agreement between the disk mass estimates from our \galpak\ model fits (described in \S~\ref{section:disk:halo}) and the SED-based ones,  except   for ID919 and ID943.
This  figure shows that our 3D disk-halo decomposition yields a disk mass consistent with the SED-derived $M_\star$, 
and thus opens the possibility to constrain disk stellar masses from rotation curves of distant galaxies for kinematically undisturbed galaxies.

The disk-halo decompositions (deprojected and `deconvolved' from instrumental effects) 
using our 3D-modeling approach with \galpak\ are shown in Figure~\ref{fig:diskhalo},
where the panels are ordered by increasing $M_\star$ as in \Fig{fig:f160w}.
The   disk/DM models used are listed in \Tab{tab:results}.
In each panel, the solid black line shows the total rotation velocity $v_\perp(r)$ corrected for asymetric drift. All velocities are `intrinsic', meaning corrected for inclination and instrumental effects, while the dot-dashed line represents the circular velocity $v_{\rm c}(r)$.
The gray band represents the URC model  as in \Fig{fig:examples}.
The solid green, red and blue lines represent the dark-matter $v_{\rm dm}(r)$, stellar $v_{\star}(r)$, and gas components $v_{\rm g}(r)$, respectively.
The dotted red lines represent the  stellar component obtained from the {\it HST}/F160W images as discussed in \S~\ref{section:mge}.

Comparing the solid with the dotted red lines in \Fig{fig:diskhalo}, one can see that there is generally  good agreement between $v_{\star}(r)$ obtained from the {\it HST} photometry and from our disk-halo decomposition with \galpak\ of the MUSE data, except again for ID919 and ID943. This comparison shows that the disk-halo decomposition obtained from the \OII\ line  agrees with the  $v_{\star}$ from the mass profile obtained on the {\it HST} photometry. {One should note that  the stellar mass $M_\star$ from SED fitting is not used as a prior in our \galpak\ fits, except for ID937 because the data for this galaxy prefers a NFW profile, which then becomes degenerate with $M_\star$.}
For the interested reader, {the potential degeneracies between $M_\star$ and $M_{\rm vir}$ are shown in \Fig{fig:corner}.}

\subsection{The stellar-to-halo mass relation} 
 
 The $M_\star-M_{\rm vir}$ relation in $\Lambda$CDM is a well-known scaling relation that reflects 
 the efficiency of feedback.  Hence, measuring this scaling relation in individual galaxies is often seen as a crucial constraint on models for galaxy formation. 
 This scaling relation can be constructed from abundance matching techniques 
 \citep[e.g.][]{ValeA_04a,MosterB_10a,BehrooziP_13b,BehrooziP_19a}.
 Observationally,   the  $z=0$ stellar-to-halo relation has been constrained by numerous authors using a variety of techniques such as weak lensing and/or clustering  \citep[e.g.][]{LeauthaudA_12a,MandelbaumR_16a}. 
Direct measurements of the $M_\star-M_{\rm vir}$ relation on individual galaxies using rotation curves  have been made on various samples of dwarfs \citep{ReadJ_17a}, spirals \citep{AllaertF_17a,KatzH_17a,LapiA_18a,PostiL_19b,DiPaoloC_19a} and early type galaxies \citep{PostiL_21a} among the most recent studies,
and these  have found a very significant scatter in this relation.

In \Fig{fig:Behroozi} (left), we show the stellar-to-halo mass ratio $M_{\star}/M_{\rm vir}$ as a function $M_\star$. The blue (gray) contours show the expectation for $z=1$ SFGs in the TNG100/50 simulations
 and the solid lines represent the $M_\star/M_{\rm vir}$ relation from \citet{BehrooziP_19a}.
\Fig{fig:Behroozi} (left) shows that our results are qualitatively in good agreement with the Behroozi relation.

\citet{RomeoA_20a} argued that   disk gravitational instabilities are the primary driver for galaxy scaling relations. Using a disk-averaged version of the \citet{ToomreA_64a} $Q$ stability criterion~\footnote{\citet{ObreschkowD_16a} used similar arguments to derive the \HI\ mass fractions.}, \citet{RomeoA_20a}
find that  
\begin{equation}
<Q_i>=\frac{j_i\hat\sigma_i}{G M_i}=A_i  
\label{eq:jRomeo:Q}
\end{equation}
 where $i=\star,\HI$ or $H_2$, $\hat\sigma_i$ is the radially averaged velocity dispersion, and $j_i$ is the total specific angular-momentum.
For $i=\star$, $A_i\approx 0.6$.

Consequently, for the stellar-halo mass relation with $i=\star$, $M_\star/M_{\rm vir}$ ought to correlate with \citep{RomeoA_20b}:
\begin{equation}
\frac{M_\star}{M_{\rm vir}}\simeq \frac{j_\star \hat{\sigma}_\star}{G M_{\rm vir}}
\label{eq:jRomeo}
\end{equation}
where $j_\star$ is the stellar specific angular momentum, $\hat \sigma_\star$ the radially averaged stellar dispersion. 
We can estimate $j_\star$ using the ionized gas kinematics, namely $\log j_\star=\log j_{\rm gas}-0.25$ as in \citet{BoucheN_21a}.
The dispersion $\hat\sigma_\star$ is not directly accessible, but we use the scaling relation with $M_\star$ ($\hat\sigma_\star\propto M_\star^{0.5}$) from \citet{RomeoA_20b} which followed from the \citet{LeroyA_08a} analysis of local galaxies. 
\Fig{fig:Behroozi} (right) shows the resulting stellar-to-halo mass ratio using $M_{\star}$ from SED and the $M_{\rm vir}$ values obtained from our disk-halo decomposition, where the inset shows the sample has $<Q_\star>\approx 0.7$, close to the expectation (\Eq{eq:jRomeo:Q}).

\begin{figure*}
\centering
\includegraphics[width=0.45\textwidth]{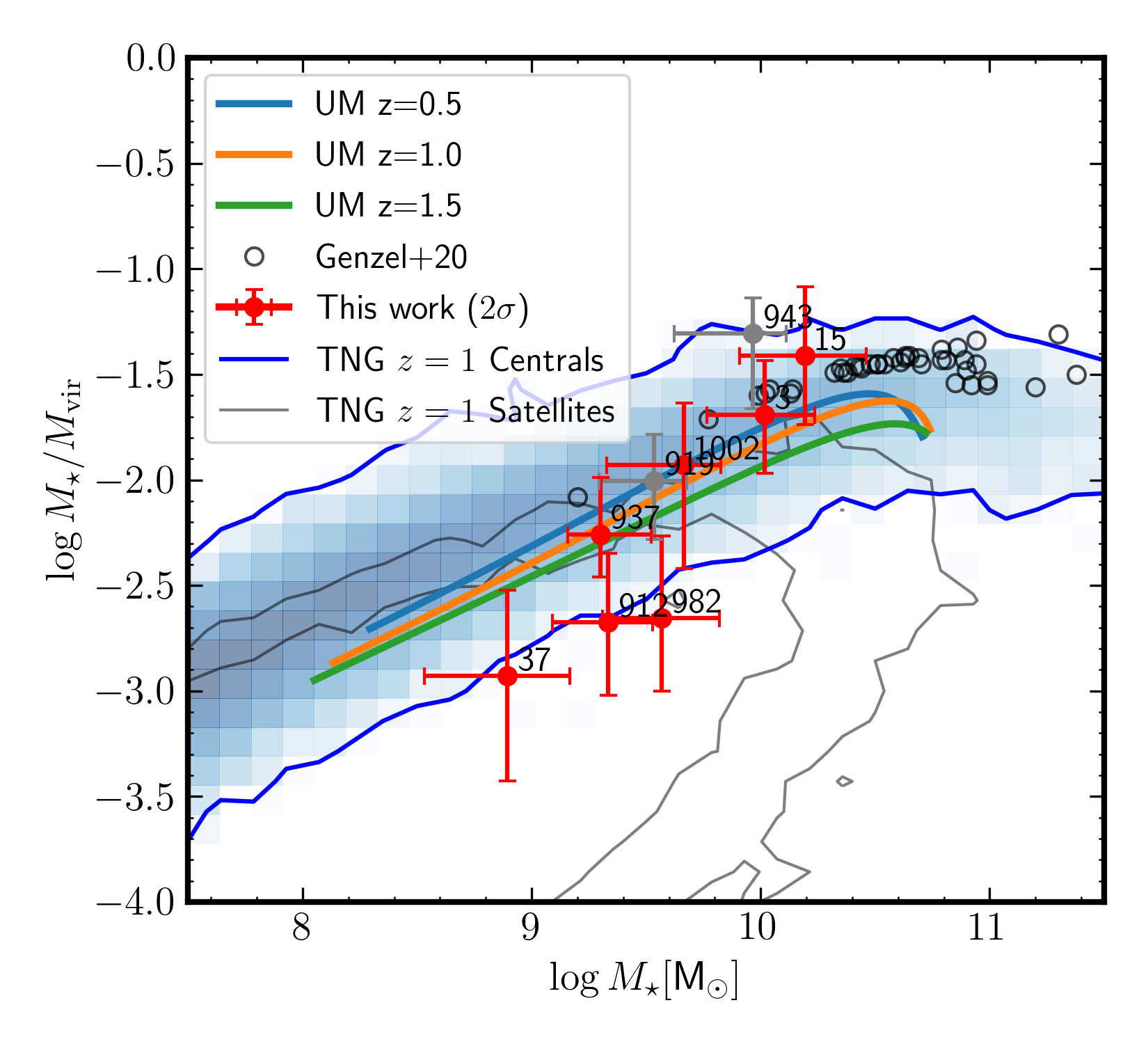}
\includegraphics[width=0.45\textwidth]{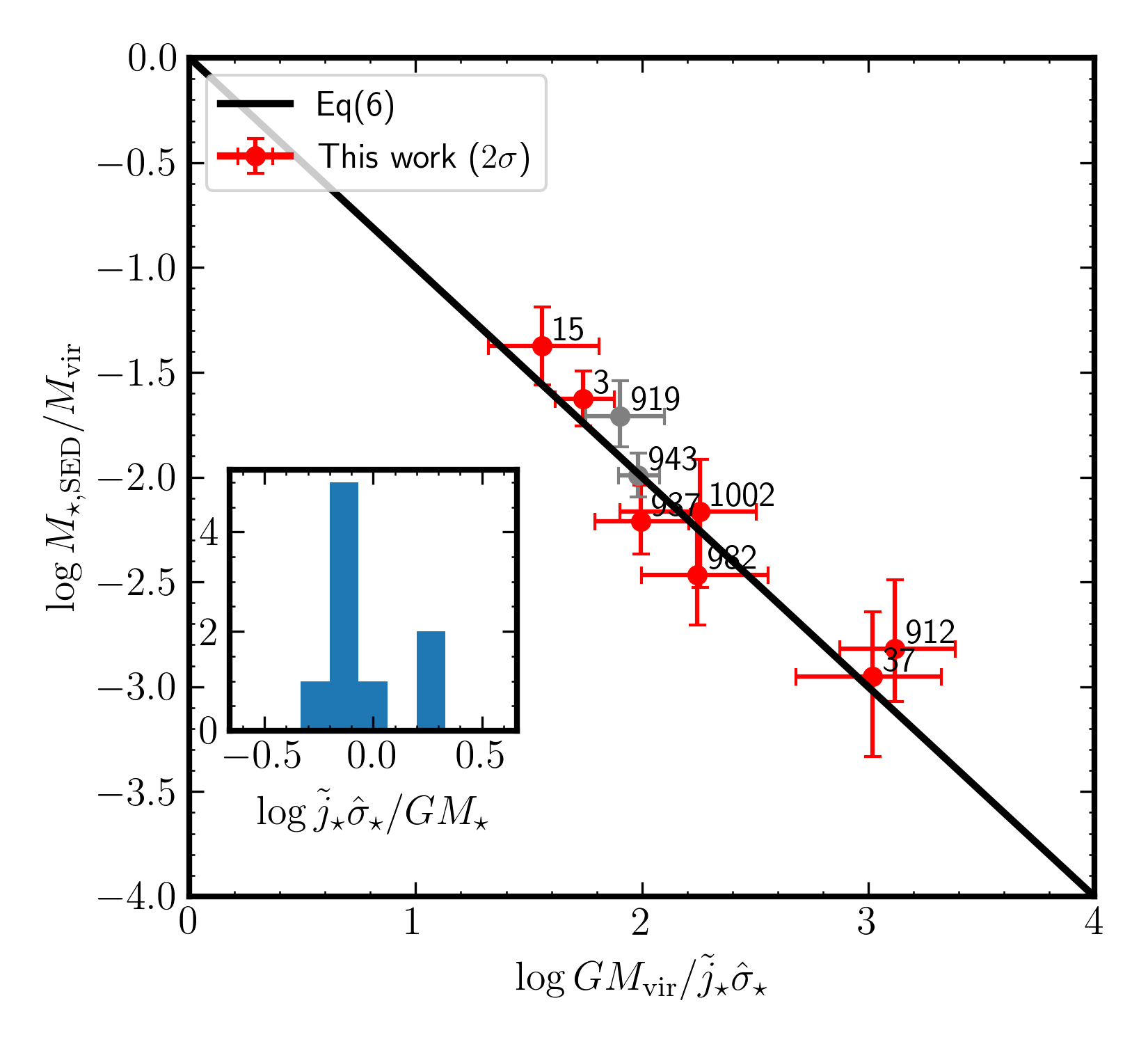}
\caption{Total stellar-to-halo fraction. {\it Left}: The total stellar-to-halo fractions $M_\star/M_{\rm vir}$ as a function of  the stellar mass $M_{\star}$ obtained from our 3D fits.
The error bars from our data are 95\%\ confidence intervals, and
 the open circles show the sample of \citet{GenzelR_20a}.
 The shaded (blue contours) histogram shows the location of SFGs in the TNG simulations for $z=1$ centrals, while the gray contours show the satellites.
The colored lines show the \citet{BehrooziP_19a} relation inferred from semi-empirical modeling at redshifts $z=0.5,1.0,1.5$, respectively.  
{\it Right}: The total stellar-to-halo fractions $M_\star/M_{\rm vir}$ as a function of $G M_{\rm vir}/j_\star\sigma_\star$
(\Eq{eq:jRomeo}) for the galaxies in our sample. 
The inset histogram shows that the sample has  $ {j_\star\hat\sigma_\star}/{G M_\star}\approx 0.7$ ($\equiv < Q_\star>$, \Eq{eq:jRomeo:Q}), see text.
}
\label{fig:Behroozi}
\end{figure*}

\subsection{DM fractions in $z=1$ SFGs}

Using the disk-halo decomposition shown in \Fig{fig:diskhalo}, we turn toward the DM fraction within $\Rhalf$, $f_{\rm DM}(<\Rhalf)$, by integrating the DM and disk mass profile 
to $\Rhalf$~\footnote{\citet{GenzelR_20a} used the ratio of velocities $f_{\rm DM}^v\equiv v^2_{\rm dm}/v_{\rm tot}^2$, whereas we use the mass ratio, $f_{\rm DM}^m$ using the \citet{UblerH_20a} notation, derived from the mass profiles.}.
\Fig{fig:fDM} shows that  $f_{\rm DM}(<\Rhalf)$ for the galaxies in our sample is larger than $50$\%\ in all cases, ranging from 60\%\ to 90\%.
The left (right) panel of \Fig{fig:fDM} shows $f_{\rm DM}(<\Rhalf)$ as a function of    $M_{\rm vir}$ ($\Sigma_{\star,1/2}$ the surface density within $\Rhalf$), respectively.  
Compared to the sample of 41 SFGs from \citet{GenzelR_20a} (open circles), our sample extends their results to the low mass regime, with $ M_{\star}<10^{10.5}~\msun$, $ M_{\rm vir}<10^{12}~\msun$ and to lower mass surface densities $\Sigma_{\star}<10^8$~${\rm M_{\odot}~kpc^{-2}}$. 

The relation between $f_{\rm DM}$ and $\Sigma_{\star,1/2}$ in \Fig{fig:fDM}  is tighter  and follows the expectation for $z=1$ SFGs in the TNG100/50 simulations (blue contour) \citep{LovellM_18a,UblerH_20a}, except at high masses. 
\citet{GenzelR_20a} already noted that the correlation with $\Sigma_{\star}$ is better than with $V_{\rm vir}$ or $M_{\rm vir}$.
This anticorrelation between the baryonic surface density and DM fraction has been noted at $z=0$ in several disk surveys \citep[e.g.][see their Fig.23]{BovyJ_13a,CourteauS_15a}.

In \S~\ref{section:discussion:fDM}, we discuss the implications of this $f_{\rm DM}-\Sigma_\star$ relation and its relation to other scaling relations.

\begin{figure*}
\centering
\includegraphics[width=0.9\textwidth]{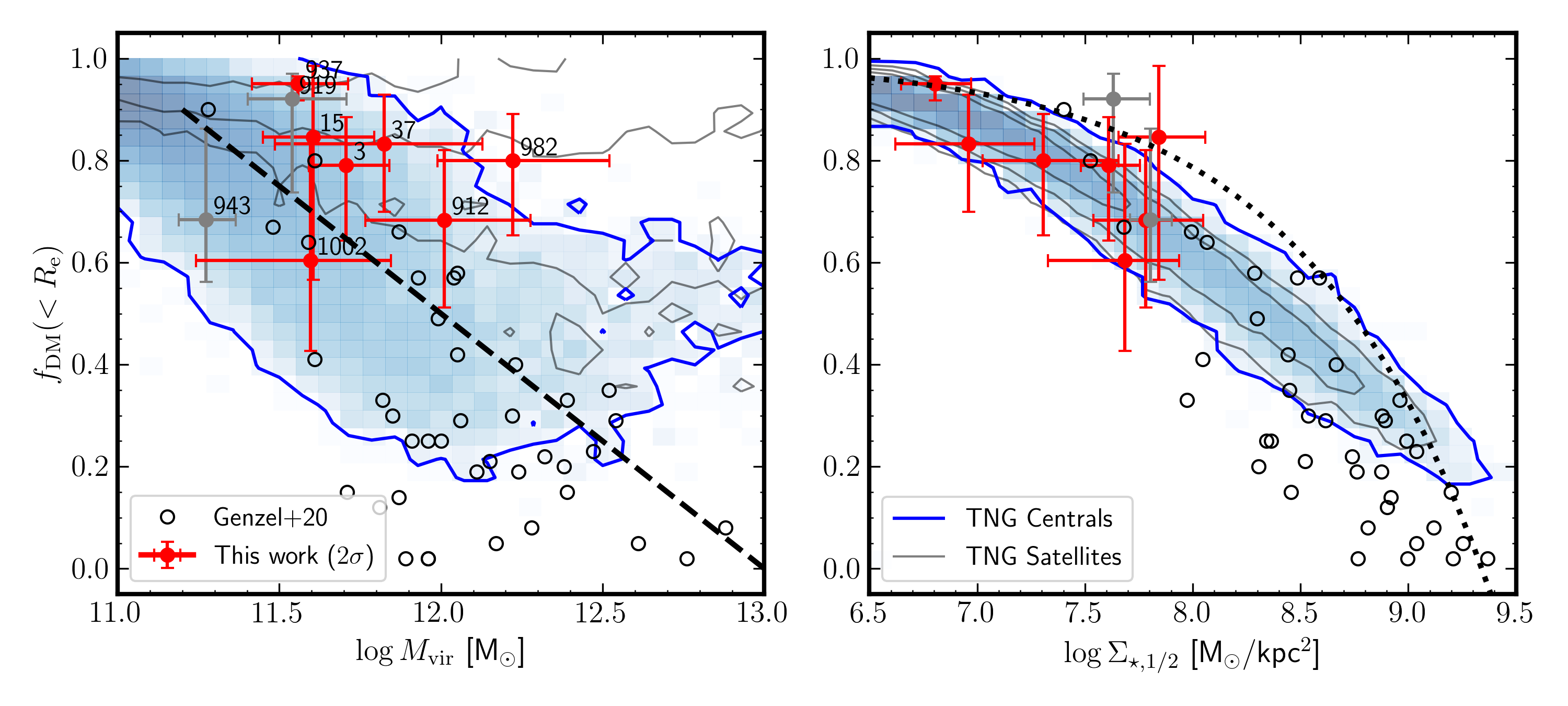}
\caption{ DM fractions for our SFGs. {\bf a)} {\it (left)}: The DM fractions within the half-light radius $\Rhalf$, $f_{\rm DM}(<\Rhalf)$, as a function of halo mass, $M_{\rm vir}$. The dashed line represent the downward trend of \citet{GenzelR_20a}.
{\bf b)}{\it (right)}: The DM fractions within $\Rhalf$ as a function of stellar mass surface density $\Sigma_{\star,1/2}$ within $\Rhalf$.
In both panels, the error bars from our data are 95\%\ confidence intervals, and
 the open circles show the sample of \citet{GenzelR_20a}.
 The shaded (blue contours) histogram shows the location of SFGs in  the TNG100 simulations for $z=1$ central SFGs, while the gray contours show the satellites.
 The dotted line represents the toy model derived from the TF  relation (Eq.~\ref{eq:toymodel}).
}
\label{fig:fDM}
\end{figure*}

\subsection{DM halo properties. The $c-M$ scaling relation}
  
Having shown (Figs~\ref{fig:mdisk}-\ref{fig:diskhalo}) that the baryonic component from our 3D fits is reliable, we now turn to the DM properties of the galaxies, and in particular to the concentration-halo mass relation ($c_{\rm vir}-M_{\rm vir}$).

The $c-M$ relation predicted from $\Lambda$CDM models  \citep[e.g.][]{BullockJ_01b,LudlowA_14a,DuttonA_14a,CorreaC_15c}  is often tested in the local universe \citep[e.g.][]{AllaertF_17a,KatzH_17a,LeierD_12a,LeierD_16a,LeierD_21a,WassermanA_18a}, but rarely beyond redshift $z=0$ except perhaps in massive clusters \citep[e.g.][]{BuoteA_07a,EttoriS_10a,SerenoM_15a,AmodeoS_16a,BivianoA_17a}. 
These generally agree with the predicted mild anticorrelation between concentration and virial mass.

 \Fig{fig:cMvir}(left) shows the  $c_{\rm vir}-M_{\rm vir}$ relation for the best 6 cases in our sample, that is
  excluding the two interacting galaxies (ID919, ID943) as well as ID15 
 because its concentration parameter remains unconstrained and degenerate with $V_{\rm vir}$ (see \Fig{fig:corner}b).  The error bars represent $2\sigma$ (95\%) and are color-coded according to the galaxy redshift.
 In \Fig{fig:cMvir}(left), the solid lines color coded with redshift represent  to the $c-M$ relation from \citet{DuttonA_14a}. 

We note that in order to fairly compare our data to such predictions from  DM-only (DMO) simulations, we show,  in \Fig{fig:cMvir},  the halo concentration parameter $c_{\rm vir}$ corrected to  a DM-only (DMO) halo following DC14~\footnote{See \citet{LazarA_20a} and \citet{FreundlichJ_20b} for  variations on this convertion.}  :
 \begin{equation}
c_{\rm vir, \rm DMO} = \frac{ c_{\rm vir,-2} }{ 1 + 0.00003 \times \exp[3.4   (\log X + 4.5)]}
\label{eq:DMO}.
\end{equation}
 { We note that the correction is important only for halos with stellar-to-halo mass ratio $\log X>-1.5$ and that most of our galaxies (7 out of 9) have $\log X<-1.5$.}

 \Fig{fig:cMvir}(right) shows the corresponding scaling relation for the scaling radius $r_s$, namely the $r_{s}-M_{\rm vir}$ relation. This relation in terms of $r_s$ is redshift independent. Several authors have shown, in various contexts (i.e., using pseudo-isothermal or \citet{BurkertA_95a} profiles),  that this quantity scales with galaxy mass or luminosity \citep[e.g.][]{SalucciP_12a,KormendyJ_16a,DiPaoloC_19a}. For illustrative purposes, we show the recent $z=0$ sequence for low surface brightness (LSB) galaxies of \citet{DiPaoloC_19a}.

 \Fig{fig:cMvir}  shows that  5 of the 6 SFGs  tend to follow the expected scaling relations for DM, the exception being ID912. One should keep in mind that cosmological simulations predict a $c-M$ relation with a significant scatter \citep[e.g.][]{CorreaC_15c}.
  To our knowledge, \Fig{fig:cMvir} is the first test of the $c-M$ relation at $z>0$ on halos with $\log M_{\rm vir}/\msun=11.5-12.5$ and our data appears to support the expectations from $\Lambda$CDM.
 
The $c_{\rm vir}-M_{\rm vir}$ or $r_s-M_{\rm vir}$ relations can be recasted as a $r_s-\rho_s$ relation  (from \Eq{eq:rho}).
\Fig{fig:rhos}(left) shows the $\rho_s-r_s$ relation and confirms the well-known anticorrelation between these two quantities with a slope of $\approx-1$ \citep[e.g.][]{SalucciP_00a,KormendyJ_04a,MartinssonT_13b,KormendyJ_16a,SpanoM_08a,SalucciP_12a,GhariA_19a,DiPaoloC_19a,LiLelli_19a},
which has been found in a wide range of  galaxies (dwarfs disks, LSBs, spirals).
These results are similar in nature, in spite of using different contexts and assumptions (namely $\rho_0$ vs $ \rho_{-2}$ or $\rho_s$). A detailed investigation of the differences related to these assumptions is beyond the scope of this paper.

As discussed in \citet{KormendyJ_04a}, this anticorrelation can be understood from the expected scaling relation of DM predicted by hierarchical clustering \citep{PeeblesP_74a} under initial density fluctuations that follow the power law  $|\delta k|^2 \propto k^n$ \citep{DjorgovskiS_92a}. \citet{DjorgovskiS_92a} showed that the size $R$, density $\rho$ of DM halos should follow $\rho\propto R^{-3(3+n)/(5+n)}$. For $n\simeq-2$ on galactic scales, $\rho\propto R^{-1}$.
This anticorrelation is also naturally present in the $\Lambda$CDM context as shown by \citet{KravtsovA_98a} with  numerical simulations.
As noted by many since \citet{KormendyJ_04a}, the anticorrelation between $\rho_s$ and $r_s$ implies a constant DM surface density $\Sigma_s\equiv\rho_s\,r_s$  \citep[e.g.][]{DonatoF_09a,SalucciP_12a,BurkertA_15a,KormendyJ_16a,KarukesE_17a,DiPaoloC_19a}.
\Fig{fig:rhos}(Right) shows the resulting DM surface density $\Sigma_s$ as a function of galaxy mass $M_d$. 
The gray band represents the range of surface densities from \citet{BurkertA_15a} for dwarfs, while the dashed line represents the range of densities
from \citet{DonatoF_09a,SalucciP_12a} for disks. \citet{KormendyJ_04a} had found a value of $\sim100$ \msun~pc$^{-2}$.

\begin{figure*}
\centering
\includegraphics[width=0.45\textwidth]{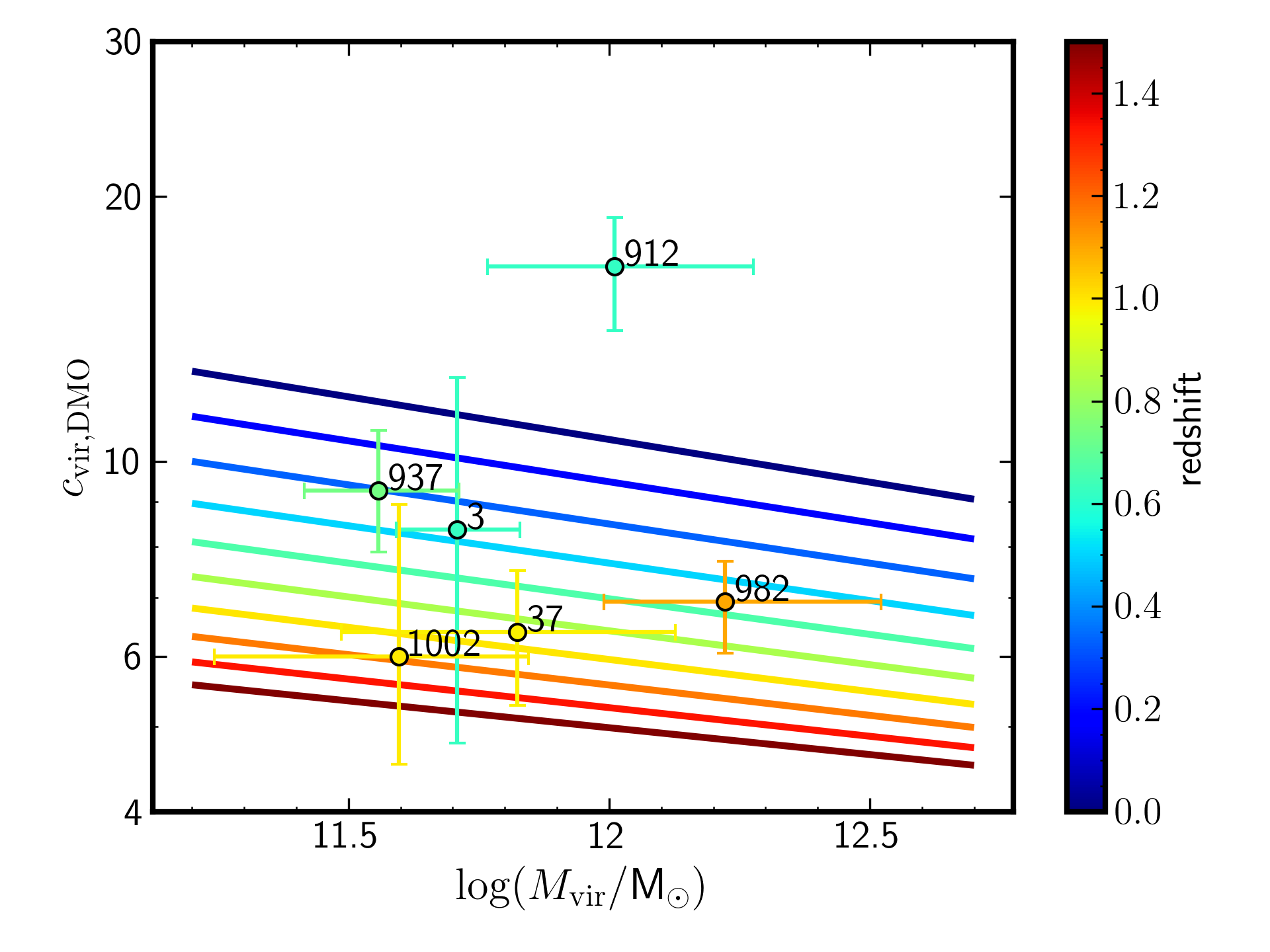}
\includegraphics[width=0.45\textwidth]{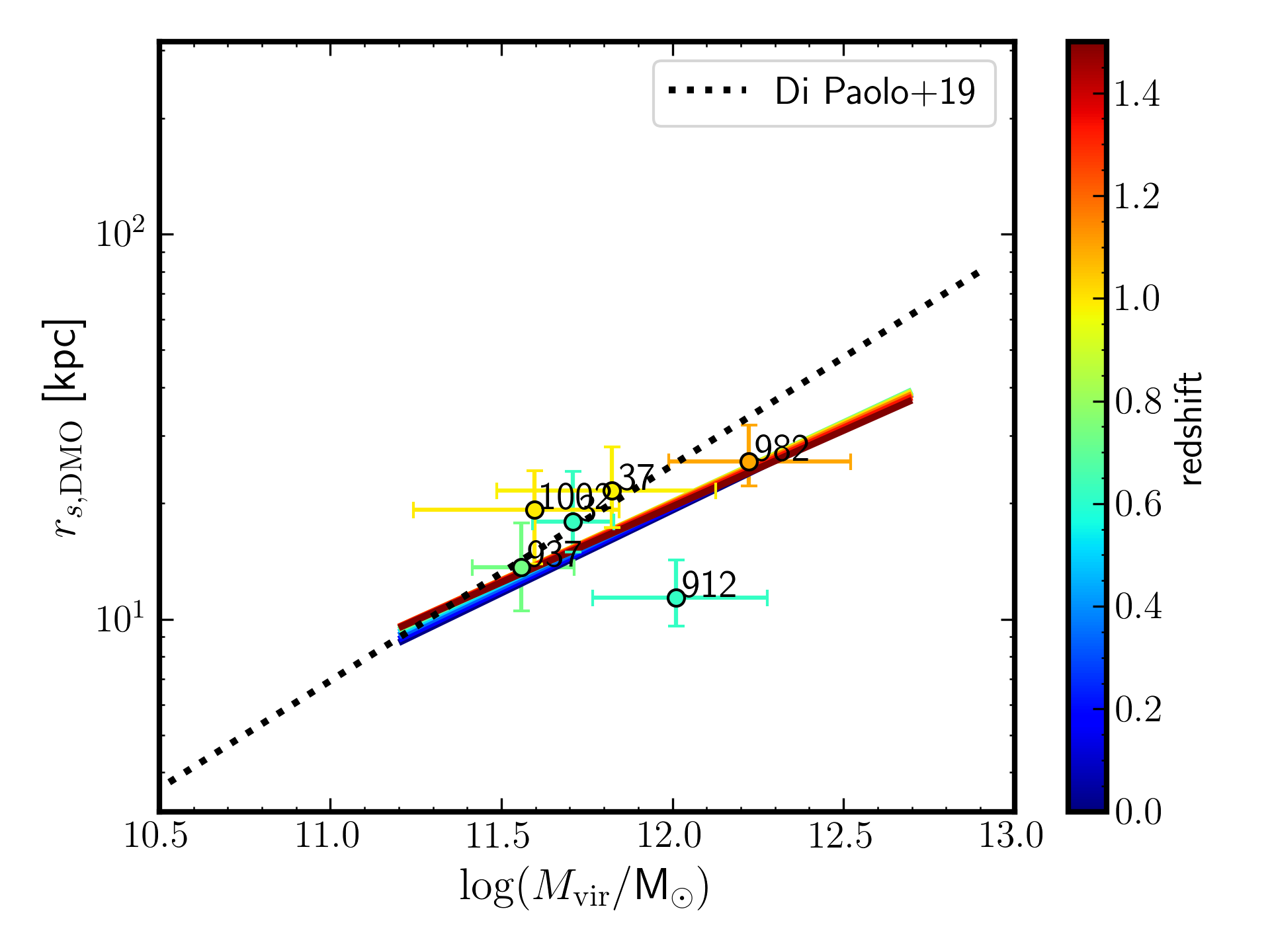}
\caption{The size of DM cores. {\it Left}: The halo concentration-halo mass relation. The concentrations $c_{\rm vir}$  for $z\simeq1$ SFGs,  derived from our 3D modeling of the \OII\ rotation curves, 
are  converted to a DM-only NFW equivalent  $c_{\rm vir, DMO}$ (see text).
{\it Right}: The DM core size $r_{s,\rm DMO}\equiv R_{\rm vir}/c_{\rm vir, DMO}$ in kpc as a function of halo mass. The dotted line represents the observed core-mass scaling relation for $z=0$ LSBs from \citet{DiPaoloC_19a} (see text).
In both panels, the solid lines represent the $c_{\rm vir}-M_{\rm vir}$ relation predicted by \citet{DuttonA_14a} for DM halos, color-coded by redshift.
The error bars are 95\%\ confidence intervals ($2\sigma$) and color-coded also by the galaxy redshift.
\label{fig:cMvir}
}
\end{figure*}

\begin{figure*}
\centering
\includegraphics[width=0.9\textwidth]{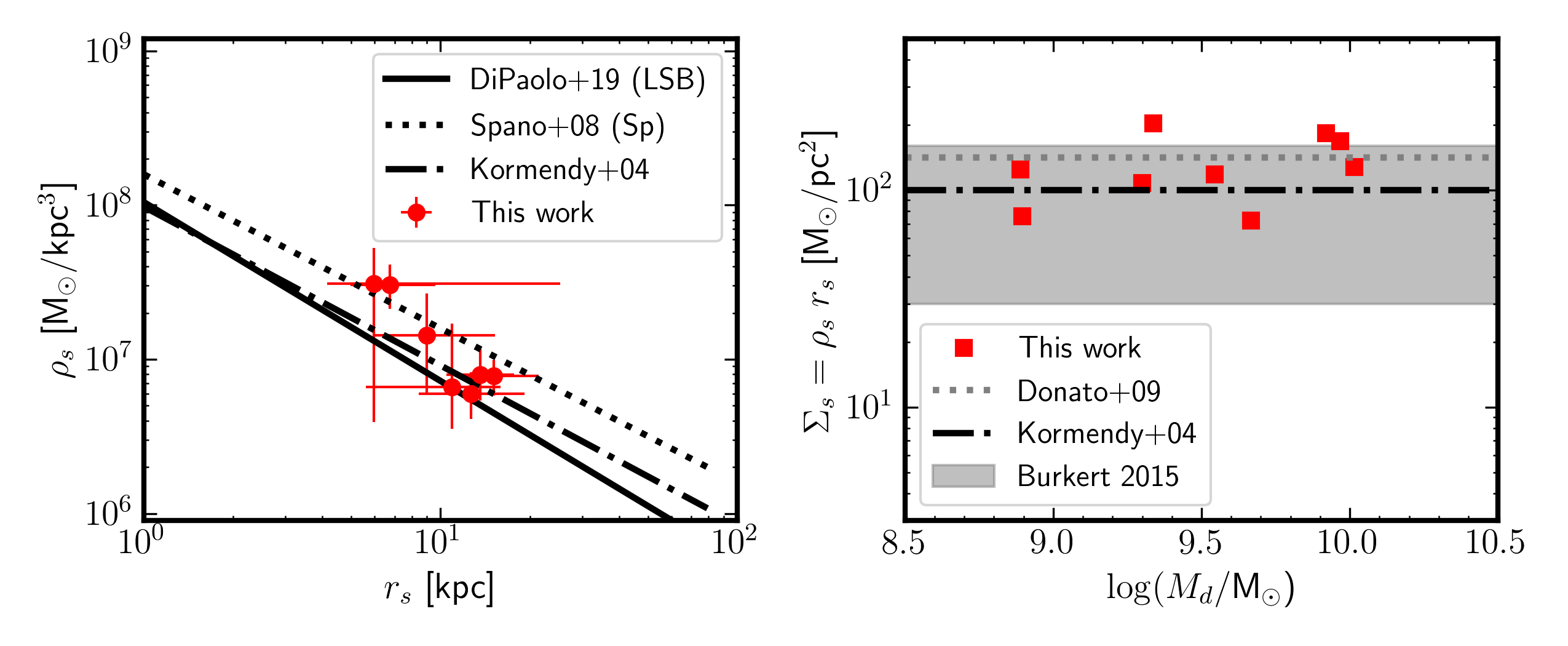}
\caption{The halo scale radius-density relation at $z=1$. {\it Left}: The $\rho_s-r_s$ scaling relation for the galaxies shown in \Fig{fig:cMvir}.
The error bars are 95\%\ confidence intervals ($2\sigma$).
For comparison, the anticorrelation of \citet{KormendyJ_04a,SpanoM_08a} and \citet{DiPaoloC_19a} are shown.
 {\it Right}: The DM surface density ($\Sigma_s\equiv\rho_s\,r_s$) as a function of galaxy mass. The anticorrelation in the left panel implies a constant DM surface density. The gray band represents the range of surface densities from \citet{BurkertA_15a} for dwarfs. The constant densities of \citet{KormendyJ_04a} and \citet{DonatoF_09a} are shown as the dotted, dot-dashed lines, respectively.
\label{fig:rhos}
}
\end{figure*}

\begin{table}
\centering
\small
\caption{Bayesian evidences for the \galpak\ fits. \label{table:evidence}
(1) Galaxy ID;
(2) Surface brightness profile; (3) Kinematic model (DM/Baryon); (4) External prior used; (5) Evidence $\ln Z$ on the deviance scale; (6) Bayesian factor between `NFW' and the  `DC14' models  (see \S~\ref{section:disk:halo}). 
}
\begin{tabular}{rrrrrrr}
ID & $I(r)$ & $v(r)$  & Prior  &  $\ln \cal Z$ & $\Delta\ln\cal Z$  \\
(1) & (2) & (3) & (4) & (5) & (6)\\
\hline 
3 & \sersic\ & DC14.MGE &   		& 17317 &0\\
3 & \sersic\ & NFW.MGE & $M_{\star,\rm SED}$ &  17312& -5 \\
15& \sersic & DC14.MGE & 		& 8019& 0\\
15& \sersic& NFW.MGE &  $M_{\star,\rm SED}$  & 8023&  +4\\
37&\sersic & DC14.MGE & 		& 9514 & 0 \\
37& \sersic& NFW.MGE &  $M_{\star,\rm SED}$ 	& 9651& +137\\
912&\sersic& DC14.MGE  &  	$i_{\star}$	& 8829 &  0 \\
912&\sersic& NFW.MGE  &   $i_{\star}$, $M_{\star,\rm SED}$		& 8931 & +102 \\
919&\sersic+B& DC14.Freeman &  $i_{\star}$     & 27552 & 0 \\
919&\sersic+B& NFW.Freeman &  $i_{\star}$, $M_{\star,\rm SED}$   &  27915 & +363\\
%
937&\sersic  & DC14.MGE         &     & 8632  & 0 \\
937&\sersic & NFW.MGE    &  $M_{\star,\rm SED}$ & 8625 & -7 \\
982&\sersic & DC14.MGE   &        & 6736 & 0 \\
982&\sersic & NFW.MGE  &     $M_{\star,\rm SED}$      & 7040 & +304\\
943&\sersic & DC14.MGE &          & 15374& 0\\
943&\sersic & NFW.MGE  &    $M_{\star,\rm SED}$ & 15372 & -2\\
%
1002&\sersic& DC14.Freeman &       & 8151 & 0\\
1002&\sersic& NFW.Freeman    &     $M_{\star,\rm SED}$    &  8155 & +4\\
\end{tabular}
\end{table}

\subsection{DM halos properties with core or cuspy profiles}
\label{section:cores}

\begin{figure*}
\centering
\includegraphics[width=0.8\textwidth]{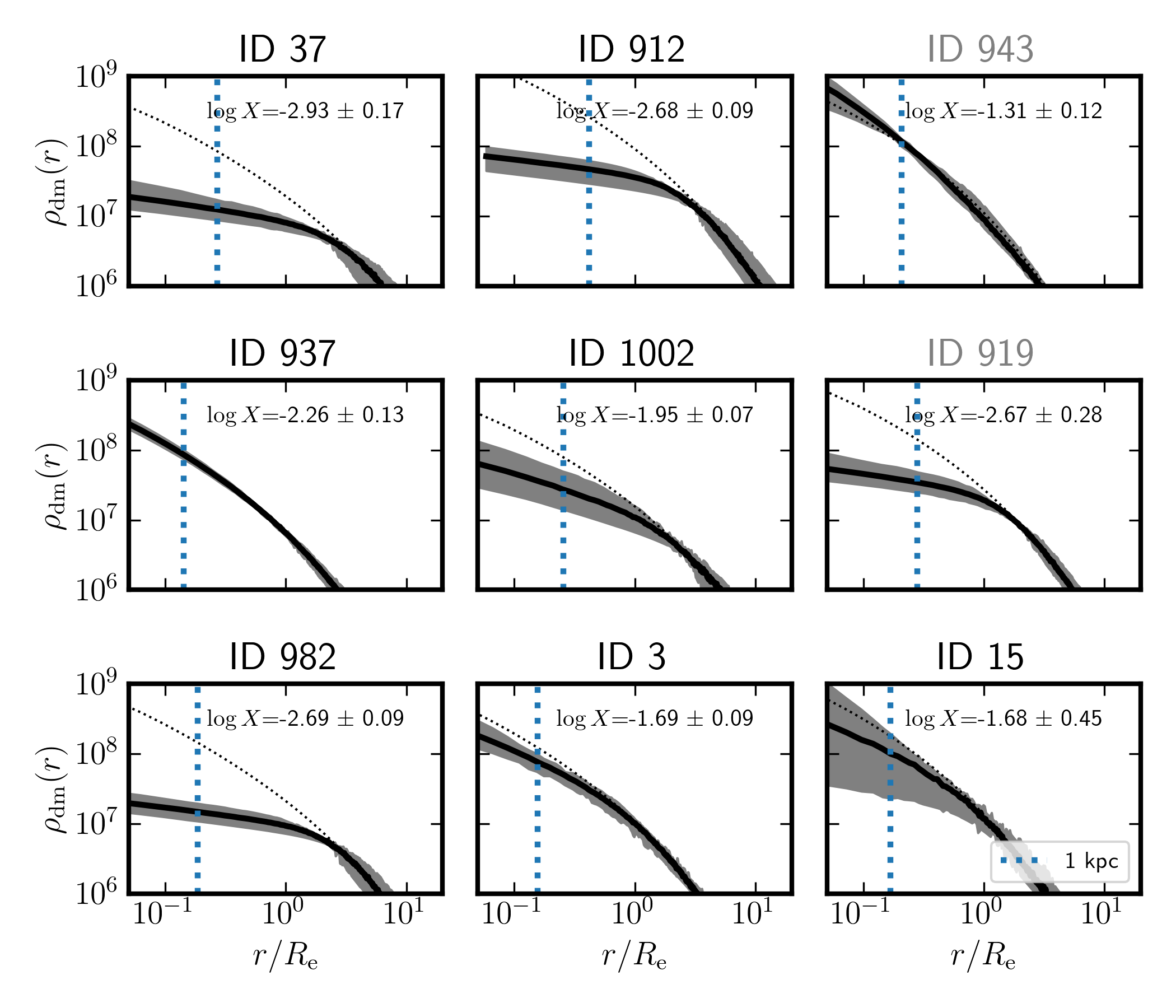}
\caption{ DM density profiles in M$_\odot/$kpc$^3$. Each panel show $\rho_{\rm dm}(r)$ as a function of $r/\Rhalf$ obtained from our disk-halo decompositions (\Fig{fig:diskhalo}).
The stellar-to-halo-mass ratio ($\log X\equiv \log M_\star/M_{\rm vir}$) is indicated.
The gray bands represent the 95\%\ confidence interval and the dotted lines represent NFW profiles.
The vertical dotted lines represent the 1~kpc physical scale, corresponding to $\approx1$  MUSE spaxel, and indicates the lower limit of our constraints.  
\label{fig:DMprofiles}
}
\end{figure*} 

We now investigate the shape of DM profiles, and in particular the inner logarithmic  slope $\gamma$  ($\rho_{\rm dm}\propto r^{-\gamma}$) in order to find evidence against or for core profiles.
There is a long history of performing this type of analysis in local dwarfs \citep[e.g.][]{KravtsovA_98a,deBlokW_01a,GoerdtT_06a,OhSH_11a,OhSH_15a,ReadJ_16b,KarukesE_17a,ReadJ_18a,ReadJ_19a, ZoutendijkB_21a},  
in spiral galaxies \citep[e.g.][]{GentileG_04a,SpanoM_08a,DonatoF_09a,MartinssonT_13a,AllaertF_17a,KatzH_17a,KorsagaM_18a,DiPaoloC_19a}
or in massive early type galaxies often aided by gravitational lensing 
\citep[e.g.][]{SuyuS_10a,NewmanA_13a,SonnenfeldA_12a,SonnenfieldA_13a,SonnenfeldA_15a,OldhamL_18a,WassermanA_18a}, but the core/cusp nature of DM is rarely 
investigated in SFGs  outside the local universe \citep[except in][]{GenzelR_20a,RizzoF_21a}
because this is a challenging task. However, owing to the high DM fractions in our sample (see \Fig{fig:fDM}), the shape the rotation curves are primarily driven by the DM profile.

 
 The DM profiles $\rho_{\rm dm}(r)$ as a function of $r/\Rhalf$ obtained from our 3D fits with the DC14 model are shown in \Fig{fig:DMprofiles}. This figure shows that the NFW profile is not compatible with the majority of the SFGs. 
\Fig{fig:DMprofiles}  shows that at least three galaxies  (IDs 37, 912, 982) show strong departures from a NFW profile, in particular they show evidence for cored DM profiles.
For these three galaxies, the logarithmic difference of the Bayes factors for the NFW profiles are $>100$ (see Table~\ref{table:evidence}), indicating very strong evidence against cuspy NFW profiles.
Our results are in good agreement with the RC41 sample of \citet{GenzelR_20a} where about half of their sample showed a preference for cored profiles (their Fig.10).

We  discuss the implications of these results in section \S~\ref{section:discussion:cores}, and in a subsequent paper we will analyze  additional DM profiles for CDM \citep[e.g.][]{EinastoJ_65a,BurkertA_95a,DekelA_17a,FreundlichJ_20b} including alternative DM models  such as ``fuzzy" axion-like DM \citep{WeinbergS_78a,BurkertA_20a},  self-interacting DM \citep[SIDM][]{SpergelD_00a,VogelsbergerM_13b}.

\section{Discussion}
\label{section:discussions}

\subsection{DM fractions in $z=1$ SFGs}
\label{section:discussion:fDM}

We return to the $f_{\rm DM}-\Sigma_\star$ relation in \Fig{fig:fDM} and its implications.
 The tight $f_{\rm DM}-\Sigma_\star$ relation can be thought of as a consequence of the tight  \citet{TullyB_77a}  relation (TFR) for disks as follows \citep[see also][]{UblerH_17a}.
 Indeed, if we approximate the DM fraction within \Rhalf\ as $f_{\rm DM}\approx V_{\rm DM}^2(\Rhalf)/V_{\rm tot}^2(\Rhalf)$ \citep{GenzelR_20a}, one has $f_{\rm DM}=(V_{\rm tot}^2-V^2_{\rm max,\star}-V^2_{\rm gas})/V_{\rm tot}^2$. Thus,
  \begin{eqnarray}
 1-f_{\rm DM}(\Rhalf)&=& \frac{V^2_{\rm max,\star}}{V_{\rm tot}^2} (1+\mu_g) \propto   \frac{G M_{\star}}{R_{\star}}   /{M_{\star}^{0.5}}\nonumber\\
 &\approx&  \frac{M_{\star}^{0.5}}{R_{\star}} (1+\mu_g)\propto   \Sigma_{\star}^{0.5} (1+\mu_g),
 \label{eq:toymodel}
 \end{eqnarray}
 where we used the stellar TFR,  $M_{\star}\propto V_{\rm tot}^4$ \citep[e.g.][]{McGaughS_05a}, the definition of gas-to-stellar mass ratio $\mu_g\equiv M_{\rm gas}/M_\star$ 
 and  the  maximum stellar rotation velocity for disks $V_{\rm max,\star}^2\propto G\,M_{\star}/R_{\rm e,\star}$.
 Eq.~\ref{eq:toymodel} shows the intimate link between the $f_{\rm DM}-\Sigma_\star$ diagram and the TFR relation.

More specifically, the TFR has $M_\star=a\,V_{\rm tot,2.2}^n$ with $n\simeq4$, $a\simeq10^{10}~\msun$  \citep{McGaughS_05a,MeyerM_08a,CresciG_09a,PelliciaD_16a,TileyA_16a,UblerH_17a,AbrilV_21a} where $V_{\rm rot,2.2}\equiv V_{\rm rot}/10^{2.2}$~km/s.
Given that  $V_{\rm max,\star}^2\equiv 0.38 \frac{G M_\star}{R_{\rm d}}$ for a \citet{FreemanK_70a} disk,     $V_{\rm max,\star}^2/V_{\rm tot}^2$ becomes
\begin{eqnarray}
\frac{V^2_{\rm max,\star}}{V_{\rm tot}^2}&=&0.38 \times1.68 a \frac{G M_\star}{a\,R_{\star} }/\left(\left(\frac{M_{\star}}{a}\right)^{1/n} 10^{2.2}\right)^2 \nonumber\\
&\approx&0.63\,  \sqrt{\pi}\,  \left(\frac{M_{\star,a}^{2(n-2)/n}}{\pi R^2_{\star}}\right)^{0.5}\, G a 10^{-4.4} \hbox{\msun~km$^{-2}$~s$^{-2}$} \nonumber\\
&\approx&  1.1  \left(\frac{M_{\star,a}^{0.94}}{\pi R^2_{\star}}\right)^{0.5}\times \left( \frac{a}{10^{10}}\right)     1.77 \hbox{kpc}
\end{eqnarray}
using $\Rhalf=1.68 R_{\rm d}$, where $M_{\star,a}\equiv M_\star/a$. For a $z\approx1$ TFR with  $n=3.8$ and $a=10^{9.8}$\msun\ \citep[e.g.][]{UblerH_17a}, \Eq{eq:toymodel} results in $1-f_{\rm DM}=\Sigma_{\star,9.8}^{0.5}(1+f_g)$, which is shown in \Fig{fig:fDM} (right) as the dotted line with $f_g=0.5$ \citep[e.g.][]{TacconiL_18a,FreundlichJ_19a}.
This exercise shows that the $f_{\rm DM}-\Sigma_\star$ relation is another manifestation of the TFR as argued in \citet{UblerH_17a}.

\subsection{Core/cusp formation}
\label{section:discussion:cores}

\begin{figure*}
\centering
\includegraphics[width=0.32\textwidth]{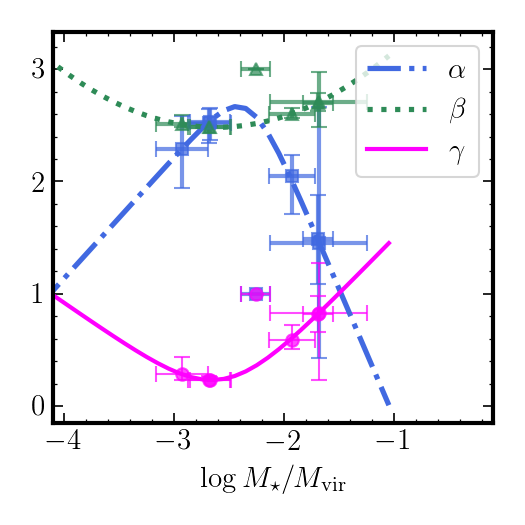} 
\includegraphics[width=0.32\textwidth]{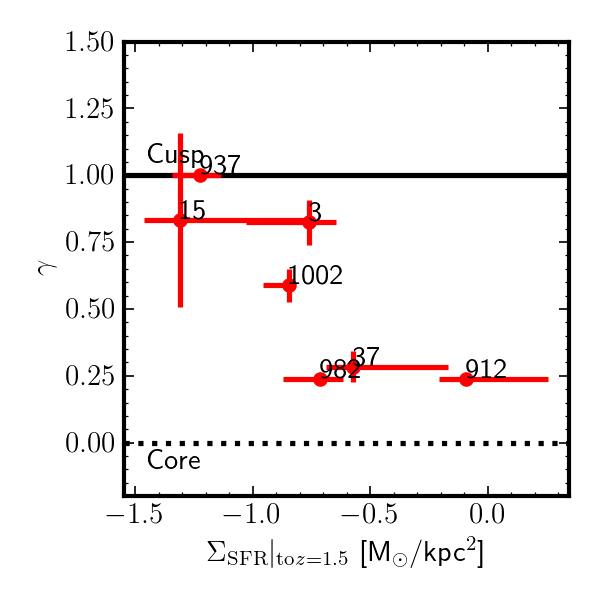} 
\includegraphics[width=0.32\textwidth]{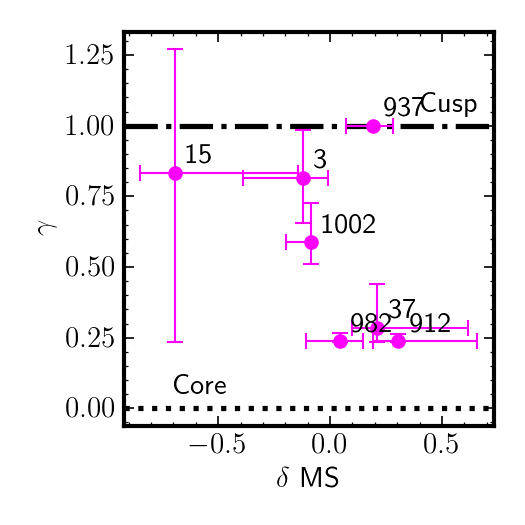} 
\caption{Relation between SFR and cores. {\it Left}: The $\alpha,\beta,\gamma$ parameters as a function of $\log M_\star/M_{\rm vir}$. 
The curves show the parameterisation of DC14 for $\alpha,\beta,\gamma$ and the solid symbols represent our SFGs, excluding ID919 and 943. 
{\it Middle}: The DM inner slope $\gamma$ as a function of the SFR surface density $\Sigma_{\rm SFR}$, scaled to $z=1.5$.
{\it Right}: The DM inner slope $\gamma$ as a function of the logarithmic offset from the MS, $\delta($MS), using the \citet{BoogaardL_18a} MS.
DM cores are present in galaxies with higher SFR and SFR surface-densities.
Error bars are $2\sigma$ (95\% CL).
}\label{fig:slope:best}
\end{figure*}

\begin{figure*}
\centering
\includegraphics[width=0.9\textwidth]{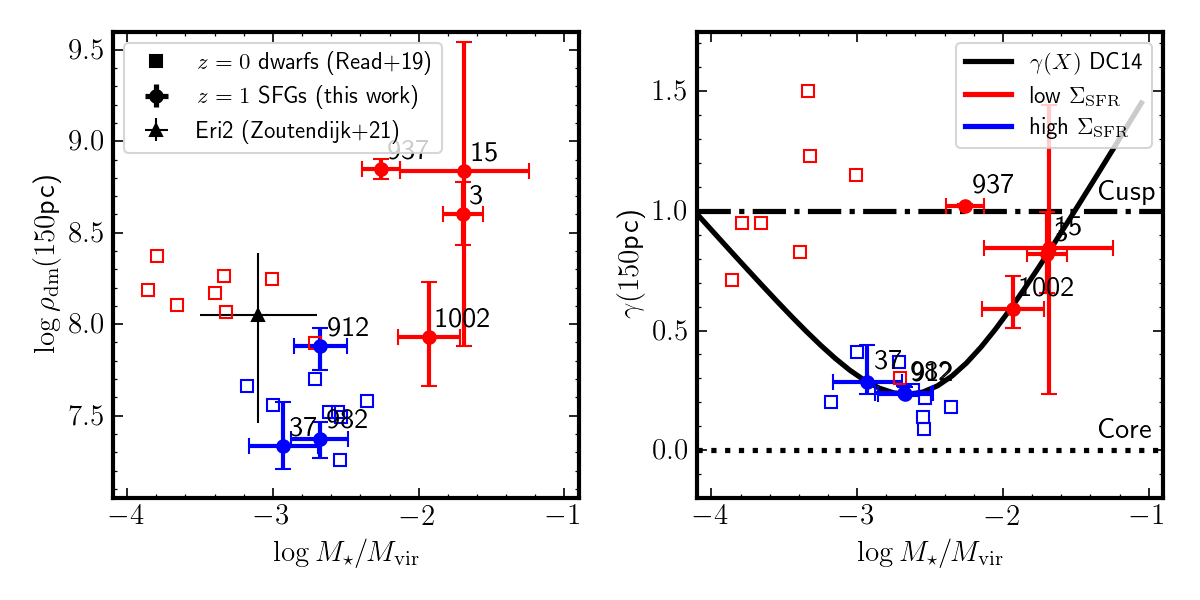} 
\caption{{\it Left}: The DM density at 150pc as a function of $\log M_\star/M_{\rm vir}$. 
The blue (red) solid circles with error bars (2$\sigma$) show our SFGs, except ID919 and 943.
{\it Right}: The DM  inner slope $\gamma$ parameter at 150~pc as a function of $\log M_\star/M_{\rm vir}$. 
The blue (red) squares represent the $z\approx0$ dwarfs from \citet{ReadJ_19a}  whose SFR was truncated less (more) than 6 Gyrs ago.
The blue (red) solid circles with error bars (2$\sigma$) show our SFGs with high (low) $\Sigma_{\rm SFR}$, respectively.
Error bars are $2\sigma$ (95\% CL).
}\label{fig:slope:Read}
\end{figure*}

Our results in \S~\ref{section:cores} (\Fig{fig:DMprofiles}) indicate a strong preference for cored DM profiles for four SFGs in our sample. Several   mechanisms have been invoked to explain the presence of cored DM profiles such  as Warm Dark Matter \citep[WDM,][]{BodeP_01a}, whose free streaming can suppress the small-scale fluctuations, axion-like ``fuzzy" DM \citep{WeinbergS_78a,HuW_00a,BurkertA_20a}, baryon-DM interactions \citep{FamaeyB_18a},  SIDM  \citep{SpergelD_00a,BurkertA_00a,VogelsbergerM_13a} or  dynamical friction  \citep{ReadJ_06a,GoerdtT_10a,OrkneyM_21a} from infalling satellites/minor mergers. 

Within the context of CDM, it has long been recognized  \citep[see review in][]{BullockJ_17a} since the original cusp/core problem first observed  in dwarfs or low-surface brightness galaxies \citep[e.g.][]{deBlokW_97a,deBlokW_01a,KravtsovA_98a} that (rapid) changes in the gravitational potential 
due to star-formation driven outflows can essentially inject energy in the DM, resulting in a flattened DM profile   \citep{NavarroJ_96b,ReadJ_05b,PontzenA_12a,TeyssierR_13a,
DiCintioA_14a,DuttonA_16a, DuttonA_20a, ChanTK_15a,ElZantA_16a,LazarA_20a,FreundlichJ_20a}.
Similarly, {DM core/cusps can also be linked to active galactic nuclei (AGN) activity \citep{PeiraniS_17a,DekelA_21a} in more massive galaxies with $M_{\rm vir}>10^{12}$\msun.}  While most of these analyses focus at cores at  $z=0$, \citet{TolletE_16a} showed that cores can form in a similar fashion as early as $z=1$.

Observationally, cores are now found up to $z\simeq2$ \citep{GenzelR_20a}, but
 the relation between outflows/star-formation and core formation has not been established,
as observations have unveiled cores in  galaxies spaning a range of halo or stellar masses \citep[e.g.][and references therein]{WassermanA_18a}
 or cusps when cores would be expected \citep[e.g][]{ShiY_21a}.
At high-redshifts, \citet{GenzelR_20a} found that cores are preferentially associated with low DM fractions.

In order to investigate the potential relation between SFR-induced feedback and DM cores, we show in \Fig{fig:slope:best}  the DM inner slope $\gamma$ as a function of SFR surface density $\Sigma_{\rm SFR}$ (left) and as a function of the offset from the main-sequence (MS) for SFGs \citep[using][]{BoogaardL_18a} (right). This figure indicates that SFGs above the MS or with high-SFR densities are preferentially found to have cores. SFGs below the MS with decaying SFR (like ID15) have low SFR densities owing to the low SFR, and show cuspy DM profiles, indicating that cusps reform when galaxies stop being active.

While the majority of research has focused on the formation of DM cores in order to match  observations at $z=0$, 
DM cusps can reform from the accretion of DM substructures
\citep{LaporteC_15a} as first argued in \citet{DekelA_03c},  or as a result of late mergers  as argued in \citet{OrkneyM_21a} for dwarfs.

In  \Fig{fig:slope:Read}, we compare our results to those of \citet{ReadJ_19a} who found that dwarfs fall in two categories, where the core/cusp presence is related to the star-formation activity.    \citet{ReadJ_19a} found  that  dwarfs  whose star-formation stopped over 6 Gyr ago show preferentially cusps (open red squares), while dwarfs with extended star-formation show shallow DM cores (open blue squares). In this figure, the  filled red (blue) circles represent our   galaxies with $\Sigma_{\rm SFR}$ smaller (larger) than $\log\Sigma_{\rm SFR}/\msun$~kpc$^{-2}=-0.7$. Our results in \Fig{fig:slope:Read}, together  with those of \citet{ReadJ_19a}, provide indirect evidence for SFR-induced core formation within the CDM scenario, where   DM   can be kinematically heated by SFR-related feedback processes.

\section{Conclusions}
\label{section:conclusions}

Using a sample of \ngal\ \OII\ emitters with the highest S/Ns in the deep (140hr) MXDF \citep{BaconR_21a} dataset,
we measure the shape of individual RCs of $z\approx 1$ SFG out to $3\times\Rhalf$
  with stellar masses ranging from $10^{8.5}$ to $10^{10.5}$ \msun, covering a range of stellar masses complementary to the analysis of \citet{GenzelR_20a}, whose sample has $M_\star>10^{10}$~\msun.

We then performed a disk-halo decomposition on the \OII\ emission lines using a 3D modeling approach that includes  stellar, dark-matter, gas (and bulge) components (\Fig{fig:diskhalo}).
The dark-matter profile is a generalized Hernquist--\citet{ZhaoH_96a}   profile using the feedback prescription of \citet{DiCintioA_14a}, which links
  the DM profile shape to the baryonic content. 

Our results are as follows. We find that 

$\bullet$ the 3D approach allows to constrain RCs to 3\Rhalf\ in individual SFGs revealing a diversity in shapes (\Fig{fig:examples}) with mostly rising and some having declining outer profiles;

$\bullet$ the disk stellar mass $M_\star$ from the \OII\ rotation curves is consistent with the SED-derived $M_\star$   (\Fig{fig:mdisk}), except for two SFGs (IDs 919, 943)  whose kinematics are strongly perturbed by a nearby companion ($<2$\arcsec);

$\bullet$ the stellar-to-DM ratio $M_\star/M_{\rm vir}$ follows the relation inferred from abundance matching \citep[e.g.][]{BehrooziP_19a}, albeit with some scatter (\Fig{fig:Behroozi});

$\bullet$ the DM fractions $f_{\rm DM}(<\Rhalf)$ are high (60-90\%) for our \ngal\ SFGs (\Fig{fig:fDM}) which have  stellar masses 
(from $10^{8.5}$\msun\ to $10^{10.5}$\msun) or surface densities ($\Sigma_\star<10^8$ \msun~kpc$^{-2}$). These DM fractions complement the low fractions of the sample of \citet{GenzelR_20a},
and globally, the $f_{\rm DM}(<\Rhalf)-\Sigma_\star$ relation is similar to the $z=0$ relation \citep[e.g.][]{CourteauS_15a}, and follows from the TFR;

$\bullet$  the fitted concentrations are consistent with the $c_{\rm vir}-M_{\rm vir}$ scaling relation predicted by DM only simulations (\Fig{fig:cMvir});

$\bullet$ the DM profiles show constant surface densities at $\sim100$ M$_\odot$/pc$^2$ (\Fig{fig:rhos});

$\bullet$  similarly to the $z>1$ samples of \citet{GenzelR_20a}, the disk-halo decomposition of our $z\approx1$ SFGs shows  cored DM profiles  for about half of the isolated galaxies  (\Fig{fig:DMprofiles}-\Fig{fig:slope:best})  in agreement with other $z=0$ studies \citep[e.g.][]{AllaertF_17a,KatzH_17a};

$\bullet$ DM cores are present in galaxies with high SFRs (above the MS), or high SFR surface density (\Fig{fig:slope:best}b-c), possibly supporting   the scenario of SN feedback-induced core formation. Galaxies below the MS or low SFR surface density have cuspy DM profiles    (\Fig{fig:slope:Read}), suggesting that cusps can reform when galaxies become passive \citep[e.g.][]{LaporteC_15a,ChanTK_15a,OrkneyM_21a}.
 
Overall, our results demonstrate the power of performing disk-halo decomposition in 3D on deep IFU data. With larger samples, it should be possible to confirm this type of
relation between cores and star-formation histories, and to test further SN feedback induced core formation within the $\Lambda$CDM framework.

\begin{acknowledgements}
We are grateful to the anonymous referee for useful comments and suggestions. 
We thank S. Genel, J. Fensch, J. Freundlich and B. Famaey for inspiring discussions.
      This  work  made  use  of  the  following  open  source
software: \galpak\ \citep{BoucheN_15b},
 \textsc{matplotlib} \citep{matplotlib}, 
\textsc{NumPy} \citep{numpy}, 
\textsc{SciPy} \citep{scipy}, 
\textsc{Colossus} \citep{DiemerB_15b},
\textsc{Astropy}  \citep{astropy2018}.

This study is based on observations collected at the European Southern
Observatory under ESO programme 1101.A-0127.
We thank the TNG collaboration for making their data available at \url{http://www.tng-project.org}.
This work has been carried out thanks to the support of the ANR 3DGasFlows (ANR-17-CE31-0017), the OCEVU Labex (ANR-11-LABX-0060). BE acknowledges financial support from
the Programme National Cosmology et Galaxies (PNCG) of CNRS/INSU with INP and IN2P3, co-funded by CEA and CNES. R.B. acknowledges support
from the ERC advanced grant 339659-MUSICOS.
SLZ acknowledges support by The Netherlands Organisation for Scientific Research~(NWO) through a TOP Grant Module~1 under project number 614.001.652.
JB acknowledges support by Fundação para a Ciência e a Tecnologia (FCT) through research grants UIDB/04434/2020 and UIDP/04434/2020 and work contract `2020.03379.CEECIND.`

\end{acknowledgements}

%
  \bibliographystyle{aa} 
   \bibliography{references} 

\begin{thebibliography}{219}
\expandafter\ifx\csname natexlab\endcsname\relax\def\natexlab#1{#1}\fi

\bibitem[{{Abril-Melgarejo} {et~al.}(2021){Abril-Melgarejo}, {Epinat},
  {Mercier}, {Contini}, {Boogaard}, {Brinchmann}, {Finley}, {Michel-Dansac},
  {Ventou}, {Amram}, {Krajnovi{\'c}}, {Mahler}, {Pineda}, \&
  {Richard}}]{AbrilV_21a}
{Abril-Melgarejo}, V., {Epinat}, B., {Mercier}, W., {et~al.} 2021, \aap, 647,
  A152

\bibitem[{Allaert {et~al.}(2017)Allaert, Gentile, \& Baes}]{AllaertF_17a}
Allaert, F., Gentile, G., \& Baes, M. 2017, \aap, 605, A55

\bibitem[{{Amodeo} {et~al.}(2016){Amodeo}, {Ettori}, {Capasso}, \&
  {Sereno}}]{AmodeoS_16a}
{Amodeo}, S., {Ettori}, S., {Capasso}, R., \& {Sereno}, M. 2016, \aap, 590,
  A126

\bibitem[{{Bacon} {et~al.}(2010){Bacon}, {Accardo}, {Adjali}, {Anwand},
  {Bauer}, {Biswas}, {Blaizot}, {Boudon}, {Brau-Nogue}, {Brinchmann},
  {Caillier}, {Capoani}, {Carollo}, {Contini}, {Couderc}, {Daguis{\'e}},
  {Deiries}, {Delabre}, {Dreizler}, {Dubois}, {Dupieux}, {Dupuy}, {Emsellem},
  {Fechner}, {Fleischmann}, {Fran{\c c}ois}, {Gallou}, {Gharsa}, {Glindemann},
  {Gojak}, {Guiderdoni}, {Hansali}, {Hahn}, {Jarno}, {Kelz}, {Koehler},
  {Kosmalski}, {Laurent}, {Le Floch}, {Lilly}, {Lizon}, {Loupias}, {Manescau},
  {Monstein}, {Nicklas}, {Olaya}, {Pares}, {Pasquini}, {P{\'e}contal-Rousset},
  {Pell{\'o}}, {Petit}, {Popow}, {Reiss}, {Remillieux}, {Renault}, {Roth},
  {Rupprecht}, {Serre}, {Schaye}, {Soucail}, {Steinmetz}, {Streicher}, {Stuik},
  {Valentin}, {Vernet}, {Weilbacher}, {Wisotzki}, \& {Yerle}}]{BaconR_10a}
{Bacon}, R., {Accardo}, M., {Adjali}, L., {et~al.} 2010, in Society of
  Photo-Optical Instrumentation Engineers (SPIE) Conference Series, Vol. 7735,
  Society of Photo-Optical Instrumentation Engineers (SPIE) Conference Series,
  8

\bibitem[{{Bacon} {et~al.}(2017){Bacon}, {Conseil}, {Mary}, {Brinchmann},
  {Shepherd}, {Akhlaghi}, {Weilbacher}, {Piqueras}, {Wisotzki}, {Lagattuta},
  {Epinat}, {Guerou}, {Inami}, {Cantalupo}, {Courbot}, {Contini}, {Richard},
  {Maseda}, {Bouwens}, {Bouch{\'e}}, {Kollatschny}, {Schaye}, {Marino},
  {Pello}, {Herenz}, {Guiderdoni}, \& {Carollo}}]{BaconR_17a}
{Bacon}, R., {Conseil}, S., {Mary}, D., {et~al.} 2017, \aap, 608, A1

\bibitem[{{Bacon} {et~al.}(2021){Bacon}, {Mary}, {Garel}, {Blaizot}, {Maseda},
  {Schaye}, {Wisotzki}, {Conseil}, {Brinchmann}, {Leclercq}, {Abril-Melgarejo},
  {Boogaard}, {Bouch{\'e}}, {Contini}, {Feltre}, {Guiderdoni}, {Herenz},
  {Kollatschny}, {Kusakabe}, {Matthee}, {Michel-Dansac}, {Nanayakkara},
  {Richard}, {Roth}, {Schmidt}, {Steinmetz}, {Tresse}, {Urrutia}, {Verhamme},
  {Weilbacher}, {Zabl}, \& {Zoutendijk}}]{BaconR_21a}
{Bacon}, R., {Mary}, D., {Garel}, T., {et~al.} 2021, \aap, 647, A107

\bibitem[{{Behroozi} {et~al.}(2019){Behroozi}, {Wechsler}, {Hearin}, \&
  {Conroy}}]{BehrooziP_19a}
{Behroozi}, P., {Wechsler}, R.~H., {Hearin}, A.~P., \& {Conroy}, C. 2019,
  \mnras, 488, 3143

\bibitem[{{Behroozi} {et~al.}(2013){Behroozi}, {Wechsler}, \&
  {Conroy}}]{BehrooziP_13b}
{Behroozi}, P.~S., {Wechsler}, R.~H., \& {Conroy}, C. 2013, \apj, 770, 57

\bibitem[{{Bershady} {et~al.}(2010){Bershady}, {Verheijen}, {Swaters},
  {Andersen}, {Westfall}, \& {Martinsson}}]{BershadyM_10a}
{Bershady}, M.~A., {Verheijen}, M. A.~W., {Swaters}, R.~A., {et~al.} 2010,
  \apj, 716, 198

\bibitem[{{Binney} \& {Tremaine}(1987)}]{BinneyJ_87a}
{Binney}, J. \& {Tremaine}, S. 1987, {Galactic dynamics} (Princeton, NJ:
  Princeton Univ. Press)

\bibitem[{{Biviano} {et~al.}(2017){Biviano}, {Moretti}, {Paccagnella},
  {Poggianti}, {Bettoni}, {Gullieuszik}, {Vulcani}, {Fasano}, {D'Onofrio},
  {Fritz}, \& {Cava}}]{BivianoA_17a}
{Biviano}, A., {Moretti}, A., {Paccagnella}, A., {et~al.} 2017, \aap, 607, A81

\bibitem[{{Blitz} \& {Rosolowsky}(2006)}]{BlitzL_06a}
{Blitz}, L. \& {Rosolowsky}, E. 2006, \apj, 650, 933

\bibitem[{{Bode} {et~al.}(2001){Bode}, {Ostriker}, \& {Turok}}]{BodeP_01a}
{Bode}, P., {Ostriker}, J.~P., \& {Turok}, N. 2001, \apj, 556, 93

\bibitem[{{Boogaard} {et~al.}(2018){Boogaard}, {Brinchmann}, {Bouch{\'e}},
  {Paalvast}, {Bacon}, {Bouwens}, {Contini}, {Gunawardhana}, {Inami}, {Marino},
  {Maseda}, {Mitchell}, {Nanayakkara}, {Richard}, {Schaye}, {Schreiber},
  {Tacchella}, {Wisotzki}, \& {Zabl}}]{BoogaardL_18a}
{Boogaard}, L.~A., {Brinchmann}, J., {Bouch{\'e}}, N., {et~al.} 2018, \aap,
  619, A27

\bibitem[{{Bouch{\'e}} {et~al.}(2016){Bouch{\'e}}, {Finley}, {Schroetter},
  {Murphy}, {Richter}, {Bacon}, {Contini}, {Richard}, {Wendt}, {Kamann},
  {Epinat}, {Cantalupo}, {Straka}, {Schaye}, {Martin}, {P{\'e}roux},
  {Wisotzki}, {Soto}, {Lilly}, {Carollo}, {Brinchmann}, \&
  {Kollatschny}}]{BoucheN_16a}
{Bouch{\'e}}, N., {Finley}, H., {Schroetter}, I., {et~al.} 2016, \apj, 820, 121

\bibitem[{{Bouch{\'e}} {et~al.}(2013){Bouch{\'e}}, {Murphy}, {Kacprzak},
  {P{\'e}roux}, {Contini}, {Martin}, \& {Dessauges-Zavadsky}}]{BoucheN_13a}
{Bouch{\'e}}, N., {Murphy}, M.~T., {Kacprzak}, G.~G., {et~al.} 2013, Science,
  341, 50

\bibitem[{{Bouch{\'e}} {et~al.}(2015{\natexlab{a}}){Bouch{\'e}}, {Carfantan},
  {Schroetter}, {Michel-Dansac}, \& {Contini}}]{BoucheN_15b}
{Bouch{\'e}}, N.~F., {Carfantan}, H., {Schroetter}, I., {Michel-Dansac}, L., \&
  {Contini}, T. 2015{\natexlab{a}}, {GalPaK 3D: Galaxy parameters and
  kinematics extraction from 3D data}, Astrophysics Source Code Library

\bibitem[{{Bouch{\'e}} {et~al.}(2015{\natexlab{b}}){Bouch{\'e}}, {Carfantan},
  {Schroetter}, {Michel-Dansac}, \& {Contini}}]{BoucheN_15a}
{Bouch{\'e}}, N.~F., {Carfantan}, H., {Schroetter}, I., {Michel-Dansac}, L., \&
  {Contini}, T. 2015{\natexlab{b}}, \aj, 150, 92

\bibitem[{{Bouch{\'e}} {et~al.}(2021){Bouch{\'e}}, {Genel}, {Pellissier},
  {Dubois}, {Contini}, {Epinat}, {Pillepich}, {Krajnovi{\'c}}, {Nelson},
  {Abril-Melgarejo}, {Richard}, {Boogaard}, {Maseda}, {Mercier}, {Bacon},
  {Steinmetz}, \& {Vogelsberger}}]{BoucheN_21a}
{Bouch{\'e}}, N.~F., {Genel}, S., {Pellissier}, A., {et~al.} 2021, \aap, 654,
  A49

\bibitem[{{Bovy} \& {Rix}(2013)}]{BovyJ_13a}
{Bovy}, J. \& {Rix}, H.-W. 2013, \apj, 779, 115

\bibitem[{{Broeils} \& {Rhee}(1997)}]{BroeilsA_97a}
{Broeils}, A.~H. \& {Rhee}, M.~H. 1997, \aap, 324, 877

\bibitem[{Bryan \& Norman(1998)}]{BryanG_98a}
Bryan, G.~L. \& Norman, M.~L. 1998, \apj, 495, 80

\bibitem[{{Buchner} {et~al.}(2014){Buchner}, {Georgakakis}, {Nandra}, {Hsu},
  {Rangel}, {Brightman}, {Merloni}, {Salvato}, {Donley}, \&
  {Kocevski}}]{BuchnerJ_14a}
{Buchner}, J., {Georgakakis}, A., {Nandra}, K., {et~al.} 2014, \aap, 564, A125

\bibitem[{{Bullock} \& {Boylan-Kolchin}(2017)}]{BullockJ_17a}
{Bullock}, J.~S. \& {Boylan-Kolchin}, M. 2017, \araa, 55, 343

\bibitem[{{Bullock} {et~al.}(2001){Bullock}, {Kolatt}, {Sigad}, {Somerville},
  {Kravtsov}, {Klypin}, {Primack}, \& {Dekel}}]{BullockJ_01b}
{Bullock}, J.~S., {Kolatt}, T.~S., {Sigad}, Y., {et~al.} 2001, \mnras, 321, 559

\bibitem[{Buote {et~al.}(2007)Buote, Gastaldello, Humphrey, Zappacosta,
  Bullock, Brighenti, \& Mathews}]{BuoteA_07a}
Buote, D.~A., Gastaldello, F., Humphrey, P.~J., {et~al.} 2007, \apj, 664, 123

\bibitem[{{Burkert}(1995)}]{BurkertA_95a}
{Burkert}, A. 1995, \apjl, 447, L25

\bibitem[{{Burkert}(2000)}]{BurkertA_00a}
{Burkert}, A. 2000, \apjl, 534, L143

\bibitem[{Burkert(2015)}]{BurkertA_15a}
Burkert, A. 2015, \apj, 808, 158

\bibitem[{{Burkert}(2020)}]{BurkertA_20a}
{Burkert}, A. 2020, \apj, 904, 161

\bibitem[{{Burkert} {et~al.}(2010){Burkert}, {Genzel}, {Bouch{\'e}}, {Cresci},
  {Khochfar}, {Sommer-Larsen}, {Sternberg}, {Naab}, {F{\"o}rster Schreiber},
  {Tacconi}, {Shapiro}, {Hicks}, {Lutz}, {Davies}, {Buschkamp}, \&
  {Genel}}]{BurkertA_10a}
{Burkert}, A., {Genzel}, R., {Bouch{\'e}}, N., {et~al.} 2010, \apj, 725, 2324

\bibitem[{{Cappellari}(2002)}]{CappellariM_02a}
{Cappellari}, M. 2002, \mnras, 333, 400

\bibitem[{{Cappellari}(2008)}]{CappellariM_08a}
{Cappellari}, M. 2008, \mnras, 390, 71

\bibitem[{{Cappellari} {et~al.}(2002){Cappellari}, {Verolme}, {van der Marel},
  {Verdoes Kleijn}, {Illingworth}, {Franx}, {Carollo}, \& {de
  Zeeuw}}]{CappellariM_02b}
{Cappellari}, M., {Verolme}, E.~K., {van der Marel}, R.~P., {et~al.} 2002,
  \apj, 578, 787

\bibitem[{{Catinella} {et~al.}(2006){Catinella}, {Giovanelli}, \&
  {Haynes}}]{CatinellaB_06a}
{Catinella}, B., {Giovanelli}, R., \& {Haynes}, M.~P. 2006, \apj, 640, 751

\bibitem[{{Chabrier}(2003)}]{ChabrierG_03a}
{Chabrier}, G. 2003, \pasp, 115, 763

\bibitem[{{Chan} {et~al.}(2015){Chan}, {Kere{\v{s}}}, {O{\~n}orbe}, {Hopkins},
  {Muratov}, {Faucher-Gigu{\`e}re}, \& {Quataert}}]{ChanTK_15a}
{Chan}, T.~K., {Kere{\v{s}}}, D., {O{\~n}orbe}, J., {et~al.} 2015, \mnras, 454,
  2981

\bibitem[{{Contini} {et~al.}(2016){Contini}, {Epinat}, {Bouch{\'e}},
  {Brinchmann}, {Boogaard}, {Ventou}, {Bacon}, {Richard}, {Weilbacher},
  {Wisotzki}, {Krajnovi{\'c}}, {Vielfaure}, {Emsellem}, {Finley}, {Inami},
  {Schaye}, {Swinbank}, {Gu{\'e}rou}, {Martinsson}, {Michel-Dansac},
  {Schroetter}, {Shirazi}, \& {Soucail}}]{ContiniT_16a}
{Contini}, T., {Epinat}, B., {Bouch{\'e}}, N., {et~al.} 2016, \aap, 591, A49

\bibitem[{{Correa} {et~al.}(2015){Correa}, {Wyithe}, {Schaye}, \&
  {Duffy}}]{CorreaC_15c}
{Correa}, C.~A., {Wyithe}, J. S.~B., {Schaye}, J., \& {Duffy}, A.~R. 2015,
  \mnras, 452, 1217

\bibitem[{{Courteau}(1997)}]{CourteauS_97a}
{Courteau}, S. 1997, \aj, 114, 2402

\bibitem[{{Courteau} \& {Dutton}(2015)}]{CourteauS_15a}
{Courteau}, S. \& {Dutton}, A.~A. 2015, \apj, 801, L20

\bibitem[{{Cresci} {et~al.}(2009){Cresci}, {Hicks}, {Genzel}, {Schreiber},
  {Davies}, {Bouch{\'e}}, {Buschkamp}, {Genel}, {Shapiro}, {Tacconi},
  {Sommer-Larsen}, {Burkert}, {Eisenhauer}, {Gerhard}, {Lutz}, {Naab},
  {Sternberg}, {Cimatti}, {Daddi}, {Erb}, {Kurk}, {Lilly}, {Renzini},
  {Shapley}, {Steidel}, \& {Caputi}}]{CresciG_09a}
{Cresci}, G., {Hicks}, E.~K.~S., {Genzel}, R., {et~al.} 2009, \apj, 697, 115

\bibitem[{{da Cunha} {et~al.}(2008){da Cunha}, {Charlot}, \&
  {Elbaz}}]{daCunha_08a}
{da Cunha}, E., {Charlot}, S., \& {Elbaz}, D. 2008, \mnras, 388, 1595

\bibitem[{{da Cunha} {et~al.}(2015){da Cunha}, {Walter}, {Smail}, {Swinbank},
  {Simpson}, {Decarli}, {Hodge}, {Weiss}, {van der Werf}, {Bertoldi},
  {Chapman}, {Cox}, {Danielson}, {Dannerbauer}, {Greve}, {Ivison}, {Karim}, \&
  {Thomson}}]{daCunha_15a}
{da Cunha}, E., {Walter}, F., {Smail}, I.~R., {et~al.} 2015, \apj, 806, 110

\bibitem[{{Dalcanton} \& {Stilp}(2010)}]{DalcantonJ_10a}
{Dalcanton}, J.~J. \& {Stilp}, A.~M. 2010, \apj, 721, 547

\bibitem[{{de Blok} \& {McGaugh}(1997)}]{deBlokW_97a}
{de Blok}, W.~J.~G. \& {McGaugh}, S.~S. 1997, \mnras, 290, 533

\bibitem[{{de Blok} {et~al.}(2001){de Blok}, {McGaugh}, \&
  {Rubin}}]{deBlokW_01a}
{de Blok}, W.~J.~G., {McGaugh}, S.~S., \& {Rubin}, V.~C. 2001, \aj, 122, 2396

\bibitem[{{Dekel} {et~al.}(2003){Dekel}, {Arad}, {Devor}, \&
  {Birnboim}}]{DekelA_03c}
{Dekel}, A., {Arad}, I., {Devor}, J., \& {Birnboim}, Y. 2003, \apj, 588, 680

\bibitem[{{Dekel} {et~al.}(2021){Dekel}, {Freundlich}, {Jiang}, {Lapiner},
  {Burkert}, {Ceverino}, {Du}, {Genzel}, \& {Primack}}]{DekelA_21a}
{Dekel}, A., {Freundlich}, J., {Jiang}, F., {et~al.} 2021, \mnras, submitted
  [\eprint{2106.01378}]

\bibitem[{{Dekel} {et~al.}(2017){Dekel}, {Ishai}, {Dutton}, \&
  {Maccio}}]{DekelA_17a}
{Dekel}, A., {Ishai}, G., {Dutton}, A.~A., \& {Maccio}, A.~V. 2017, \mnras,
  468, 1005

\bibitem[{{Di Cintio} {et~al.}(2014){Di Cintio}, Brook, Dutton, Macci\`{o},
  Stinson, \& Knebe}]{DiCintioA_14a}
{Di Cintio}, A., Brook, C.~B., Dutton, A.~a., {et~al.} 2014, \mnras, 441, 2986

\bibitem[{{Di Teodoro} \& {Fraternali}(2015)}]{DiTeodoroE_15a}
{Di Teodoro}, E.~M. \& {Fraternali}, F. 2015, \mnras, 451, 3021

\bibitem[{{Diemer}(2015)}]{DiemerB_15b}
{Diemer}, B. 2015, {Colossus: COsmology, haLO, and large-Scale StrUcture
  toolS}, Astrophysics Source Code Library

\bibitem[{Di Paolo {et~al.}(2019)Di Paolo, Salucci, \& Erkurt}]{DiPaoloC_19a}
Di Paolo, C., Salucci, P., \& Erkurt, A. 2019, \mnras, 490, 5451

\bibitem[{{Djorgovski}(1992)}]{DjorgovskiS_92a}
{Djorgovski}, S.~G. 1992, in Astronomical Society of the Pacific Conference
  Series, Vol.~24, Cosmology and Large-Scale Structure in the Universe, ed.
  R.~R. {de Carvalho}, 19

\bibitem[{{Donato} {et~al.}(2009){Donato}, {Gentile}, {Salucci}, {Frigerio
  Martins}, {Wilkinson}, {Gilmore}, {Grebel}, {Koch}, \& {Wyse}}]{DonatoF_09a}
{Donato}, F., {Gentile}, G., {Salucci}, P., {et~al.} 2009, \mnras, 397, 1169

\bibitem[{{Duffy} {et~al.}(2008){Duffy}, {Schaye}, {Kay}, \& {Dalla
  Vecchia}}]{DuffyM_08a}
{Duffy}, A.~R., {Schaye}, J., {Kay}, S.~T., \& {Dalla Vecchia}, C. 2008,
  \mnras, 390, L64

\bibitem[{Dutton {et~al.}(2020)Dutton, Buck, Macci\`{o}, Dixon, Blank, \&
  Obreja}]{DuttonA_20a}
Dutton, A.~A., Buck, T., Macci\`{o}, A.~V., {et~al.} 2020, \mnras, 499, 2648

\bibitem[{Dutton \& Macci\`{o}(2014)}]{DuttonA_14a}
Dutton, A.~A. \& Macci\`{o}, A.~V. 2014, \mnras, 441, 3359

\bibitem[{Dutton {et~al.}(2016)Dutton, Macciò, Dekel, Wang, Stinson, Obreja,
  Di Cintio, Brook, Buck, \& Kang}]{DuttonA_16a}
Dutton, A.~A., Macciò, A.~V., Dekel, A., {et~al.} 2016, \mnras, 461, 2658

\bibitem[{{Einasto}(1965)}]{EinastoJ_65a}
{Einasto}, J. 1965, Trudy Astrofizicheskogo Instituta Alma-Ata, 5, 87

\bibitem[{{Eke} {et~al.}(2001){Eke}, {Navarro}, \& {Steinmetz}}]{EkeV_01a}
{Eke}, V.~R., {Navarro}, J.~F., \& {Steinmetz}, M. 2001, \apj, 554, 114

\bibitem[{El-Zant {et~al.}(2016)El-Zant, Freundlich, \& Combes}]{ElZantA_16a}
El-Zant, A.~A., Freundlich, J., \& Combes, F. 2016, \mnras, 461, 1745

\bibitem[{{Emsellem} {et~al.}(1994{\natexlab{a}}){Emsellem}, {Monnet}, \&
  {Bacon}}]{EmsellemE_94b}
{Emsellem}, E., {Monnet}, G., \& {Bacon}, R. 1994{\natexlab{a}}, \aap, 285, 723

\bibitem[{{Emsellem} {et~al.}(1994{\natexlab{b}}){Emsellem}, {Monnet}, {Bacon},
  \& {Nieto}}]{EmsellemE_94a}
{Emsellem}, E., {Monnet}, G., {Bacon}, R., \& {Nieto}, J.~L.
  1994{\natexlab{b}}, \aap, 285, 739

\bibitem[{{Epinat} {et~al.}(2012){Epinat}, {Tasca}, {Amram}, {Contini}, {Le
  F{\`e}vre}, {Queyrel}, {Vergani}, {Garilli}, {Kissler-Patig}, {Moultaka},
  {Paioro}, {Tresse}, {Bournaud}, {L{\'o}pez-Sanjuan}, \&
  {Perret}}]{EpinatB_12a}
{Epinat}, B., {Tasca}, L., {Amram}, P., {et~al.} 2012, \aap, 539, A92

\bibitem[{{Ettori} {et~al.}(2010){Ettori}, {Gastaldello}, {Leccardi},
  {Molendi}, {Rossetti}, {Buote}, \& {Meneghetti}}]{EttoriS_10a}
{Ettori}, S., {Gastaldello}, F., {Leccardi}, A., {et~al.} 2010, \aap, 524, A68

\bibitem[{{Famaey} {et~al.}(2018){Famaey}, {Khoury}, \& {Penco}}]{FamaeyB_18a}
{Famaey}, B., {Khoury}, J., \& {Penco}, R. 2018, \jcap, 2018, 038

\bibitem[{Feroz {et~al.}(2009)Feroz, Hobson, \& Bridges}]{multinest}
Feroz, F., Hobson, M.~P., \& Bridges, M. 2009, \mnras, 398, 1601

\bibitem[{{F{\"o}rster Schreiber} {et~al.}(2006){F{\"o}rster Schreiber},
  {Genzel}, {Lehnert}, {Bouch{\'e}}, {Verma}, {Erb}, {Shapley}, {Steidel},
  {Davies}, {Lutz}, {Nesvadba}, {Tacconi}, {Eisenhauer}, {Abuter}, {Gilbert},
  {Gillessen}, \& {Sternberg}}]{ForsterSchreiberN_06a}
{F{\"o}rster Schreiber}, N.~M., {Genzel}, R., {Lehnert}, M.~D., {et~al.} 2006,
  \apj, 645, 1062

\bibitem[{{F{\"o}rster Schreiber} {et~al.}(2018){F{\"o}rster Schreiber},
  {Renzini}, {Mancini}, {Genzel}, {Bouch{\'e}}, {Cresci}, {Hicks}, {Lilly},
  {Peng}, {Burkert}, {Carollo}, {Cimatti}, {Daddi}, {Davies}, {Genel}, {Kurk},
  {Lang}, {Lutz}, {Mainieri}, {McCracken}, {Mignoli}, {Naab}, {Oesch},
  {Pozzetti}, {Scodeggio}, {Shapiro Griffin}, {Shapley}, {Sternberg},
  {Tacchella}, {Tacconi}, {Wuyts}, \& {Zamorani}}]{ForsterSchreiberN_18a}
{F{\"o}rster Schreiber}, N.~M., {Renzini}, A., {Mancini}, C., {et~al.} 2018,
  \apjs, 238, 21

\bibitem[{{F{\"o}rster Schreiber} \& {Wuyts}(2020)}]{ForsterSchreiberN_20a}
{F{\"o}rster Schreiber}, N.~M. \& {Wuyts}, S. 2020, \araa, 58, 661

\bibitem[{{Frank} {et~al.}(2016){Frank}, {de Blok}, {Walter}, {Leroy}, \&
  {Carignan}}]{FrankB_16a}
{Frank}, B.~S., {de Blok}, W.~J.~G., {Walter}, F., {Leroy}, A., \& {Carignan},
  C. 2016, \aj, 151, 94

\bibitem[{{Fraternali} {et~al.}(2021){Fraternali}, {Karim}, {Magnelli},
  {G{\'o}mez-Guijarro}, {Jim{\'e}nez-Andrade}, \& {Posses}}]{FraternaliF_21a}
{Fraternali}, F., {Karim}, A., {Magnelli}, B., {et~al.} 2021, \aap, 647, A194

\bibitem[{{Freeman}(1970)}]{FreemanK_70a}
{Freeman}, K.~C. 1970, \apj, 160, 811

\bibitem[{{Freundlich} {et~al.}(2019){Freundlich}, {Combes}, {Tacconi},
  {Genzel}, {Garcia-Burillo}, {Neri}, {Contini}, {Bolatto}, {Lilly},
  {Salom{\'e}}, {Bicalho}, {Boissier}, {Boone}, {Bouch{\'e}}, {Bournaud},
  {Burkert}, {Carollo}, {Cooper}, {Cox}, {Feruglio}, {F{\"o}rster Schreiber},
  {Juneau}, {Lippa}, {Lutz}, {Naab}, {Renzini}, {Saintonge}, {Sternberg},
  {Walter}, {Weiner}, {Wei{\ss}}, \& {Wuyts}}]{FreundlichJ_19a}
{Freundlich}, J., {Combes}, F., {Tacconi}, L.~J., {et~al.} 2019, \aap, 622,
  A105

\bibitem[{{Freundlich} {et~al.}(2020{\natexlab{a}}){Freundlich}, {Dekel},
  {Jiang}, {Ishai}, {Cornuault}, {Lapiner}, {Dutton}, \&
  {Macci{\`o}}}]{FreundlichJ_20a}
{Freundlich}, J., {Dekel}, A., {Jiang}, F., {et~al.} 2020{\natexlab{a}},
  \mnras, 491, 4523

\bibitem[{{Freundlich} {et~al.}(2020{\natexlab{b}}){Freundlich}, {Jiang},
  {Dekel}, {Cornuault}, {Ginzburg}, {Koskas}, {Lapiner}, {Dutton}, \&
  {Macci{\`o}}}]{FreundlichJ_20b}
{Freundlich}, J., {Jiang}, F., {Dekel}, A., {et~al.} 2020{\natexlab{b}},
  \mnras, 499, 2912

\bibitem[{Gelman {et~al.}(2014)Gelman, Hwang, \& Vehtari}]{GelmanA_14a}
Gelman, A., Hwang, J., \& Vehtari, A. 2014, Statistics and Computing, 24, 997

\bibitem[{{Gentile} {et~al.}(2004){Gentile}, {Salucci}, {Klein}, {Vergani}, \&
  {Kalberla}}]{GentileG_04a}
{Gentile}, G., {Salucci}, P., {Klein}, U., {Vergani}, D., \& {Kalberla}, P.
  2004, \mnras, 351, 903

\bibitem[{{Genzel} {et~al.}(2008){Genzel}, {Burkert}, {Bouch{\'e}}, {Cresci},
  {F{\"o}rster Schreiber}, {Shapley}, {Shapiro}, {Tacconi}, {Buschkamp},
  {Cimatti}, {Daddi}, {Davies}, {Eisenhauer}, {Erb}, {Genel}, {Gerhard},
  {Hicks}, {Lutz}, {Naab}, {Ott}, {Rabien}, {Renzini}, {Steidel}, {Sternberg},
  \& {Lilly}}]{GenzelR_08a}
{Genzel}, R., {Burkert}, A., {Bouch{\'e}}, N., {et~al.} 2008, \apj, 687, 59

\bibitem[{Genzel {et~al.}(2020)Genzel, Price, {\"{U}}bler, {F{\"{o}}rster
  Schreiber}, Shimizu, Tacconi, Bender, Burkert, Contursi, Coogan, Davies,
  Davies, Dekel, Herrera-Camus, Lee, Lutz, Naab, Neri, Nestor, Renzini, Saglia,
  Schuster, Sternberg, Wisnioski, \& Wuyts}]{GenzelR_20a}
Genzel, R., Price, S.~H., {\"{U}}bler, H., {et~al.} 2020, \apj, 902, 98

\bibitem[{{Genzel} {et~al.}(2017){Genzel}, {Schreiber}, {{\"U}bler}, {Lang},
  {Naab}, {Bender}, {Tacconi}, {Wisnioski}, {Wuyts}, {Alexander}, {Beifiori},
  {Belli}, {Brammer}, {Burkert}, {Carollo}, {Chan}, {Davies}, {Fossati},
  {Galametz}, {Genel}, {Gerhard}, {Lutz}, {Mendel}, {Momcheva}, {Nelson},
  {Renzini}, {Saglia}, {Sternberg}, {Tacchella}, {Tadaki}, \&
  {Wilman}}]{GenzelR_17a}
{Genzel}, R., {Schreiber}, N.~M.~F., {{\"U}bler}, H., {et~al.} 2017, \nat, 543,
  397

\bibitem[{{Genzel} {et~al.}(2006){Genzel}, {Tacconi}, {Eisenhauer},
  {F{\"o}rster Schreiber}, {Cimatti}, {Daddi}, {Bouch{\'e}}, {Davies},
  {Lehnert}, {Lutz}, {Nesvadba}, {Verma}, {Abuter}, {Shapiro}, {Sternberg},
  {Renzini}, {Kong}, {Arimoto}, \& {Mignoli}}]{GenzelR_06a}
{Genzel}, R., {Tacconi}, L.~J., {Eisenhauer}, F., {et~al.} 2006, \nat, 442, 786

\bibitem[{{Ghari} {et~al.}(2019){Ghari}, {Famaey}, {Laporte}, \&
  {Haghi}}]{GhariA_19a}
{Ghari}, A., {Famaey}, B., {Laporte}, C., \& {Haghi}, H. 2019, \aap, 623, A123

\bibitem[{{Goerdt} {et~al.}(2010){Goerdt}, {Dekel}, {Sternberg}, {Ceverino},
  {Teyssier}, \& {Primack}}]{GoerdtT_10a}
{Goerdt}, T., {Dekel}, A., {Sternberg}, A., {et~al.} 2010, \mnras, 407, 613

\bibitem[{Goerdt {et~al.}(2006)Goerdt, Moore, Read, Stadel, \&
  Zemp}]{GoerdtT_06a}
Goerdt, T., Moore, B., Read, J.~I., Stadel, J., \& Zemp, M. 2006, Monthly
  Notices of the Royal Astronomical Society, 368, 1073

\bibitem[{Hernquist(1990)}]{HernquistL_90a}
Hernquist, L. 1990, \apj, 356, 359

\bibitem[{{Ho} {et~al.}(2017){Ho}, {Martin}, {Kacprzak}, \&
  {Churchill}}]{HoS_17a}
{Ho}, S.~H., {Martin}, C.~L., {Kacprzak}, G.~G., \& {Churchill}, C.~W. 2017,
  \apj, 835, 267

\bibitem[{{Ho} {et~al.}(2019){Ho}, {Martin}, \& {Turner}}]{HoS_19a}
{Ho}, S.~H., {Martin}, C.~L., \& {Turner}, M.~L. 2019, \apj, 875, 54

\bibitem[{{Hu} {et~al.}(2000){Hu}, {Barkana}, \& {Gruzinov}}]{HuW_00a}
{Hu}, W., {Barkana}, R., \& {Gruzinov}, A. 2000, \prl, 85, 1158

\bibitem[{{Hunter}(2007)}]{matplotlib}
{Hunter}, J.~D. 2007, Computing in Science and Engineering, 9, 90

\bibitem[{Ianjamasimanana {et~al.}(2018)Ianjamasimanana, Walter, Blok, Heald,
  \& Brinks}]{Ianja_18a}
Ianjamasimanana, R., Walter, F., Blok, W.~J., Heald, G.~H., \& Brinks, E. 2018,
  \aj, 155, 233

\bibitem[{{Inami} {et~al.}(2017){Inami}, {Bacon}, {Brinchmann}, {Richard},
  {Contini}, {Conseil}, {Hamer}, {Akhlaghi}, {Bouch{\'e}}, {Cl{\'e}ment},
  {Desprez}, {Drake}, {Hashimoto}, {Leclercq}, {Maseda}, {Michel-Dansac},
  {Paalvast}, {Tresse}, {Ventou}, {Kollatschny}, {Boogaard}, {Finley},
  {Marino}, {Schaye}, \& {Wisotzki}}]{InamiH_17a}
{Inami}, H., {Bacon}, R., {Brinchmann}, J., {et~al.} 2017, \aap, 608, A2

\bibitem[{{Jaffe}(1983)}]{JaffeW_83a}
{Jaffe}, W. 1983, \mnras, 202, 995

\bibitem[{{Jeffreys}(1961)}]{JeffreysH_61a}
{Jeffreys}, H. 1961, {The Theory of Probability} (Oxford U. Press)

\bibitem[{Jenkins \& Peacock(2018)}]{JenkinsC_18a}
Jenkins, C.~R. \& Peacock, J.~A. 2018, \mnras, 413, 2895

\bibitem[{{Jones} {et~al.}(2001){Jones}, {Oliphant}, {Peterson},
  {et~al.}}]{scipy}
{Jones}, E., {Oliphant}, T., {Peterson}, P., {et~al.} 2001

\bibitem[{{Karukes} \& {Salucci}(2017)}]{KarukesE_17a}
{Karukes}, E.~V. \& {Salucci}, P. 2017, \mnras, 465, 4703

\bibitem[{Kass \& Raftery(1995)}]{KassR_95a}
Kass, R.~E. \& Raftery, A.~E. 1995, Journal of the American Statistical
  Association, 90, 773

\bibitem[{Katz {et~al.}(2017)Katz, Lelli, Mcgaugh, Cintio, Brook, \&
  Schombert}]{KatzH_17a}
Katz, H., Lelli, F., Mcgaugh, S.~S., {et~al.} 2017, \mnras, 466, 1648

\bibitem[{{Kennicutt}(1998)}]{KennicuttR_98a}
{Kennicutt}, R.~C. 1998, \araa, 36, 189

\bibitem[{{Kormendy} \& {Freeman}(2004)}]{KormendyJ_04a}
{Kormendy}, J. \& {Freeman}, K.~C. 2004, in Dark Matter in Galaxies, ed.
  S.~{Ryder}, D.~{Pisano}, M.~{Walker}, \& K.~{Freeman}, Vol. 220, 377

\bibitem[{{Kormendy} \& {Freeman}(2016)}]{KormendyJ_16a}
{Kormendy}, J. \& {Freeman}, K.~C. 2016, \apj, 817, 84

\bibitem[{{Korsaga} {et~al.}(2018){Korsaga}, {Carignan}, {Amram}, {Epinat}, \&
  {Jarrett}}]{KorsagaM_18a}
{Korsaga}, M., {Carignan}, C., {Amram}, P., {Epinat}, B., \& {Jarrett}, T.~H.
  2018, \mnras, 478, 50

\bibitem[{{Korsaga} {et~al.}(2019){Korsaga}, {Epinat}, {Amram}, {Carignan},
  {Adamczyk}, \& {Sorgho}}]{KorsagaM_19b}
{Korsaga}, M., {Epinat}, B., {Amram}, P., {et~al.} 2019, \mnras, 490, 2977

\bibitem[{Kravtsov {et~al.}(1998)Kravtsov, Klypin, Bullock, \&
  Primack}]{KravtsovA_98a}
Kravtsov, A.~V., Klypin, A.~A., Bullock, J.~S., \& Primack, J.~R. 1998, The
  Astrophysical Journal, 502, 48

\bibitem[{Lang {et~al.}(2017)Lang, {Forster Schreiber}, Genzel, Wuyts,
  Wisnioski, Beifiori, Belli, Bender, \& al.}]{LangP_17a}
Lang, P., {Forster Schreiber}, N.~M., Genzel, R., {et~al.} 2017, ApJ, 1

\bibitem[{Lapi {et~al.}(2018)Lapi, Salucci, \& Danese}]{LapiA_18a}
Lapi, A., Salucci, P., \& Danese, L. 2018, The Astrophysical Journal, 859, 2

\bibitem[{{Laporte} \& {Pe{\~{n}}arrubia}(2015)}]{LaporteC_15a}
{Laporte}, C.~F. \& {Pe{\~{n}}arrubia}, J. 2015, \mnras, 449, L90

\bibitem[{Lazar {et~al.}(2020)Lazar, Bullock, Boylan-Kolchin, Chan, Hopkins,
  Graus, Wetzel, El-Badry, Wheeler, Straight, Keres, Faucher-Gigu{\`{e}}re,
  Fitts, \& Garrison-Kimmel}]{LazarA_20a}
Lazar, A., Bullock, J.~S., Boylan-Kolchin, M., {et~al.} 2020, \mnras, 497, 2393

\bibitem[{{Leauthaud} {et~al.}(2012){Leauthaud}, {Tinker}, {Bundy}, {Behroozi},
  {Massey}, {Rhodes}, {George}, {Kneib}, {Benson}, {Wechsler}, {Busha},
  {Capak}, {Cort{\^e}s}, {Ilbert}, {Koekemoer}, {Le F{\`e}vre}, {Lilly},
  {McCracken}, {Salvato}, {Schrabback}, {Scoville}, {Smith}, \&
  {Taylor}}]{LeauthaudA_12a}
{Leauthaud}, A., {Tinker}, J., {Bundy}, K., {et~al.} 2012, \apj, 744, 159

\bibitem[{{Leier} {et~al.}(2021){Leier}, {Ferreras}, {Negri}, \&
  {Saha}}]{LeierD_21a}
{Leier}, D., {Ferreras}, I., {Negri}, A., \& {Saha}, P. 2021, arXiv e-prints,
  arXiv:2105.05856

\bibitem[{{Leier} {et~al.}(2012){Leier}, {Ferreras}, \& {Saha}}]{LeierD_12a}
{Leier}, D., {Ferreras}, I., \& {Saha}, P. 2012, \mnras, 424, 104

\bibitem[{{Leier} {et~al.}(2016){Leier}, {Ferreras}, {Saha}, {Charlot},
  {Bruzual}, \& {La Barbera}}]{LeierD_16a}
{Leier}, D., {Ferreras}, I., {Saha}, P., {et~al.} 2016, \mnras, 459, 3677

\bibitem[{{Lelli} {et~al.}(2016){Lelli}, {McGaugh}, \&
  {Schombert}}]{LelliF_16a}
{Lelli}, F., {McGaugh}, S.~S., \& {Schombert}, J.~M. 2016, \aj, 152, 157

\bibitem[{{Leroy} {et~al.}(2008){Leroy}, {Walter}, {Brinks}, {Bigiel}, {de
  Blok}, {Madore}, \& {Thornley}}]{LeroyA_08a}
{Leroy}, A.~K., {Walter}, F., {Brinks}, E., {et~al.} 2008, \aj, 136, 2782

\bibitem[{{Li} {et~al.}(2020){Li}, {Lelli}, {McGaugh}, \&
  {Schombert}}]{LiLelli_20a}
{Li}, P., {Lelli}, F., {McGaugh}, S., \& {Schombert}, J. 2020, \apjs, 247, 31

\bibitem[{{Li} {et~al.}(2019){Li}, {Lelli}, {McGaugh}, {Starkman}, \&
  {Schombert}}]{LiLelli_19a}
{Li}, P., {Lelli}, F., {McGaugh}, S.~S., {Starkman}, N., \& {Schombert}, J.~M.
  2019, \mnras, 482, 5106

\bibitem[{{Lima Neto} {et~al.}(1999){Lima Neto}, {Gerbal}, \&
  {M{\'a}rquez}}]{LimaNetoG_99a}
{Lima Neto}, G.~B., {Gerbal}, D., \& {M{\'a}rquez}, I. 1999, \mnras, 309, 481

\bibitem[{{Lovell} {et~al.}(2018){Lovell}, {Pillepich}, {Genel}, {Nelson},
  {Springel}, {Pakmor}, {Marinacci}, {Weinberger}, {Torrey}, {Vogelsberger},
  {Alabi}, \& {Hernquist}}]{LovellM_18a}
{Lovell}, M.~R., {Pillepich}, A., {Genel}, S., {et~al.} 2018, \mnras, 481, 1950

\bibitem[{{Ludlow} {et~al.}(2014){Ludlow}, {Navarro}, {Angulo},
  {Boylan-Kolchin}, {Springel}, {Frenk}, \& {White}}]{LudlowA_14a}
{Ludlow}, A.~D., {Navarro}, J.~F., {Angulo}, R.~E., {et~al.} 2014, \mnras, 441,
  378

\bibitem[{Mandelbaum {et~al.}(2016)Mandelbaum, Wang, Zu, White, Henriques, \&
  More}]{MandelbaumR_16a}
Mandelbaum, R., Wang, W., Zu, Y., {et~al.} 2016, \mnras, 457, 3200

\bibitem[{{Martinsson} {et~al.}(2016){Martinsson}, {Verheijen}, {Bershady},
  {Westfall}, {Andersen}, \& {Swaters}}]{MartinssonT_16a}
{Martinsson}, T. P.~K., {Verheijen}, M. A.~W., {Bershady}, M.~A., {et~al.}
  2016, \aap, 585, A99

\bibitem[{Martinsson {et~al.}(2013)Martinsson, Verheijen, Westfall, Bershady,
  Andersen, \& Swaters}]{MartinssonT_13a}
Martinsson, T. P.~K., Verheijen, M. a.~W., Westfall, K.~B., {et~al.} 2013,
  \aap, 557, A131

\bibitem[{{Martinsson} {et~al.}(2013){Martinsson}, {Verheijen}, {Westfall},
  {Bershady}, {Schechtman-Rook}, {Andersen}, \& {Swaters}}]{MartinssonT_13b}
{Martinsson}, T. P.~K., {Verheijen}, M. A.~W., {Westfall}, K.~B., {et~al.}
  2013, \aap, 557, A130

\bibitem[{{Maseda} {et~al.}(2017){Maseda}, {Brinchmann}, {Franx}, {Bacon},
  {Bouwens}, {Schmidt}, {Boogaard}, {Contini}, {Feltre}, {Inami},
  {Kollatschny}, {Marino}, {Richard}, {Verhamme}, \& {Wisotzki}}]{MasedaM_17a}
{Maseda}, M.~V., {Brinchmann}, J., {Franx}, M., {et~al.} 2017, \aap, 608, A4

\bibitem[{{McGaugh}(2005)}]{McGaughS_05a}
{McGaugh}, S.~S. 2005, \apj, 632, 859

\bibitem[{{Meurer} {et~al.}(1996){Meurer}, {Carignan}, {Beaulieu}, \&
  {Freeman}}]{MeurerG_96a}
{Meurer}, G.~R., {Carignan}, C., {Beaulieu}, S.~F., \& {Freeman}, K.~C. 1996,
  \aj, 111, 1551

\bibitem[{{Meyer} {et~al.}(2008){Meyer}, {Zwaan}, {Webster}, {Schneider}, \&
  {Staveley-Smith}}]{MeyerM_08a}
{Meyer}, M.~J., {Zwaan}, M.~A., {Webster}, R.~L., {Schneider}, S., \&
  {Staveley-Smith}, L. 2008, \mnras, 391, 1712

\bibitem[{{Monnet} {et~al.}(1992){Monnet}, {Bacon}, \&
  {Emsellem}}]{MonnetG_92a}
{Monnet}, G., {Bacon}, R., \& {Emsellem}, E. 1992, \aap, 253, 366

\bibitem[{{Moster} {et~al.}(2010){Moster}, {Somerville}, {Maulbetsch}, {van den
  Bosch}, {Macci{\`o}}, {Naab}, \& {Oser}}]{MosterB_10a}
{Moster}, B.~P., {Somerville}, R.~S., {Maulbetsch}, C., {et~al.} 2010, \apj,
  710, 903

\bibitem[{Navarro {et~al.}(1996)Navarro, Eke, \& Frenk}]{NavarroJ_96b}
Navarro, J.~F., Eke, V.~R., \& Frenk, C.~S. 1996, \mnras, 283, L72

\bibitem[{{Navarro} {et~al.}(1997){Navarro}, {Frenk}, \&
  {White}}]{NavarroJ_97a}
{Navarro}, J.~F., {Frenk}, C.~S., \& {White}, S.~D.~M. 1997, \apj, 490, 493

\bibitem[{{Neeleman} {et~al.}(2020){Neeleman}, {Prochaska}, {Kanekar}, \&
  {Rafelski}}]{NeelemanM_20a}
{Neeleman}, M., {Prochaska}, J.~X., {Kanekar}, N., \& {Rafelski}, M. 2020,
  \nat, 581, 269

\bibitem[{{Nelson} {et~al.}(2019){Nelson}, {Pillepich}, {Springel}, {Pakmor},
  {Weinberger}, {Genel}, {Torrey}, {Vogelsberger}, {Marinacci}, \&
  {Hernquist}}]{NelsonD_19a}
{Nelson}, D., {Pillepich}, A., {Springel}, V., {et~al.} 2019, \mnras, 490, 3234

\bibitem[{{Nelson} {et~al.}(2016){Nelson}, {van Dokkum}, {F{\"o}rster
  Schreiber}, {Franx}, {Brammer}, {Momcheva}, {Wuyts}, {Whitaker}, {Skelton},
  {Fumagalli}, {Hayward}, {Kriek}, {Labb{\'e}}, {Leja}, {Rix}, {Tacconi}, {van
  der Wel}, {van den Bosch}, {Oesch}, {Dickey}, \& {Ulf Lange}}]{NelsonE_16a}
{Nelson}, E.~J., {van Dokkum}, P.~G., {F{\"o}rster Schreiber}, N.~M., {et~al.}
  2016, \apj, 828, 27

\bibitem[{{Newman} {et~al.}(2013){Newman}, {Treu}, {Ellis}, \&
  {Sand}}]{NewmanA_13a}
{Newman}, A.~B., {Treu}, T., {Ellis}, R.~S., \& {Sand}, D.~J. 2013, \apj, 765,
  25

\bibitem[{{Obreschkow} {et~al.}(2016){Obreschkow}, {Glazebrook}, {Kilborn}, \&
  {Lutz}}]{ObreschkowD_16a}
{Obreschkow}, D., {Glazebrook}, K., {Kilborn}, V., \& {Lutz}, K. 2016, \apjl,
  824, L26

\bibitem[{{Oh} {et~al.}(2011){Oh}, {de Blok}, {Brinks}, {Walter}, \&
  {Kennicutt}}]{OhSH_11a}
{Oh}, S.-H., {de Blok}, W.~J.~G., {Brinks}, E., {Walter}, F., \& {Kennicutt},
  Robert~C., J. 2011, \aj, 141, 193

\bibitem[{{Oh} {et~al.}(2015){Oh}, {Hunter}, {Brinks}, {Elmegreen}, {Schruba},
  {Walter}, {Rupen}, {Young}, {Simpson}, {Johnson}, {Herrmann}, {Ficut-Vicas},
  {Cigan}, {Heesen}, {Ashley}, \& {Zhang}}]{OhSH_15a}
{Oh}, S.-H., {Hunter}, D.~A., {Brinks}, E., {et~al.} 2015, \aj, 149, 180

\bibitem[{Oldham \& Auger(2018)}]{OldhamL_18a}
Oldham, L.~J. \& Auger, M.~W. 2018, \mnras, 476, 133

\bibitem[{{Oort}(1932)}]{OortJ_32a}
{Oort}, J.~H. 1932, \bain, 6, 249

\bibitem[{{Orkney} {et~al.}(2021){Orkney}, {Read}, {Rey}, {Nasim}, {Pontzen},
  {Agertz}, {Kim}, {Delorme}, \& {Dehnen}}]{OrkneyM_21a}
{Orkney}, M. D.~A., {Read}, J.~I., {Rey}, M.~P., {et~al.} 2021, \mnras, 504,
  3509

\bibitem[{{Peebles}(1974)}]{PeeblesP_74a}
{Peebles}, P.~J.~E. 1974, \apjl, 189, L51

\bibitem[{{Peebles} \& {Partridge}(1967)}]{PeeblesP_67a}
{Peebles}, P.~J.~E. \& {Partridge}, R.~B. 1967, \apj, 148, 713

\bibitem[{{Peebles} \& {Yu}(1970)}]{PeeblesP_70a}
{Peebles}, P.~J.~E. \& {Yu}, J.~T. 1970, \apj, 162, 815

\bibitem[{{Peirani} {et~al.}(2017){Peirani}, {Dubois}, {Volonteri},
  {Devriendt}, {Bundy}, {Silk}, {Pichon}, {Kaviraj}, {Gavazzi}, \&
  {Habouzit}}]{PeiraniS_17a}
{Peirani}, S., {Dubois}, Y., {Volonteri}, M., {et~al.} 2017, \mnras, 472, 2153

\bibitem[{{Pelliccia} {et~al.}(2017){Pelliccia}, {Tresse}, {Epinat}, {Ilbert},
  {Scoville}, {Amram}, {Lemaux}, \& {Zamorani}}]{PelliciaD_16a}
{Pelliccia}, D., {Tresse}, L., {Epinat}, B., {et~al.} 2017, \aap, 599, A25

\bibitem[{{Peng} {et~al.}(2002){Peng}, {Ho}, {Impey}, \& {Rix}}]{PengC_02a}
{Peng}, C.~Y., {Ho}, L.~C., {Impey}, C.~D., \& {Rix}, H.-W. 2002, \aj, 124, 266

\bibitem[{{Persic} {et~al.}(1996){Persic}, {Salucci}, \& {Stel}}]{PersicM_96a}
{Persic}, M., {Salucci}, P., \& {Stel}, F. 1996, \mnras, 281, 27

\bibitem[{{Pillepich} {et~al.}(2019){Pillepich}, {Nelson}, {Springel},
  {Pakmor}, {Torrey}, {Weinberger}, {Vogelsberger}, {Marinacci}, {Genel}, {van
  der Wel}, \& {Hernquist}}]{PillepichA_19a}
{Pillepich}, A., {Nelson}, D., {Springel}, V., {et~al.} 2019, \mnras, 490, 3196

\bibitem[{Pineda {et~al.}(2017)Pineda, Hayward, Springel, \&
  de~Oliveira}]{PinedaJ_17a}
Pineda, J.~C., Hayward, C.~C., Springel, V., \& de~Oliveira, C.~M. 2017,
  \mnras, 466, 63

\bibitem[{{Planck Collaboration XIII} {et~al.}(2016){Planck Collaboration
  XIII}, {Ade}, {Aghanim}, {Arnaud}, {Ashdown}, {Aumont}, {Baccigalupi},
  {Banday}, {Barreiro}, {Bartlett}, {Bartolo}, {Battaner}, {Battye}, {Benabed},
  {Beno{\^\i}t}, {Benoit-L{\'e}vy}, {Bernard}, {Bersanelli}, {Bielewicz},
  {Bock}, {Bonaldi}, {Bonavera}, {Bond}, {Borrill}, {Bouchet}, {Boulanger},
  {Bucher}, {Burigana}, {Butler}, {Calabrese}, {Cardoso}, {Catalano},
  {Challinor}, {Chamballu}, {Chary}, {Chiang}, {Chluba}, {Christensen},
  {Church}, {Clements}, {Colombi}, {Colombo}, {Combet}, {Coulais}, {Crill},
  {Curto}, {Cuttaia}, {Danese}, {Davies}, {Davis}, {de Bernardis}, {de Rosa},
  {de Zotti}, {Delabrouille}, {D{\'e}sert}, {Di Valentino}, {Dickinson},
  {Diego}, {Dolag}, {Dole}, {Donzelli}, {Dor{\'e}}, {Douspis}, {Ducout},
  {Dunkley}, {Dupac}, {Efstathiou}, {Elsner}, {En{\ss}lin}, {Eriksen},
  {Farhang}, {Fergusson}, {Finelli}, {Forni}, {Frailis}, {Fraisse},
  {Franceschi}, {Frejsel}, {Galeotta}, {Galli}, {Ganga}, {Gauthier}, {Gerbino},
  {Ghosh}, {Giard}, {Giraud-H{\'e}raud}, {Giusarma}, {Gjerl{\o}w},
  {Gonz{\'a}lez-Nuevo}, {G{\'o}rski}, {Gratton}, {Gregorio}, {Gruppuso},
  {Gudmundsson}, {Hamann}, {Hansen}, {Hanson}, {Harrison}, {Helou},
  {Henrot-Versill{\'e}}, {Hern{\'a}ndez-Monteagudo}, {Herranz}, {Hildebrand t},
  {Hivon}, {Hobson}, {Holmes}, {Hornstrup}, {Hovest}, {Huang}, {Huffenberger},
  {Hurier}, {Jaffe}, {Jaffe}, {Jones}, {Juvela}, {Keih{\"a}nen}, {Keskitalo},
  {Kisner}, {Kneissl}, {Knoche}, {Knox}, {Kunz}, {Kurki-Suonio}, {Lagache},
  {L{\"a}hteenm{\"a}ki}, {Lamarre}, {Lasenby}, {Lattanzi}, {Lawrence}, {Leahy},
  {Leonardi}, {Lesgourgues}, {Levrier}, {Lewis}, {Liguori}, {Lilje},
  {Linden-V{\o}rnle}, {L{\'o}pez-Caniego}, {Lubin}, {Mac{\'\i}as-P{\'e}rez},
  {Maggio}, {Maino}, {Mandolesi}, {Mangilli}, {Marchini}, {Maris}, {Martin},
  {Martinelli}, {Mart{\'\i}nez-Gonz{\'a}lez}, {Masi}, {Matarrese}, {McGehee},
  {Meinhold}, {Melchiorri}, {Melin}, {Mendes}, {Mennella}, {Migliaccio},
  {Millea}, {Mitra}, {Miville-Desch{\^e}nes}, {Moneti}, {Montier}, {Morgante},
  {Mortlock}, {Moss}, {Munshi}, {Murphy}, {Naselsky}, {Nati}, {Natoli},
  {Netterfield}, {N{\o}rgaard-Nielsen}, {Noviello}, {Novikov}, {Novikov},
  {Oxborrow}, {Paci}, {Pagano}, {Pajot}, {Paladini}, {Paoletti}, {Partridge},
  {Pasian}, {Patanchon}, {Pearson}, {Perdereau}, {Perotto}, {Perrotta},
  {Pettorino}, {Piacentini}, {Piat}, {Pierpaoli}, {Pietrobon}, {Plaszczynski},
  {Pointecouteau}, {Polenta}, {Popa}, {Pratt}, {Pr{\'e}zeau}, {Prunet},
  {Puget}, {Rachen}, {Reach}, {Rebolo}, {Reinecke}, {Remazeilles}, {Renault},
  {Renzi}, {Ristorcelli}, {Rocha}, {Rosset}, {Rossetti}, {Roudier},
  {Rouill{\'e} d'Orfeuil}, {Rowan-Robinson}, {Rubi{\~n}o-Mart{\'\i}n},
  {Rusholme}, {Said}, {Salvatelli}, {Salvati}, {Sandri}, {Santos},
  {Savelainen}, {Savini}, {Scott}, {Seiffert}, {Serra}, {Shellard}, {Spencer},
  {Spinelli}, {Stolyarov}, {Stompor}, {Sudiwala}, {Sunyaev}, {Sutton},
  {Suur-Uski}, {Sygnet}, {Tauber}, {Terenzi}, {Toffolatti}, {Tomasi},
  {Tristram}, {Trombetti}, {Tucci}, {Tuovinen}, {T{\"u}rler}, {Umana},
  {Valenziano}, {Valiviita}, {Van Tent}, {Vielva}, {Villa}, {Wade}, {Wandelt},
  {Wehus}, {White}, {White}, {Wilkinson}, {Yvon}, {Zacchei}, \&
  {Zonca}}]{Planck2015}
{Planck Collaboration XIII}, {Ade}, P.~A.~R., {Aghanim}, N., {et~al.} 2016,
  \aap, 594, A13

\bibitem[{{Pontzen} \& {Governato}(2012)}]{PontzenA_12a}
{Pontzen}, A. \& {Governato}, F. 2012, \mnras, 421, 3464

\bibitem[{{Posti} \& {Fall}(2021)}]{PostiL_21a}
{Posti}, L. \& {Fall}, S.~M. 2021, \aap, 649, A119

\bibitem[{{Posti} {et~al.}(2018){Posti}, {Fraternali}, {Di Teodoro}, \&
  {Pezzulli}}]{PostiL_18b}
{Posti}, L., {Fraternali}, F., {Di Teodoro}, E.~M., \& {Pezzulli}, G. 2018,
  \aap, 612, L6

\bibitem[{{Posti} {et~al.}(2019){Posti}, {Fraternali}, \&
  {Marasco}}]{PostiL_19b}
{Posti}, L., {Fraternali}, F., \& {Marasco}, A. 2019, \aap, 626, A56

\bibitem[{{Rafelski} {et~al.}(2015){Rafelski}, {Teplitz}, {Gardner}, {Coe},
  {Bond}, {Koekemoer}, {Grogin}, {Kurczynski}, {McGrath}, {Bourque}, {Atek},
  {Brown}, {Colbert}, {Codoreanu}, {Ferguson}, {Finkelstein}, {Gawiser},
  {Giavalisco}, {Gronwall}, {Hanish}, {Lee}, {Mehta}, {de Mello},
  {Ravindranath}, {Ryan}, {Scarlata}, {Siana}, {Soto}, \&
  {Voyer}}]{RafelskiM_15a}
{Rafelski}, M., {Teplitz}, H.~I., {Gardner}, J.~P., {et~al.} 2015, \aj, 150, 31

\bibitem[{{Read} {et~al.}(2016{\natexlab{a}}){Read}, {Agertz}, \&
  {Collins}}]{ReadJ_16a}
{Read}, J.~I., {Agertz}, O., \& {Collins}, M.~L.~M. 2016{\natexlab{a}}, \mnras,
  459, 2573

\bibitem[{Read \& Gilmore(2005)}]{ReadJ_05b}
Read, J.~I. \& Gilmore, G. 2005, \mnras, 356, 107

\bibitem[{Read {et~al.}(2006)Read, Goerdt, Moore, Pontzen, Stadel, \&
  Lake}]{ReadJ_06a}
Read, J.~I., Goerdt, T., Moore, B., {et~al.} 2006, \mnras, 373, 1451

\bibitem[{{Read} {et~al.}(2016{\natexlab{b}}){Read}, {Iorio}, {Agertz}, \&
  {Fraternali}}]{ReadJ_16b}
{Read}, J.~I., {Iorio}, G., {Agertz}, O., \& {Fraternali}, F.
  2016{\natexlab{b}}, \mnras, 462, 3628

\bibitem[{{Read} {et~al.}(2017){Read}, {Iorio}, {Agertz}, \&
  {Fraternali}}]{ReadJ_17a}
{Read}, J.~I., {Iorio}, G., {Agertz}, O., \& {Fraternali}, F. 2017, \mnras,
  467, 2019

\bibitem[{Read {et~al.}(2018)Read, Walker, \& Steger}]{ReadJ_18a}
Read, J.~I., Walker, M.~G., \& Steger, P. 2018, \mnras, 481, 860

\bibitem[{Read {et~al.}(2019)Read, Walker, \& Steger}]{ReadJ_19a}
Read, J.~I., Walker, M.~G., \& Steger, P. 2019, \mnras, 484, 1401

\bibitem[{Rix {et~al.}(1997)Rix, Guhathakurta, Colless, \& Ing}]{RixH_97a}
Rix, H.-W., Guhathakurta, P., Colless, M., \& Ing, K. 1997, \mnras, 285, 779

\bibitem[{{Rizzo} {et~al.}(2021){Rizzo}, {Vegetti}, {Fraternali}, {Stacey}, \&
  {Powell}}]{RizzoF_21a}
{Rizzo}, F., {Vegetti}, S., {Fraternali}, F., {Stacey}, H.~R., \& {Powell}, D.
  2021, \mnras, 507, 3952

\bibitem[{Rizzo {et~al.}(2020)Rizzo, Vegetti, Powell, Fraternali, McKean,
  Stacey, \& White}]{RizzoF_20a}
Rizzo, F., Vegetti, S., Powell, D., {et~al.} 2020, \nat, 584, 201

\bibitem[{Robert {et~al.}(2009)Robert, Chopin, \& Rousseau}]{RobertsC_09a}
Robert, C.~P., Chopin, N., \& Rousseau, J. 2009, Statistical Science, 24, 141

\bibitem[{{Romeo}(2020)}]{RomeoA_20a}
{Romeo}, A.~B. 2020, \mnras, 491, 4843

\bibitem[{{Romeo} {et~al.}(2020){Romeo}, {Agertz}, \& {Renaud}}]{RomeoA_20b}
{Romeo}, A.~B., {Agertz}, O., \& {Renaud}, F. 2020, \mnras, 499, 5656

\bibitem[{{Rubin} \& {Ford}(1970)}]{RubinV_70a}
{Rubin}, V.~C. \& {Ford}, W.~Kent, J. 1970, \apj, 159, 379

\bibitem[{{Salucci} \& {Burkert}(2000)}]{SalucciP_00a}
{Salucci}, P. \& {Burkert}, A. 2000, \apjl, 537, L9

\bibitem[{{Salucci} {et~al.}(2012){Salucci}, {Wilkinson}, {Walker}, {Gilmore},
  {Grebel}, {Koch}, {Frigerio Martins}, \& {Wyse}}]{SalucciP_12a}
{Salucci}, P., {Wilkinson}, M.~I., {Walker}, M.~G., {et~al.} 2012, \mnras, 420,
  2034

\bibitem[{{Schaye}(2001)}]{SchayeJ_01b}
{Schaye}, J. 2001, \apjl, 562, L95

\bibitem[{{Scott} {et~al.}(2013){Scott}, {Cappellari}, {Davies}, {Verdoes
  Kleijn}, {Bois}, {Alatalo}, {Blitz}, {Bournaud}, {Bureau}, {Crocker},
  {Davis}, {de Zeeuw}, {Duc}, {Emsellem}, {Khochfar}, {Krajnovi{\'c}},
  {Kuntschner}, {McDermid}, {Morganti}, {Naab}, {Oosterloo}, {Sarzi}, {Serra},
  {Weijmans}, \& {Young}}]{ScottN_13a}
{Scott}, N., {Cappellari}, M., {Davies}, R.~L., {et~al.} 2013, \mnras, 432,
  1894

\bibitem[{{Sereno} {et~al.}(2015){Sereno}, {Giocoli}, {Ettori}, \&
  {Moscardini}}]{SerenoM_15a}
{Sereno}, M., {Giocoli}, C., {Ettori}, S., \& {Moscardini}, L. 2015, \mnras,
  449, 2024

\bibitem[{{S{\'e}rsic}(1963)}]{SersicJ_63a}
{S{\'e}rsic}, J.~L. 1963, BAAA, 6, 41

\bibitem[{{Sharma} {et~al.}(2021){Sharma}, {Salucci}, {Harrison}, {van de Ven},
  \& {Lapi}}]{SharmaG_21a}
{Sharma}, G., {Salucci}, P., {Harrison}, C.~M., {van de Ven}, G., \& {Lapi}, A.
  2021, \mnras, 503, 1753

\bibitem[{Shi {et~al.}(2021)Shi, Zhang, Wang, Chen, Gu, Yu, \& Li}]{ShiY_21a}
Shi, Y., Zhang, Z.-Y., Wang, J., {et~al.} 2021, The Astrophysical Journal, 909,
  20

\bibitem[{{Sonnenfeld} {et~al.}(2012){Sonnenfeld}, {Treu}, {Gavazzi},
  {Marshall}, {Auger}, {Suyu}, {Koopmans}, \& {Bolton}}]{SonnenfeldA_12a}
{Sonnenfeld}, A., {Treu}, T., {Gavazzi}, R., {et~al.} 2012, \apj, 752, 163

\bibitem[{{Sonnenfeld} {et~al.}(2013){Sonnenfeld}, {Treu}, {Gavazzi}, {Suyu},
  {Marshall}, {Auger}, \& {Nipoti}}]{SonnenfieldA_13a}
{Sonnenfeld}, A., {Treu}, T., {Gavazzi}, R., {et~al.} 2013, \apj, 777, 98

\bibitem[{Sonnenfeld {et~al.}(2015)Sonnenfeld, Treu, Marshall, Suyu, Gavazzi,
  Auger, \& Nipoti}]{SonnenfeldA_15a}
Sonnenfeld, A., Treu, T., Marshall, P.~J., {et~al.} 2015, \apj, 800, 94

\bibitem[{{Spano} {et~al.}(2008){Spano}, {Marcelin}, {Amram}, {Carignan},
  {Epinat}, \& {Hernandez}}]{SpanoM_08a}
{Spano}, M., {Marcelin}, M., {Amram}, P., {et~al.} 2008, \mnras, 383, 297

\bibitem[{Spergel \& Steinhardt(2000)}]{SpergelD_00a}
Spergel, D.~N. \& Steinhardt, P.~J. 2000, Phys. Rev. Lett., 84, 3760

\bibitem[{{Springel} {et~al.}(2006){Springel}, {Frenk}, \&
  {White}}]{SpringelV_06a}
{Springel}, V., {Frenk}, C.~S., \& {White}, S.~D.~M. 2006, \nat, 440, 1137

\bibitem[{{Suyu} {et~al.}(2010){Suyu}, {Marshall}, {Auger}, {Hilbert},
  {Blandford}, {Koopmans}, {Fassnacht}, \& {Treu}}]{SuyuS_10a}
{Suyu}, S.~H., {Marshall}, P.~J., {Auger}, M.~W., {et~al.} 2010, \apj, 711, 201

\bibitem[{{Tacconi} {et~al.}(2018){Tacconi}, {Genzel}, {Saintonge}, {Combes},
  {Garc{\'\i}a-Burillo}, {Neri}, {Bolatto}, {Contini}, {F{\"o}rster Schreiber},
  {Lilly}, {Lutz}, {Wuyts}, {Accurso}, {Boissier}, {Boone}, {Bouch{\'e}},
  {Bournaud}, {Burkert}, {Carollo}, {Cooper}, {Cox}, {Feruglio}, {Freundlich},
  {Herrera-Camus}, {Juneau}, {Lippa}, {Naab}, {Renzini}, {Salome}, {Sternberg},
  {Tadaki}, {{\"U}bler}, {Walter}, {Weiner}, \& {Weiss}}]{TacconiL_18a}
{Tacconi}, L.~J., {Genzel}, R., {Saintonge}, A., {et~al.} 2018, \apj, 853, 179

\bibitem[{{Teyssier} {et~al.}(2013){Teyssier}, {Pontzen}, {Dubois}, \&
  {Read}}]{TeyssierR_13a}
{Teyssier}, R., {Pontzen}, A., {Dubois}, Y., \& {Read}, J.~I. 2013, \mnras,
  429, 3068

\bibitem[{{The Astropy Collaboration} {et~al.}(2018){The Astropy
  Collaboration}, {Price-Whelan}, {Sip{\H o}cz}, {G{\"u}nther}, {Lim},
  {Crawford}, {Conseil}, {Shupe}, {Craig}, {Dencheva}, {Ginsburg},
  {VanderPlas}, {Bradley}, {P{\'e}rez-Su{\'a}rez}, {de Val-Borro}, {Aldcroft},
  {Cruz}, {Robitaille}, {Tollerud}, {Ardelean}, {Babej}, {Bachetti}, {Bakanov},
  {Bamford}, {Barentsen}, {Barmby}, {Baumbach}, {Berry}, {Biscani}, {Boquien},
  {Bostroem}, {Bouma}, {Brammer}, {Bray}, {Breytenbach}, {Buddelmeijer},
  {Burke}, {Calderone}, {Cano Rodr{\'{\i}}guez}, {Cara}, {Cardoso},
  {Cheedella}, {Copin}, {Crichton}, {D{\'A}vella}, {Deil}, {Depagne},
  {Dietrich}, {Donath}, {Droettboom}, {Earl}, {Erben}, {Fabbro}, {Ferreira},
  {Finethy}, {Fox}, {Garrison}, {Gibbons}, {Goldstein}, {Gommers}, {Greco},
  {Greenfield}, {Groener}, {Grollier}, {Hagen}, {Hirst}, {Homeier}, {Horton},
  {Hosseinzadeh}, {Hu}, {Hunkeler}, {Ivezi{\'c}}, {Jain}, {Jenness}, {Kanarek},
  {Kendrew}, {Kern}, {Kerzendorf}, {Khvalko}, {King}, {Kirkby}, {Kulkarni},
  {Kumar}, {Lee}, {Lenz}, {Littlefair}, {Ma}, {Macleod}, {Mastropietro},
  {McCully}, {Montagnac}, {Morris}, {Mueller}, {Mumford}, {Muna}, {Murphy},
  {Nelson}, {Nguyen}, {Ninan}, {N{\"o}the}, {Ogaz}, {Oh}, {Parejko}, {Parley},
  {Pascual}, {Patil}, {Patil}, {Plunkett}, {Prochaska}, {Rastogi}, {Reddy
  Janga}, {Sabater}, {Sakurikar}, {Seifert}, {Sherbert}, {Sherwood-Taylor},
  {Shih}, {Sick}, {Silbiger}, {Singanamalla}, {Singer}, {Sladen}, {Sooley},
  {Sornarajah}, {Streicher}, {Teuben}, {Thomas}, {Tremblay}, {Turner},
  {Terr{\'o}n}, {van Kerkwijk}, {de la Vega}, {Watkins}, {Weaver}, {Whitmore},
  {Woillez}, \& {Zabalza}}]{astropy2018}
{The Astropy Collaboration}, {Price-Whelan}, A.~M., {Sip{\H o}cz}, B.~M.,
  {et~al.} 2018, ArXiv e-prints [\eprint[arXiv]{1801.02634}]

\bibitem[{{Tiley} {et~al.}(2016){Tiley}, {Stott}, {Swinbank}, {Bureau},
  {Harrison}, {Bower}, {Johnson}, {Bunker}, {Jarvis}, {Magdis}, {Sharples},
  {Smail}, {Sobral}, \& {Best}}]{TileyA_16a}
{Tiley}, A.~L., {Stott}, J.~P., {Swinbank}, A.~M., {et~al.} 2016, \mnras, 460,
  103

\bibitem[{Tiley {et~al.}(2019)Tiley, Swinbank, Harrison, Smail, Turner,
  Schaller, Stott, Sobral, Theuns, Sharples, Gillman, Bower, Bunker, Best,
  Richard, Bacon, Bureau, Cirasuolo, \& Magdis}]{TileyA_19b}
Tiley, A.~L., Swinbank, A.~M., Harrison, C.~M., {et~al.} 2019, \mnras, 485, 934

\bibitem[{Tollet {et~al.}(2016)Tollet, Macci, Dutton, Stinson, Wang, Penzo,
  Gutcke, Buck, Kang, Brook, Cintio, Keller, \& Wadsley}]{TolletE_16a}
Tollet, E., Macci, A.~V., Dutton, A.~A., {et~al.} 2016, \mnras, 456, 3542

\bibitem[{{Toomre}(1964)}]{ToomreA_64a}
{Toomre}, A. 1964, \apj, 139, 1217

\bibitem[{{Tully} \& {Fisher}(1977)}]{TullyB_77a}
{Tully}, R.~B. \& {Fisher}, J.~R. 1977, \aap, 54, 661

\bibitem[{{\"{U}}bler {et~al.}(2017){\"{U}}bler, {F{\"{o}}rster Schreiber},
  Genzel, Wisnioski, Wuyts, Lang, Naab, Burkert, van Dokkum, Tacconi, Wilman,
  Fossati, Mendel, Beifiori, Belli, Bender, Brammer, Chan, Davies, Fabricius,
  Galametz, Lutz, Momcheva, Nelson, Saglia, Seitz, \& Tadaki}]{UblerH_17a}
{\"{U}}bler, H., {F{\"{o}}rster Schreiber}, N.~M., Genzel, R., {et~al.} 2017,
  \apj, 842, 121

\bibitem[{\"{U}bler {et~al.}(2020)\"{U}bler, Genel, Sternberg, Genzel, Price,
  Schreiber, Shimizu, Pillepich, Nelson, Burkert, Davies, Hernquist, Lang,
  Lutz, Pakmor, \& Tacconi}]{UblerH_20a}
\"{U}bler, H., Genel, S., Sternberg, A., {et~al.} 2020, \mnras, staa3464

\bibitem[{{\"{U}}bler {et~al.}(2019){\"{U}}bler, Genzel, Wisnioski, Schreiber,
  Shimizu, Price, Tacconi, Belli, Wilman, Fossati, Mendel, Davies, Beifiori,
  Bender, Brammer, Burkert, Chan, Davies, Fabricius, Galametz, Herrera-Camus,
  Lang, Lutz, Momcheva, Naab, Nelson, Saglia, Tadaki, van Dokkum, \&
  Wuyts}]{UblerH_19a}
{\"{U}}bler, H., Genzel, R., Wisnioski, E., {et~al.} 2019, \apj, 880, 48

\bibitem[{{Vale} \& {Ostriker}(2004)}]{ValeA_04a}
{Vale}, A. \& {Ostriker}, J.~P. 2004, \mnras, 353, 189

\bibitem[{van~den Bosch {et~al.}(2000)van~den Bosch, Robertson, DAlcanton, \&
  de~Blok}]{VandenBoschF_00a}
van~den Bosch, F.~C., Robertson, B.~E., DAlcanton, J., \& de~Blok, W. J.~G.
  2000, ApJ, 119, 1579

\bibitem[{{Van Der Walt} {et~al.}(2011){Van Der Walt}, {Colbert}, \&
  {Varoquaux}}]{numpy}
{Van Der Walt}, S., {Colbert}, S.~C., \& {Varoquaux}, G. 2011, Computing in
  Science and Engineering, 13, 22

\bibitem[{{van der Wel} {et~al.}(2014){van der Wel}, {Chang}, {Bell}, {Holden},
  {Ferguson}, {Giavalisco}, {Rix}, {Skelton}, {Whitaker}, {Momcheva},
  {Brammer}, {Kassin}, {Martig}, {Dekel}, {Ceverino}, {Koo}, {Mozena}, {van
  Dokkum}, {Franx}, {Faber}, \& {Primack}}]{vanderWelA_14a}
{van der Wel}, A., {Chang}, Y.-Y., {Bell}, E.~F., {et~al.} 2014, \apjl, 792, L6

\bibitem[{{Vogelsberger} {et~al.}(2013){Vogelsberger}, {Genel}, {Sijacki},
  {Torrey}, {Springel}, \& {Hernquist}}]{VogelsbergerM_13a}
{Vogelsberger}, M., {Genel}, S., {Sijacki}, D., {et~al.} 2013, \mnras, 436,
  3031

\bibitem[{{Vogelsberger} \& {Zavala}(2013)}]{VogelsbergerM_13b}
{Vogelsberger}, M. \& {Zavala}, J. 2013, \mnras, 430, 1722

\bibitem[{Wang {et~al.}(2020)Wang, Catinella, Saintonge, Pan, Serra, \&
  Shao}]{WangJ_20a}
Wang, J., Catinella, B., Saintonge, A., {et~al.} 2020, \apj, 890, 63

\bibitem[{{Wang} {et~al.}(2016){Wang}, {Koribalski}, {Serra}, {van der Hulst},
  {Roychowdhury}, {Kamphuis}, \& {Chengalur}}]{WangJ_16a}
{Wang}, J., {Koribalski}, B.~S., {Serra}, P., {et~al.} 2016, \mnras, 460, 2143

\bibitem[{{Wasserman} {et~al.}(2018){Wasserman}, {Romanowsky}, {Brodie}, {van
  Dokkum}, {Conroy}, {Villaume}, {Forbes}, {Strader}, {Alabi}, \&
  {Bellstedt}}]{WassermanA_18a}
{Wasserman}, A., {Romanowsky}, A.~J., {Brodie}, J., {et~al.} 2018, \apj, 863,
  130

\bibitem[{{Wechsler} {et~al.}(2002){Wechsler}, {Bullock}, {Primack},
  {Kravtsov}, \& {Dekel}}]{WechslerR_02a}
{Wechsler}, R.~H., {Bullock}, J.~S., {Primack}, J.~R., {Kravtsov}, A.~V., \&
  {Dekel}, A. 2002, \apj, 568, 52

\bibitem[{{Weijmans} {et~al.}(2008){Weijmans}, {Krajnovi{\'c}}, {van de Ven},
  {Oosterloo}, {Morganti}, \& {de Zeeuw}}]{WeijmansAM_08a}
{Weijmans}, A.-M., {Krajnovi{\'c}}, D., {van de Ven}, G., {et~al.} 2008,
  \mnras, 383, 1343

\bibitem[{Weinberg(1978)}]{WeinbergS_78a}
Weinberg, S. 1978, Phys. Rev. Lett., 40, 223

\bibitem[{{Willmer}(2018)}]{WillmerC_18a}
{Willmer}, C. N.~A. 2018, \apjs, 236, 47

\bibitem[{Wilman {et~al.}(2020)Wilman, Fossati, Mendel, Saglia, Wisnioski,
  Wuyts, Schreiber, Beifiori, Bender, Belli, {\"{U}}bler, Lang, Chan, Davies,
  Nelson, Genzel, Tacconi, Galametz, Davies, Lutz, Price, Burkert, Tadaki,
  Herrera-Camus, Brammer, Momcheva, \& van Dokkum}]{WilmanD_20a}
Wilman, D.~J., Fossati, M., Mendel, J.~T., {et~al.} 2020, \apj, 892, 1

\bibitem[{{Wisnioski} {et~al.}(2015){Wisnioski}, {F{\"o}rster Schreiber},
  {Wuyts}, {Wuyts}, {Bandara}, {Wilman}, {Genzel}, {Bender}, {Davies},
  {Fossati}, {Lang}, {Mendel}, {Beifiori}, {Brammer}, {Chan}, {Fabricius},
  {Fudamoto}, {Kulkarni}, {Kurk}, {Lutz}, {Nelson}, {Momcheva}, {Rosario},
  {Saglia}, {Seitz}, {Tacconi}, \& {van Dokkum}}]{WisnioskiE_15a}
{Wisnioski}, E., {F{\"o}rster Schreiber}, N.~M., {Wuyts}, S., {et~al.} 2015,
  \apj, 799, 209

\bibitem[{Wuyts {et~al.}(2016)Wuyts, Schreiber, Wisnioski, Genzel, Burkert,
  Bandara, Beifiori, Belli, Bender, Brammer, Chan, Davies, Fossati, Galametz,
  Kulkarni, Lang, Lutz, Mendel, Momcheva, Naab, Nelson, Saglia, Seitz, Tacconi,
  Tadaki, {\"{U}}bler, van Dokkum, Wilman, \& Wuyts}]{WuytsS_16a}
Wuyts, S., Schreiber, N. M.~F., Wisnioski, E., {et~al.} 2016, \apj, 831, 149

\bibitem[{{Zabl} {et~al.}(2019){Zabl}, {Bouch{\'e}}, {Schroetter}, {Wendt},
  {Finley}, {Schaye}, {Conseil}, {Contini}, {Marino}, {Mitchell}, {Muzahid},
  {Pezzulli}, \& {Wisotzki}}]{ZablJ_19a}
{Zabl}, J., {Bouch{\'e}}, N.~F., {Schroetter}, I., {et~al.} 2019, \mnras, 485,
  1961

\bibitem[{{Zhao}(1996)}]{ZhaoH_96a}
{Zhao}, H. 1996, \mnras, 278, 488

\bibitem[{{Zoutendijk} {et~al.}(2021){Zoutendijk}, {Brinchmann}, {Bouch{\'e}},
  {den Brok}, {Krajnovi{\'c}}, {Kuijken}, {Maseda}, \&
  {Schaye}}]{ZoutendijkB_21a}
{Zoutendijk}, S.~L., {Brinchmann}, J., {Bouch{\'e}}, N.~F., {et~al.} 2021,
  \aap, in press [\eprint{2101.00253}]

\bibitem[{{Zwicky}(1933)}]{ZwickyF_33a}
{Zwicky}, F. 1933, Helvetica Physica Acta, 6, 110

\end{thebibliography}
%
   
   \appendix

   \section{Asymmetric drift}
   \label{appendix:asym}

Generally speaking, $F_{\rm g}=GM{\rm d}m/r^2$ is   balanced by the centripedal force $F_c$ and the outward force $F_p(=-\frac{{\rm d}P}{{\rm d}r}{\rm d}r\,{\rm d}A$) from the pressure gradient such that $F_{g}=\frac{v_\perp^2}{r}+F_p$. Using ${\rm d}m=\rho(r) {\rm d}r {\rm d}A$, the gravitational potential $F_{\rm g}$ is balanced with 
\begin{eqnarray}
v_{\rm c}^2&\equiv&\frac{GM}{r}={v_\perp^2}-r\frac{{\rm d}P}{{\rm d}r}{\rm d}r \,{\rm d}A\frac{1}{{\rm d}m}\nonumber\\
v_{\rm c}^2&=& {v_\perp^2}-\frac{1}{\rho}\frac{{\rm d} P}{{\rm d} \ln r}\nonumber\\
v_{\rm c}^2&=&  {v_\perp^2}-\sigma_{ r}^2(r)\frac{{\rm d}\ln P}{{\rm d}\ln r} \qquad \hbox{using  $P=\rho\sigma_{ r}^2$} \label{eq:adrift:gen}
\end{eqnarray}
where $\rho$ is the gas density and $\sigma_r$ is the gas dispersion in the radial direction. For the ISM gas, it is often assumed that the dispersion is isotropic, that is $\sigma_r=\sigma_\perp$.

This pressure correction for rotation curves (often referred to as asymmetric drift [AD] correction) is at $z=0$ larger for the stellar component than for the gas component \citep[e.g.][]{MartinssonT_13b}, but at high-redshifts, the gas velocity dispersion is larger \citep[e.g.][]{GenzelR_08a,UblerH_19a} and this correction becomes important
\citep[e.g.][]{BurkertA_10a,ForsterSchreiberN_20a}.

 In most general terms \Eq{eq:adrift:gen} can be expanded as:
\begin{eqnarray}
v_{\rm c}^2&=& {v_\perp^2}-\sigma_{ r}^2(r) \left[2\frac{{\rm d} \ln \sigma_{\rm r}}{{\rm d} \ln r}+\frac{{\rm d}\ln \rho}{{\rm d}\ln r} \right]  
\end{eqnarray}
as in \citet{DalcantonJ_10a}.

For a density profile $\rho=\Sigma/h_z$, the most general expression is
\begin{eqnarray}
v_{\rm c}^2&=& {v_\perp^2}-\sigma_{ r}^2(r) \left[2\frac{{\rm d} \ln \sigma_{ r}}{{\rm d} \ln r}+\frac{{\rm d}\ln\Sigma}{{\rm d}\ln r}-\frac{{\rm d} \ln h_z}{{\rm d} \ln r}\right] \label{eq:AD:Meurer}
\end{eqnarray}
as in \citet{MeurerG_96a}.

Assuming a constant disk thickness $h_z$, that is $\rho\propto\Sigma$, and a constant dispersion profile $\sigma_{\rm r}$, one has
\begin{eqnarray}
v_{\rm c}^2 &=&v_\perp^2-\sigma_r^2 \left[ \frac{{\rm d}\ln \Sigma}{{\rm d}\ln r}  \right] \label{eq:AD:Sigma}
\end{eqnarray}
 which becomes
\begin{eqnarray}
&=&{v_\perp^2}+ \sigma_{ r}^2\left(\frac{r}{r_d}\right) \nonumber
\end{eqnarray}
for exponential disks with $\Sigma\propto \exp(-r/r_d)$.
 
This is close to what one gets for  a turbulent disk with an exponential surface density $\Sigma$, where the ISM pressue $P_{\rm turb}$
 is found to follow $P_{\rm turb}\propto\Sigma_{\rm sfr}^{0.66}$ \citep{DalcantonJ_10a}, and using the \citet{KennicuttR_98a} relation ($\Sigma_{\rm sfr}=\Sigma_{\rm gas}^{1.4}$), one has 
\begin{eqnarray}
v_{\rm c}^2 &=&{v_\perp^2}+0.92 \sigma_{\rm r}^2\left(\frac{r}{r_d}\right)\label{eq:AD:Dalcanton}
\end{eqnarray}
 as Eq.17 in \citet{DalcantonJ_10a}.
 
 \citet{BurkertA_10a} used the hydro-static equilibrium condition for $h_z$, finding that the disk thickness $h_z\propto \exp(r/r_d)$, that is a flaring disk thickness, leading to
\begin{eqnarray}
v_{\rm c}^2&=& {v_\perp^2}-\sigma_r^2\left(\frac{{\rm d}\ln \Sigma}{{\rm d}\ln r} - \frac{{\rm d}\ln h_z}{{\rm d}\ln r} \right)\nonumber\\
&=&v_{\perp}^2+2\sigma^2 \left(\frac{r}{r_d}\right). \label{eq:AD:Genzel}
\end{eqnarray}


Similarly, \citet{WeijmansAM_08a} expanded the formalism of \citet{BinneyJ_87a}, taking into account
the full Jeans equation for spheroids and thin disks.  Neglecting the last term of Eq.A17 of \citet{WeijmansAM_08a},   \citet{SharmaG_21a} used  the AD correction  :
\begin{eqnarray}
v^2_c&=&v_\perp^2-\sigma_r^2\left[\frac{{\rm d}\ln \Sigma}{{\rm d}\ln r}+\frac{{\rm d}\ln\sigma^2_r}{{\rm d}\ln r}+\frac{1}{2}(1-\alpha_r)\right]\nonumber\\
&=&v_\perp^2-\sigma_r^2\left[\frac{{\rm d}\ln \Sigma}{{\rm d}\ln r}+\frac{1}{2}(1-\alpha_r)\right]\label{eq:AD:Weijmans}
\end{eqnarray}
where $\alpha_r=\frac{{\rm d}\ln v_\perp}{{\rm d}\ln r}$is the slope of the velocity profile,  and the second equation applies for a constant $\sigma_r$ profile. For an exponential disk, this becomes
 \begin{eqnarray}
v_c^2&=&v_\perp^2+\sigma_r^2\left(\frac{r}{r_d}-\frac{1}{2}(1-\alpha_r)\right) . \nonumber
\end{eqnarray}

\citet{PostiL_18b} used \Eq{eq:AD:Sigma} with a constant scale height ad an exponentially declining dispersion profile $\sigma_r(r)\equiv \sigma_0\exp(-r/2r_d)$ and found with \Eq{eq:AD:Meurer}
\begin{eqnarray}
v_c^2&=&v_\perp^2+\sigma_0^2\frac{3 r}{2 r_d}\exp(-r/2 r_d)
\label{eq:AD:Martinsson}
\end{eqnarray}
for exponential disks.

 In    \Fig{fig:asym:comp}, we compare the various AD   prescriptions such as   the `Sigma' (\Eq{eq:AD:Sigma}), 
  the \citet{DalcantonJ_10a} (\Eq{eq:AD:Dalcanton}), 
  the \citet{BurkertA_10a} (\Eq{eq:AD:Genzel}),
  the \citet{WeijmansAM_08a} (\Eq{eq:AD:Weijmans}), 
  and the \citet{PostiL_18b} (\Eq{eq:AD:Martinsson}) prescription.
   
   \begin{figure}
   \centering
   \includegraphics[width=8cm]{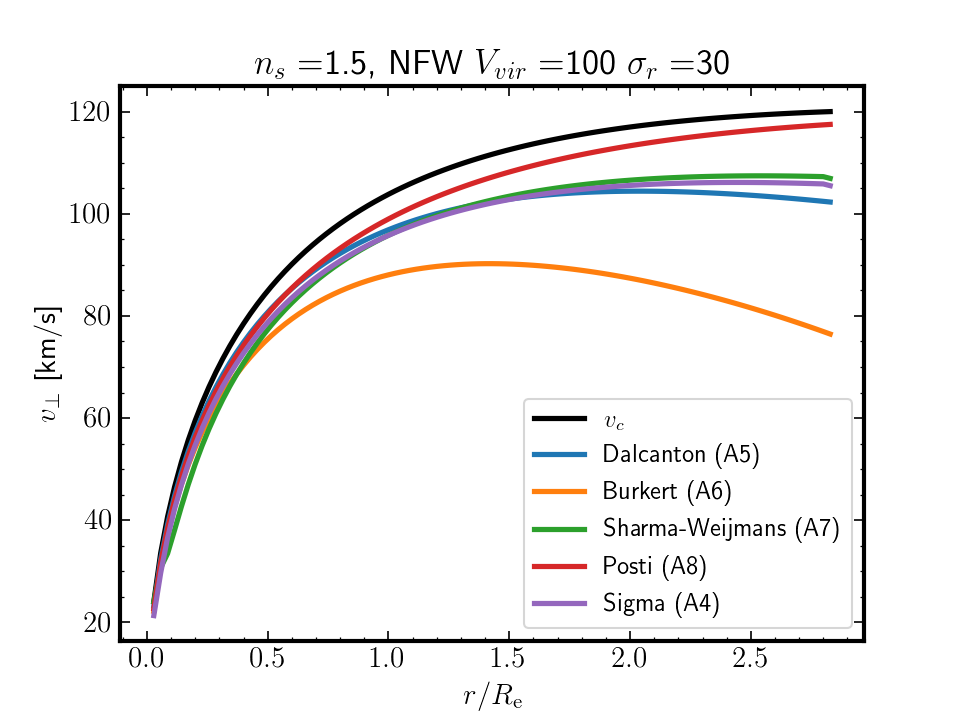}
   \caption{The rotation velocity $v_{\perp}$ with various asymmetric drift correction for a galaxy with a \sersic\ index $n_s=1.5$, a NFW profile with $V_{\rm vir}=100$ km/s, $c_{\rm vir}=10$, and a velocity dispersion of $\sigma=30$ km/s. }
   \label{fig:asym:comp}
   \end{figure}
   
   \section{Parameter degeneracies}

\Fig{fig:corner:MMvir}  shows the potential correlations between $M_\star$ and $M_{\rm vir}$ from
the \galpak\ DC14 fits.   For ID937, which is better fitted by a NFW profile, the fit was restricted to the SED-derived stellar mass $M_\star$. One sees that, generally, there seems to be a good agreement between the \galpak-derived $M_\star$ with the SED-derived ones, except for ID943, ID919.
The halo mass for ID1002  and the concentration for ID15 are poorly constrained.

\Fig{fig:corner:cM} shows the potential correlations between $c_{\rm vir}$ and $M_{\rm vir}$
from the \galpak\ fits. As in the left panel, the triangles represent the best-fit values.
This shows figure shows that the halo concentration is well determined except for ID15 and ID943.

\begin{figure*}
\centering
\includegraphics[width=0.85\textwidth]{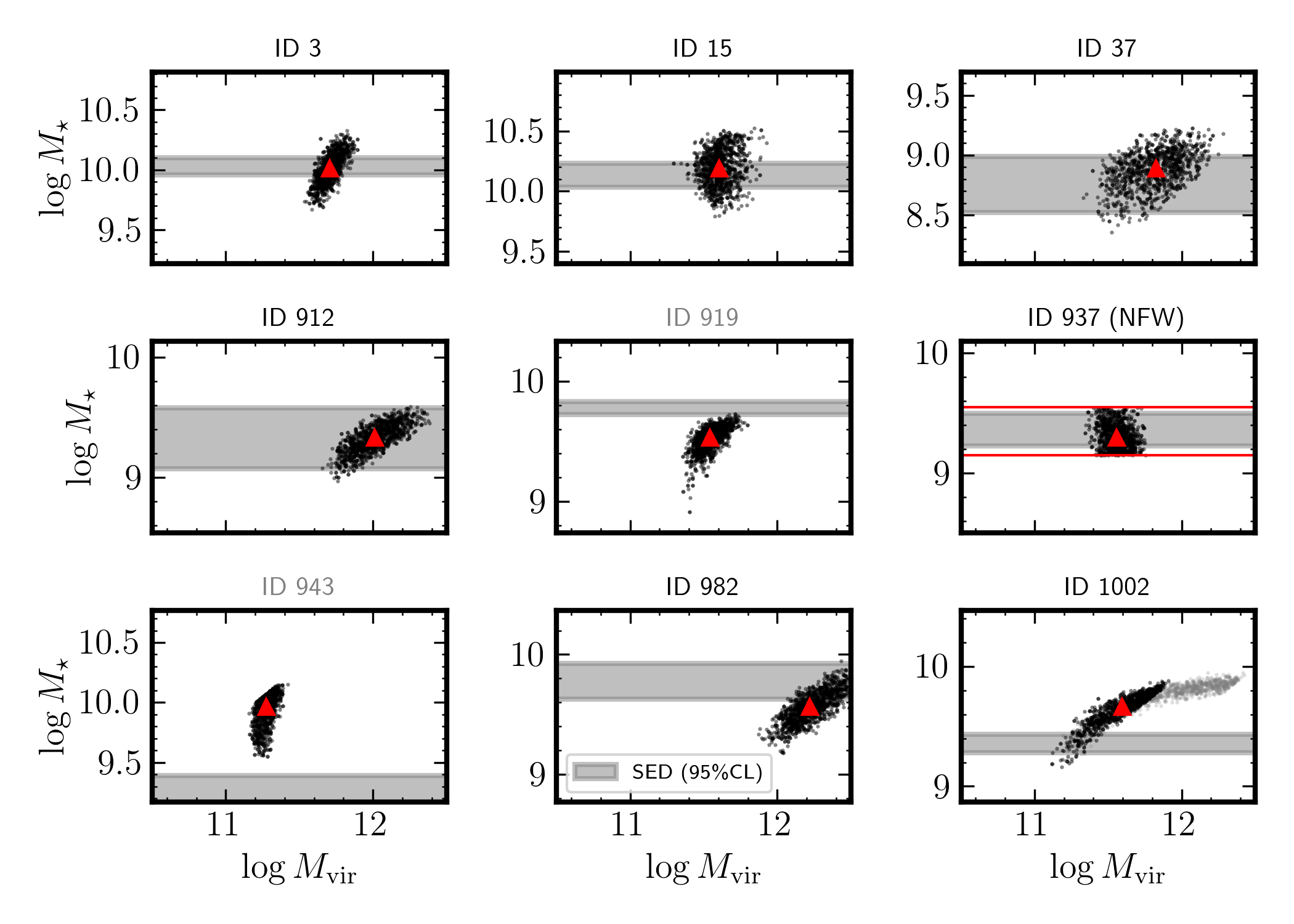}
\caption{The $M_\star-M_{\rm vir}$ potential correlations from the \galpak\ fits, derived from the $\log X-V_{\rm vir}$ fits.  The horizontal gray band represents the SED-based $M_\star$ and its $2\sigma$ uncertainty. The triangles represent the best-fit values. For ID937, the fit was restricted to the SED $M_\star$  value.
There is  good agreement between the \galpak-derived $M_\star$ with the SED-derived ones, except for ID943, ID919.
}
\label{fig:corner}\label{fig:corner:MMvir}
\end{figure*}   

\begin{figure*}
\centering
\includegraphics[width=0.85\textwidth]{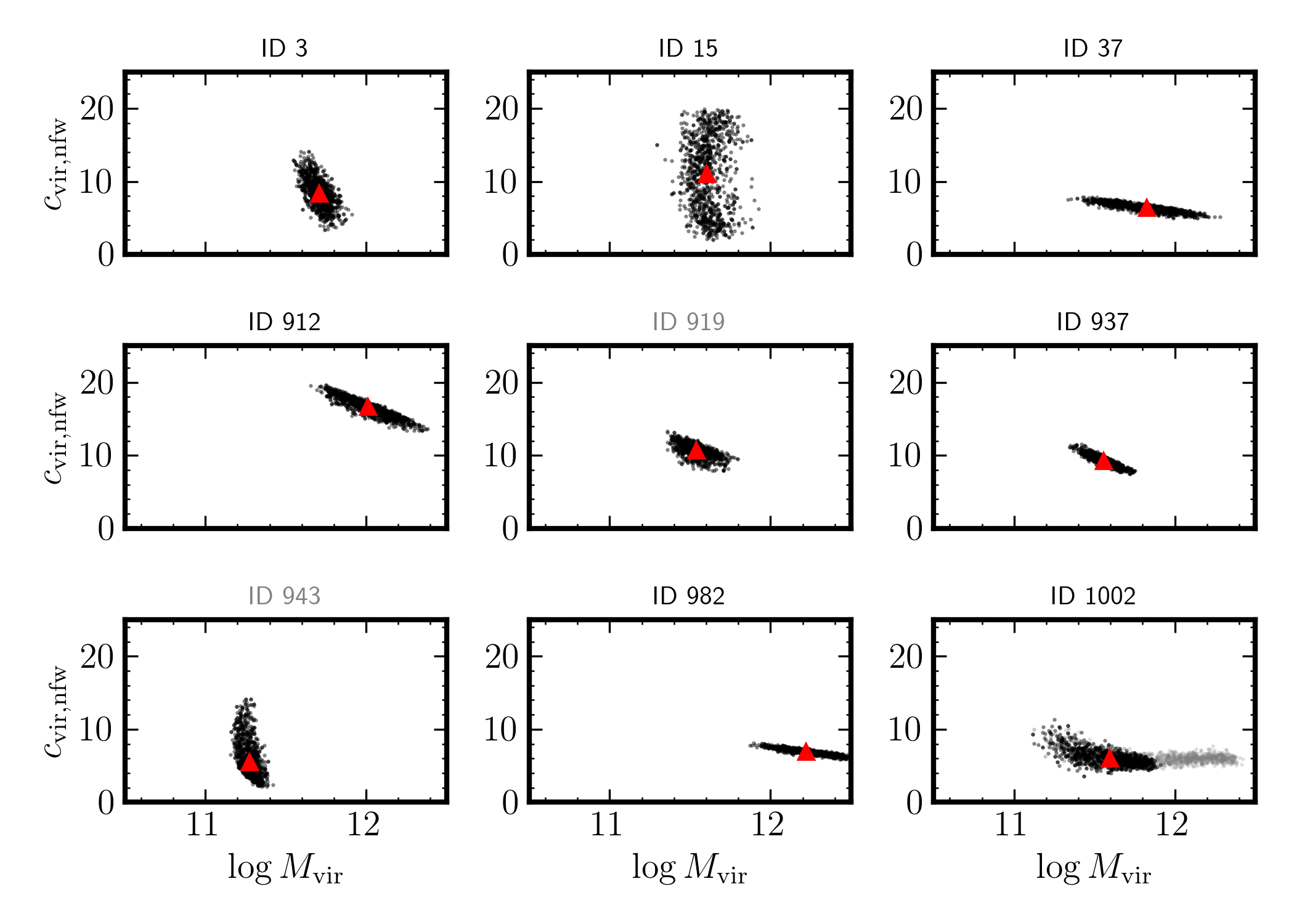}
\caption{The $c_{\rm vir}-M_{\rm vir}$ correlations from the MCMC chain. The triangles represent the best-fit values. One sees that the halo mass for ID1002  and the concentration for IDs15, 943 are poorly constrained.}
\label{fig:corner}\label{fig:corner:cM}
\end{figure*}

   \section{Relaxing constraints on $\alpha,\beta,\gamma$}

   In \Fig{fig:DC14:Zhao}, we show the derived $\alpha,\beta,\gamma$ parameters  using \citet{ZhaoH_96a} DM profile with the 3 parameters (inner \&\ outer slopes, transition sharpness) free. The solid, dotted, dot-dashed lines represent the inner ($\gamma$), outer ($\beta$) slopes of the DM profile and the sharpness of the transition $\alpha$ as a function of $\log X=M_\star/M_{\rm vir}$ from the DC14 (their Eq. 3). Here, the ratio $M_\star/M_{\rm vir}$ is degenerate with the profile parameters, so we used the SED stellar mass $M_{\star,\rm SED}$ as prior.
   
   This figure shows that the DC14 model is not far from the actual DM profiles and indicates that the \citet{DekelA_17a} profile with $\alpha,\beta,\gamma=(0.5,3.5,\gamma)$ \citep{FreundlichJ_20b} would be disfavored.
   

 \begin{figure*}
 \centering
\includegraphics[width=0.32\textwidth]{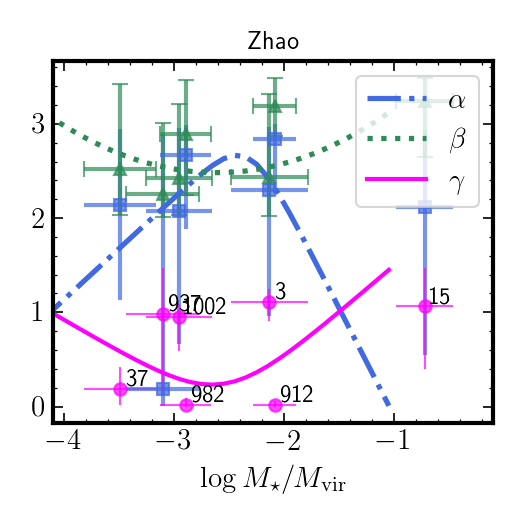}
\includegraphics[width=0.32\textwidth]{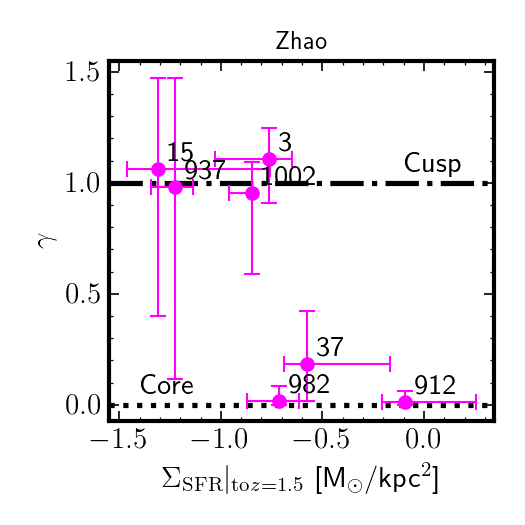}
\includegraphics[width=0.32\textwidth]{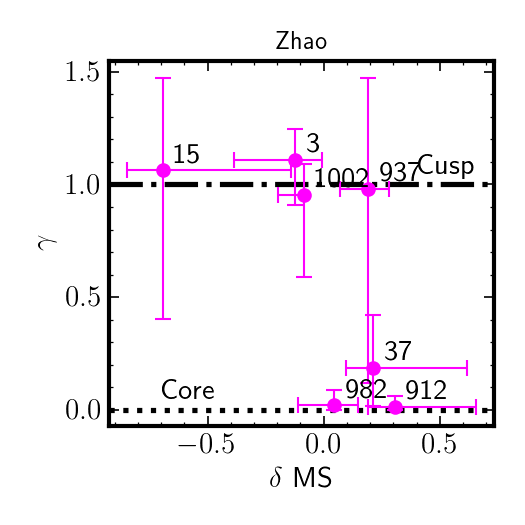}
\caption{Same as \Fig{fig:slope:best}, but with the `Zhao'  DM profiles with free $\alpha,\beta,\gamma$.}\label{fig:slope:Zhao}\label{fig:DC14:Zhao}
\end{figure*}


   

%
%
\end{document}